%% file: main.tex
\documentclass[preprint]{cimento}
\usepackage{graphicx}
\usepackage{amssymb}
\usepackage{amsmath}
\usepackage{url}
\usepackage{rotating}
\usepackage{multirow}
\usepackage{rotfloat}

\include{defs}

\begin{document}



\title{The search for neutrinoless double beta decay}

\author{J.J.~G\'omez-Cadenas\from{ins:ific},
J.~Mart\'in-Albo\from{ins:ific},
M.~Mezzetto\from{ins:infn}, 	
F.~Monrabal\from{ins:ific} \atque
M.~Sorel\from{ins:ific}\thanks{Corresponding author: sorel@ific.uv.es}}

\instlist{
\inst{ins:ific} Instituto de F\'isica Corpuscular (IFIC), CSIC \& Univ.\ de Valencia, Valencia, Spain
\inst{ins:infn} Istituto Nazionale di Fisica Nucleare (INFN), Sezione di Padova, Padova, Italy}

\PACSes{
\PACSit{23.40.-s}{$\beta$ decay; double $\beta$ decay; electron and muon capture}
\PACSit{14.60.Pq}{Neutrino mass and mixing}
}

\maketitle


\begin{abstract}
In the last two decades the search for neutrinoless double beta decay has evolved into one of the highest priorities for understanding neutrinos and the origin of mass. The main reason for this paradigm shift has been the discovery of neutrino oscillations, which clearly established the existence of massive neutrinos. An additional motivation for conducting such searches comes from the existence of an unconfirmed, but not refuted, claim of evidence for neutrinoless double decay in $^{76}\text{Ge}$. As a consequence, a new generation of experiments, employing different detection techniques and $\beta\beta$ isotopes, is being actively promoted by experimental groups across the world. In addition, nuclear theorists are making remarkable progress in the calculation of the neutrinoless double beta decay nuclear matrix elements, thus eliminating a substantial part of the theoretical uncertainties affecting the particle physics interpretation of this process. In this report, we review the main aspects of the double beta decay process and some of the most relevant experiments. The picture that emerges is one where searching for neutrinoless double beta decay is recognized to have both far-reaching theoretical implications and promising prospects for experimental observation in the near future.

\end{abstract}

\tableofcontents


\section{Introduction} \label{sec:intro} 
\input{src/intro.tex}

\section{Massive neutrinos} \label{sec:massivenu}
\input{src/massivenus.tex}

\section{Neutrinoless double beta decay} \label{sec:bb0nu}
\input{src/bb0nu.tex}

\section{Calculating nuclear matrix elements} \label{sec:nme}
\input{src/nme.tex}

\section{Ingredients for the ultimate \bbonu\ experiment} \label{sec:ingredients}
\input{src/ingredients.tex}

\section{A selection of new-generation experimental proposals} \label{sec:experiments}
\input{src/exps.tex}

\section{Conclusions}
\input{src/conclusions.tex}


\acknowledgments
The authors acknowledge support by the Spanish Ministerio de Ciencia e Innovaci\'on (MICINN) under grants CONSOLIDER-INGENIO 2010 CSD2008-0037 (CUP) and FPA2009-13697-C04-04. We also thank Alessandro Bettini and Martin Hirsch for useful discussions and comments on the manuscript.


\bibliography{references}
\bibliographystyle{varenna}

\end{document}

%% file: defs.tex
\newcommand{\bb}{\ensuremath{\beta\beta}}
\newcommand{\bbonu}{\ensuremath{\beta\beta0\nu}}
\newcommand{\bbtnu}{\ensuremath{\beta\beta2\nu}}

\newcommand{\Qbb}{\ensuremath{Q_{\beta\beta}}}

\newcommand{\mbb}{\ensuremath{m_{\bb}}}
\newcommand{\Tonu}{\ensuremath{T^{0\nu}_{1/2}}}
\newcommand{\Gonu}{\ensuremath{G^{0\nu}(Q,Z)}}

\newcommand{\CA}{\ensuremath{^{48}\mathrm{Ca}}}
\newcommand{\GE}{\ensuremath{^{76}\mathrm{Ge}}}
\newcommand{\SE}{\ensuremath{^{82}\mathrm{Se}}}
\newcommand{\MO}{\ensuremath{^{100}\mathrm{Mo}}}
\newcommand{\CD}{\ensuremath{^{116}\mathrm{Cd}}}
\newcommand{\TE}{\ensuremath{^{130}\mathrm{Te}}}
\newcommand{\XE}{\ensuremath{^{136}\mathrm{Xe}}}
\newcommand{\ND}{\ensuremath{^{150}\mathrm{Nd}}}

\newcommand{\URANIUM}{\ensuremath{^{238}\mathrm{U}}}
\newcommand{\THORIUM}{\ensuremath{^{232}\mathrm{Th}}}
\newcommand{\TL}{\ensuremath{^{208}\mathrm{Tl}}}
\newcommand{\BI}{\ensuremath{^{214}\mathrm{Bi}}}

\newcommand{\ckky}{counts/(keV~\ensuremath{\cdot}~kg~\ensuremath{\cdot}~year)}

\newcommand{\ckkbby}{counts/(keV~\ensuremath{\cdot}~kg\ensuremath{_{\beta\beta}}~\ensuremath{\cdot}~year)}

\newcommand{\Mbb}{\ensuremath{M_{\beta\beta}}}

\newcommand{\kgbb}{kg\ensuremath{_{\beta\beta}}}

%% file: src/intro.tex
Neutrinoless double beta decay (\bbonu) is a hypothetical nuclear transition in which two neutrons undergo $\beta$-decay simultaneously and without the emission of neutrinos. If realized in Nature, this transition would be extremely rare: the most constraining lower bound on the neutrinoless double beta decay half-life, in \GE , is \mbox{$T^{0\nu}_{1/2} > 1.9 \times 10^{25}$} years (at 90\% confidence level, from \cite{KlapdorKleingrothaus:2000sn}). The importance of \bbonu\ searches goes beyond its intrinsic interest, as it is the only practical way to reveal experimentally that neutrinos are Majorana particles. If $\nu$ is a field describing a neutrino, stating that the neutrino is a Majorana particle is equivalent to saying that the charge-conjugated field --- that is, a field with all charges reversed --- also describes the same particle: $\nu=\nu^c$. If such Majorana condition is not fulfilled, we speak instead of Dirac neutrinos. 

The theoretical implications of experimentally establishing \bbonu\ would be profound. In a broad sense, Majorana neutrinos would constitute a new form of matter, given that no Majorana fermions have been observed so far. Also, \bbonu\ observation would prove that total lepton number is not conserved in physical phenomena, a fact that could be linked to the cosmic asymmetry between matter and antimatter.  Finally, Majorana neutrinos would mean that a new physics scale must exist and is accessible in an indirect way through neutrino masses. 

In addition to theoretical prejudice in favor of Majorana neutrinos, there are other reasons to hope that experimental observation of \bbonu\ is at hand. Neutrinos are now known to be massive particles, thanks to neutrino oscillation experiments. If \bbonu\ is mediated by the standard light Majorana neutrino exchange mechanism, a non-zero neutrino mass would almost certainly translate into a non-zero \bbonu\ rate. While neutrino oscillation phenomenology is not enough \emph{per se} to provide a firm prediction for what such a rate should be, it does give us hope that a sufficiently fast one to be observable may be realized in Nature. Furthermore, \bbonu\ may have been observed \emph{already}: there is an extremely intriguing, albeit controversial, claim for \bbonu\ observation in \GE\ that is awaiting unambiguous confirmation by future \bbonu\ experiments.

The profound theoretical implications of massive Majorana neutrinos, and the possibility that an experimental observation is at hand, has triggered a new generation of $\bb0\nu$~experiments. At the time of writing this report, this new generation of experiments spans at least half a dozen isotopes, and an equally rich selection of experimental techniques, ranging from the well-established germanium calorimeters, to xenon time projection chambers. Some of the experiments are already running or will run very soon. Some of them are still in their R\&D period. Some of them push to the limit the technique they use, in particular concerning the target mass. Others are easier to scale up. All of them claim to be sensitive to very light neutrino masses, by assuming that they can do one to three orders of magnitude better in background suppression and by significantly increasing their target mass, compared to previous experiments. In this report we review the state-of-the-art of this exciting and rapidly changing field. 

This review is organized as follows. The introductory material is covered in sects.~\ref{sec:massivenu} and \ref{sec:bb0nu}. The key particle physics concepts involving massive Majorana neutrinos and neutrinoless double beta decay are laid out here. The current experimental knowledge on neutrino masses, lepton number violating processes in general, and \bbonu\ in particular, is also described in sects.~\ref{sec:massivenu} and \ref{sec:bb0nu}. Sections \ref{sec:nme}, \ref{sec:ingredients} and \ref{sec:experiments} cover more advanced topics. The theoretical aspects of the nuclear physics of \bbonu\ are discussed in sect.~\ref{sec:nme}. Sections \ref{sec:ingredients} and \ref{sec:experiments} deal with experimental aspects of \bbonu, and can be read without knowledge of sect.~\ref{sec:nme}. An attempt at a pedagogical discussion of experimental ingredients affecting \bbonu\ searches is made in sect.~\ref{sec:ingredients}. Section \ref{sec:experiments} adds a description of selected new-generation experimental proposals, together with a comparison of their physics case.

%% file: src/massivenus.tex
\subsection{\label{subsec:massivenus_whereweare}Current knowledge of neutrino mass and mixing}

Neutrinos are the lightest known elementary fermions. Neutrinos do not carry any electrical charge, do not undergo strong interactions, and are observable only via weak interactions. In the Standard Model of elementary particles, neutrinos are paired with charged leptons in weak isodoublets. Experimentally, we know that only three light (that is, of mass $<m_Z/2$, where $m_Z$ is the $Z$ boson mass) \emph{active} neutrino families exist.

More recently, neutrino oscillation experiments have unambiguously demonstrated that neutrinos are massive particles (see, for example, \cite{GonzalezGarcia:2007ib}). Because of the interferometric nature of neutrino oscillations, such experiments can only measure neutrino squared mass differences and not the absolute neutrino mass scales. Solar and reactor experiments have measured one mass splitting, the so-called \emph{solar mass splitting}, to be: $\Delta m^2_{	\rm sol}\equiv m^2_2-m^2_1= (7.58^{+0.22}_{-0.26})\times 10^{-5}\ {\rm eV}^2$. Atmospheric and accelerator-based experiments have measured a different mass splitting, the so-called \emph{atmospheric mass splitting}, to be: $|\Delta m^2_{\rm atm}|\equiv |m^2_3-(m^2_1+m^2_2)/2|= (2.35^{+0.12}_{-0.09})\times 10^{-3}\ {\rm eV}^2\gg \Delta m^2_{\rm sol}$. In the standard 3-neutrino oscillations paradigm, those are the only two independent mass splittings available. The best-fit values and 1$\sigma$ ranges quoted were obtained from a recent global 3-neutrino fit \cite{Fogli:2011qn}.

The observation of neutrino flavor oscillations also imply that the neutrino states participating in the weak interactions (\emph{flavor eigenstates}) are different from the neutrino states controlling free particle evolution (\emph{mass eigenstates}). In other words, the three weak eigenstates $|\nu_{\alpha}\rangle,\ \alpha=e,\mu,\tau$, can be expressed as a linear combination of the three mass eigenstates $|\nu_{i}\rangle,\ i=1,2,3$:
\begin{equation}
|\nu_{\alpha}\rangle = \sum_i U_{\alpha i}^{\ast}|\nu_i\rangle
\label{eq:neutrinomixing}
\end{equation}
\noindent where $U$ is a $3\times3$, unitary, \emph{neutrino mixing matrix}, that is different from unity. Equation \ref{eq:neutrinomixing} implies the violation of the individual lepton flavors $L_{\alpha}$, but not necessarily the violation of total lepton number $L\equiv \sum_{\alpha} L_{\alpha} = L_e+L_{\mu}+L_{\tau}$. The $3\times 3$ neutrino mixing matrix is usually parametrized in terms of 3 Euler angles $(\vartheta_{12},\ \vartheta_{13},\ \vartheta_{23})$ and 3 phases $(\delta,\ \alpha_{21},\ \alpha_{31})$ (see, for example, \cite{Nakamura:2010zzi}). If the massive neutrinos are \emph{Dirac particles} (see sect.~\ref{subsec:massivenus_identity}), only the \emph{Dirac phase} $\delta$ is physical and can be responsible for CP violation in the lepton sector. If the massive neutrinos are \emph{Majorana particles} (sect.~\ref{subsec:massivenus_identity}), the two additional \emph{Majorana phases} $(\alpha_{21},\ \alpha_{31}$) are also potentially observable.

 Neutrino oscillation experiments have measured with reasonably good accuracy the flavor content of the neutrino mass states participating in 3-neutrino mixing. Atmospheric and accelerator-based neutrino oscillation experiments are mostly consistent with $\nu_{\mu}\to\nu_{\tau}$ oscillations only (see, however, \cite{Abe:2011sj}). They have therefore measured the muon flavor content of the $\nu_3$ mass state to be $|U_{\mu 3}|^2\simeq 0.5$, and that such mass state has little (if non-zero) electron flavor content, $|U_{e3}|^2\simeq 0$. On the other hand, solar neutrino oscillation experiments are consistent with $\nu_e\to\nu_{\mu}$ and/or $\nu_e\to\nu_{\tau}$ oscillations. They have measured $|U_{e2}|^2\simeq 1/3$. The remaining elements of the leptonic mixing matrix can approximately be derived, given $(|U_{\mu 3}|,|U_{e3}|,|U_{e2}|)$, assuming unitarity. More precisely, according to \cite{Fogli:2011qn}, the best-fit values and 1$\sigma$ ranges in the neutrino mixing parameters measured via neutrino oscillations are: $|U_{e3}|^2=0.025\pm 0.07$, $|U_{\mu3}|^2/(1-|U_{e3}|^2)=0.42^{+0.08}_{-0.03}$ and $|U_{e2}|^2/(1-|U_{e3}|^2)=0.312^{+0.017}_{-0.016}$. The value of the Dirac CP-violating phase $\delta$, also potentially observable in neutrino oscillation experiments, is currently unknown.

\begin{figure}[t!b!]
\begin{center}
\includegraphics[width=0.80\textwidth]{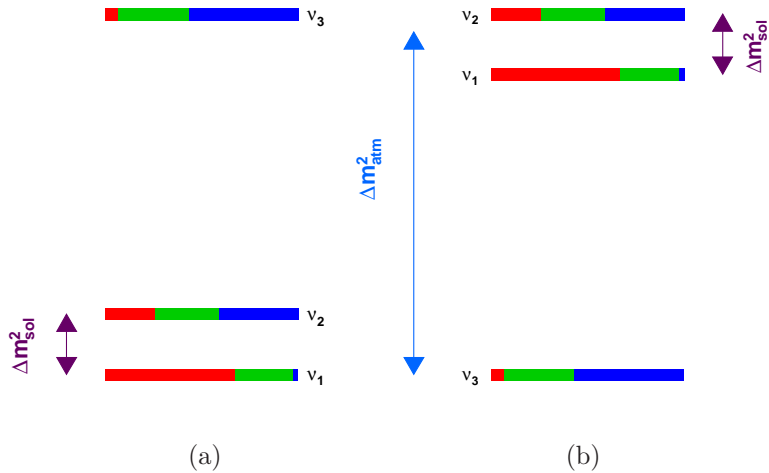}
\hspace{0.33\textwidth}{(a)}
\hspace{0.33\textwidth}{(b)}
\end{center}
\caption{Knowledge on neutrino masses and mixings from neutrino oscillation experiments. Panels (a) and (b) show the normal and inverted mass orderings, respectively. Neutrino masses increase from bottom to top. The electron, muon and tau flavor content of each neutrino mass eigenstate is shown via the red, green and blue fractions, respectively.}  \label{fig:numass_ordering}
\end{figure}

The current knowledge on neutrino masses and mixings provided by neutrino oscillation experiments is summarized in fig.~\ref{fig:numass_ordering}. The diagram shows the two possible mass orderings that are compatible with neutrino oscillation data, with increasing neutrino masses from bottom to top. In addition, the electron, muon and tau flavor content of each mass eigenstate is also shown, according to the best-fit values in reference \cite{Fogli:2011qn}.

 To complete our knowledge on neutrino masses, two pieces of information remain to be known: the neutrino mass ordering and the absolute value of the lightest neutrino mass.
 
Concerning the neutrino mass ordering, current neutrino oscillation results cannot differentiate between two possibilities, usually referred to as \emph{normal} and \emph{inverted orderings}‚ (see fig.~\ref{fig:numass_ordering}). In the former, the gap between the two lightest mass eigenstates corresponds to the small mass difference, indicated by solar experiments ($\Delta m^2_{\rm sol}$), while in the second case the gap between the two lightest states corresponds to the large mass difference, inferred from atmospheric experiments ($\Delta m^2_{\rm atm}$). While we do not know at present whether $\nu_3$ is heavier or lighter than $\nu_1$, we do know that $\nu_2$ is heavier than $\nu_1$, thanks to matter effects affecting the propagation of neutrinos inside the Sun. The exploitation of the same type of matter effect on future accelerator-based neutrino experiments may allow us to experimentally establish the neutrino mass ordering in the future, provided that $|U_{e3}|>0$. In the particular case in which the neutrino mass differences are very small compared with its absolute scale, we speak of the \emph{degenerate} spectrum.

The absolute value of the neutrino mass scale can instead be probed via neutrinoless double beta decay searches, cosmological observations and beta decay experiments. Only upper bounds on the neutrino mass, of order $\sim$1 eV, currently exist. Constraints on the lightest neutrino mass coming from neutrinoless double beta decay will be discussed in sect.~\ref{subsec:bb0nu_lightmajoranaexchange}. In the following, we briefly summarize cosmological and beta decay constraints.


Primordial neutrinos have a profound impact on cosmology since they affect both the expansion history of the Universe and the growth of perturbations (see, for instance, reference \cite{Lesgourgues:2006nd}). Cosmological observations  can probe the sum of the three neutrino masses:
\begin{equation}
m_{{\rm cosmo}}\equiv \sum_{i=1}^3 m_i 
\label{eq:mcosmo}
\end{equation}

\begin{figure}[t!b!]
\begin{center}
\includegraphics[width=0.45\textwidth]{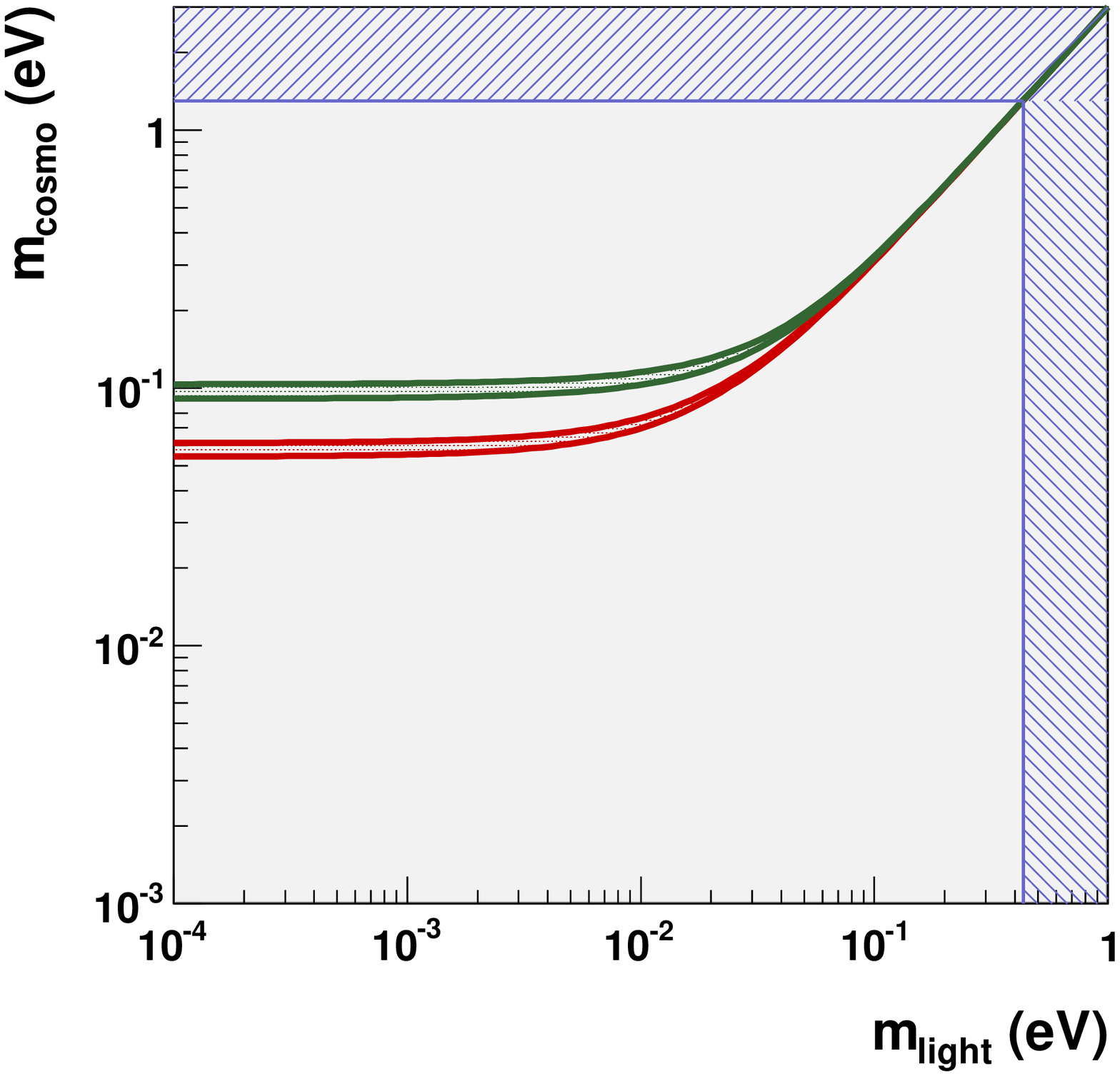} \hspace{0.04\textwidth}
\includegraphics[width=0.45\textwidth]{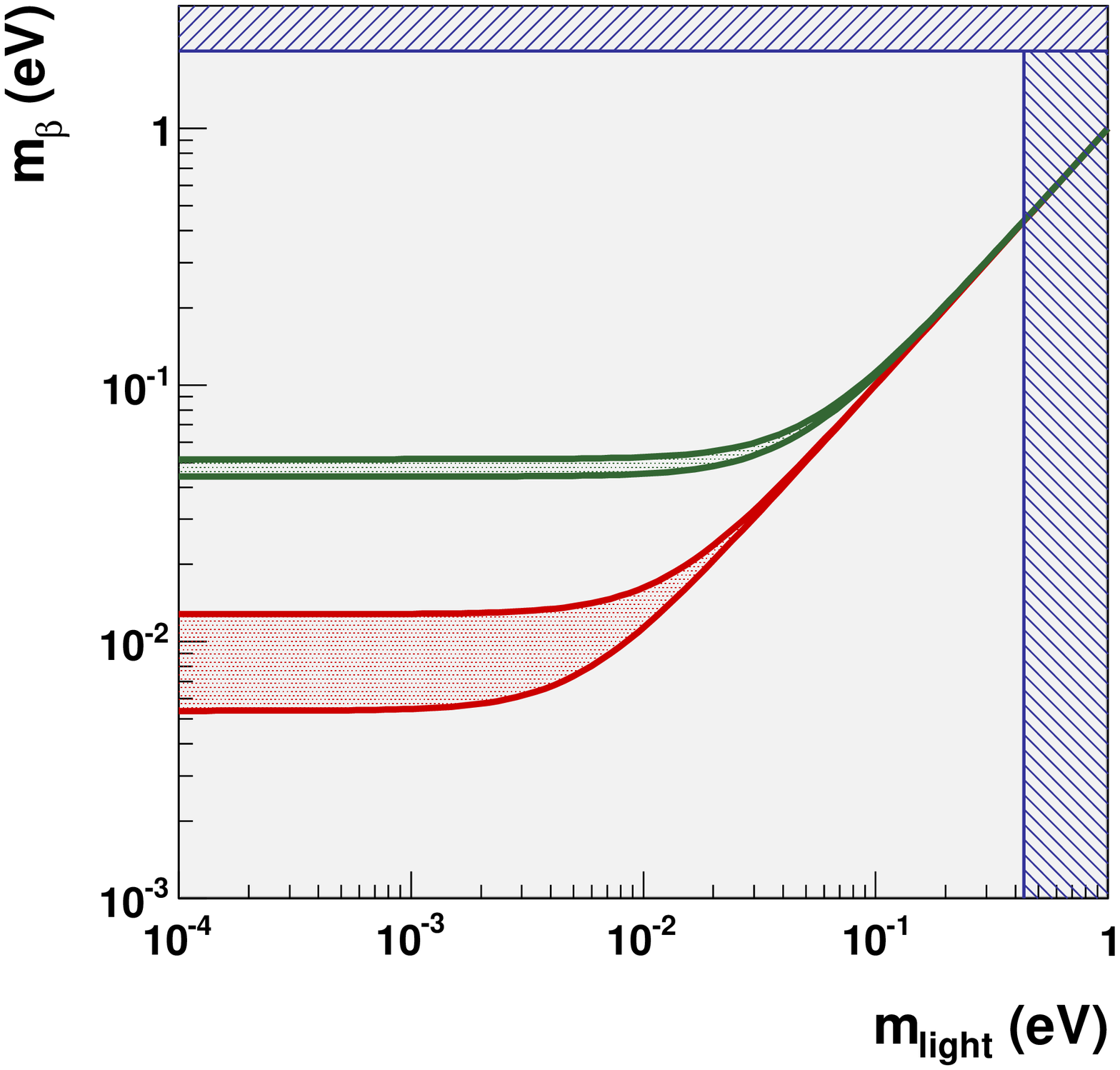}
\end{center}
\hspace{0.25\textwidth}{\small (a)}
\hspace{0.45\textwidth}{\small (b)}
\caption{\label{fig:mass_constraints_cosmo_beta}Constraints on the lightest neutrino mass $m_{\rm light}$ coming from a) cosmological and b) $\beta$ decay experiments. The red and green bands correspond to the normal and inverted orderings, respectively. The $m_{{\rm cosmo}}$ upper bound in panel (a) is from \cite{Komatsu:2010fb}, and translates into a $m_{{\rm light}}$ upper limit shown via the vertical band in the same panel. The cosmological constraint on $m_{{\rm light}}$ is also shown in panel (b), together with the upper limit on $m_{\beta}$ from tritium $\beta$ decay experiments \cite{Nakamura:2010zzi}.}
\end{figure}

Cosmological data are currently compatible with massless neutrinos. Several upper limit values on $m_{{\rm cosmo}}$ can be found in the literature, depending on the details of the cosmological datasets and of the cosmological model that were used in the analysis. A conservative upper limit on $m_{{\rm cosmo}}$ of 1.3 eV at 95\% confidence level \cite{Komatsu:2010fb} is obtained when CMB measurements from the Wilkinson Microwave Anisotropy Probe (WMAP) are combined with measurements of the distribution of galaxies (SDSSII-BAO) and of the Hubble constant $H_0$ (HST), in the framework of a cold dark matter model with dark energy whose equation of state is allowed to differ from $-1$. 
The relationship between $m_{{\rm cosmo}}$, defined in eq.~(\ref{eq:mcosmo}), and the lightest neutrino mass $m_{{\rm light}}$ ---that is, $m_1$ ($m_3$) in the case of normal (inverted) ordering--- is shown in fig.~\ref{fig:mass_constraints_cosmo_beta}a. The two bands correspond to the normal and inverted orderings, respectively. The width of the bands is given by the 3$\sigma$ ranges in the mass oscillation parameters $\Delta m^2_{{\rm sol}}$ and $\Delta m^2_{{\rm atm}}$ \cite{Fogli:2011qn}. The horizontal band in fig.~\ref{fig:mass_constraints_cosmo_beta}(a) is the upper limit on $m_{{\rm cosmo}}$. In this quasi-degenerate regime, this upper bound implies that $m_{{\rm light}}\simeq m_{{\rm cosmo}}/3 \lesssim 0.43\ {\rm eV}$ at 95\% CL, as shown by the vertical band in fig.~\ref{fig:mass_constraints_cosmo_beta}(a).

The neutrino mass scale can also be probed in laboratory-based experiments (see, for example, \cite{Otten:2008zz}). The differential electron energy spectrum in nuclear $\beta$ decay experiments is affected both by the neutrino masses and by the mixings defining the electron neutrino state in terms of mass eigenstates. In this case, the mass combination probed is given by:
\begin{equation}
m_{\beta}^2\equiv\sum_{i=1}^3|U_{ei}|^2m_i^2
\label{eq:mbeta}
\end{equation}

The relationship between $m_{\beta}$ in eq.~(\ref{eq:mbeta}) and $m_{{\rm light}}$ is shown in fig.~\ref{fig:mass_constraints_cosmo_beta}(b). Again, the results of a recent global fit to neutrino oscillation data \cite{Fogli:2011qn} are used to determine the $3\sigma$ bands for both the normal and inverted orderings. From the experimental point of view, the region of interest for the study of neutrino properties is located near the $\beta$ endpoint. The most sensitive searches conducted so far are based upon the decay of tritium, via $^3{\rm H}\to^3{\rm He}^+e^-\bar{\nu}_e$, mostly because of the very low $\beta$ endpoint energy of this element (18.6 keV). As for cosmology, $\beta$ decay searches of neutrino mass have so far yielded negative results. The horizontal band in fig.~\ref{fig:mass_constraints_cosmo_beta}(b) comes from the two current best limits from the Troitsk \cite{Lobashev:1999tp} and Mainz \cite{Kraus:2004zw} experiments. The combined limit is $m_{\beta}<2$ eV at 95\% CL \cite{Nakamura:2010zzi}. The resulting constraint on $m_{{\rm light}}$ is less stringent than the cosmological one. The KATRIN experiment \cite{Otten:2008zz} should be able to improve the $m_{\beta}$ (and therefore $m_{{\rm light}}$) sensitivity by roughly an order of magnitude in the forthcoming years, thanks to its better statistics, energy resolution, and background rejection.


\subsection{\label{subsec:massivenus_identity}The origin of neutrino mass: Dirac \emph{versus} Majorana neutrinos}

Are neutrinos their own antiparticles? If the answer is no, we speak of \emph{Dirac neutrinos}. If the answer is yes, we speak of \emph{Majorana neutrinos}. Both possibilities exist for the neutrino, being electrically neutral and not carrying any other charge-like conserved quantum number. Whether neutrinos are Majorana or Dirac particles depends on the nature of the physics that give them mass, given that the two characters are physically indistinguishable for massless neutrinos.

In the Standard Model, only the negative chirality component $\Psi_L$ of a fermion field $\Psi=\Psi_R+\Psi_L$ is involved in the weak interactions. A negative (positive) chirality field $\Psi_{L(R)}$ is a field that obeys the relations $P_{L(R)}\Psi_{L(R)}=\Psi_{L(R)}$ and $P_{R(L)}\Psi_{L(R)}=0$, where $P_L=(1-\gamma_5)/2$ and $P_R=(1+\gamma_5)/2$ are the positive and negative chiral projection operators. 

For massless neutrinos (see, for example, \cite{Hernandez:2010mi}), only the negative chirality neutrino field $\nu_L$ is needed in the theory, regardless of the Dirac/Majorana nature of the neutrino discussed below, since neutrinos only participate in the weak interactions. This field describes \emph{negative helicity} neutrino states $|\nu_L\rangle$ and \emph{positive helicity} antineutrino states\footnote{As customarily done, we use the subscript ``L'' to denote both negative helicity states $|\nu_L\rangle$ and negative chirality fields $\nu_L$, since the terms left-handed helicity states and left-handed chirality fields are also commonly used. Similarly, we denote positive helicity states and positive chirality fields with the subscript ``R'', as in ``right-handed''.}. The positive and negative helicity states are eigenstates of the helicity operator $h\equiv \vec{\sigma}\cdot\hat{p}$ with eigenvalues $\pm 1/2$, respectively, where $\vec{\sigma}$ is the neutrino spin and $\hat{p}$ the neutrino momentum direction. The fact that $\nu_L$ annihilates particles of negative helicity, and creates antiparticles with positive helicity, is not inconsistent with Lorentz invariance, given that the helicity is the same in any reference frame for a fermion travelling at the speed of light. In the Standard Model with massless neutrinos, positive helicity neutrinos and negative helicity antineutrinos do not exist. As a consequence, and since a negative helicity state transforms into a positive helicity state under the parity transformation, the chiral nature of the weak interaction (differentiating negative from positive chirality) implies that parity is maximally violated in the weak interactions.

For relativistic neutrinos of non-zero mass $m$, the neutrino field participating in the weak interactions has still negative chirality, $\nu_L$, but there are sub-leading corrections to the particle annihilation/creation rules described above. The state $|\nu_L\rangle$ that is annihilated by the negative chirality field $\nu_{L}$ is now a linear superposition of the $-1/2$ and $+1/2$ helicity states. The $+1/2$ helicity state enters into the superposition with a coefficient $\propto m/E$, where $E$ is the neutrino energy, and is therefore highly suppressed. 

Neutrino mass terms can be added to the Standard Model Lagrangian in two ways (see, for example, \cite{Giunti:2003qt}). The first way is in direct analogy to the Dirac masses of quarks and charged leptons, by adding the positive chirality component $\nu_R$ of the Dirac neutrino field, describing predominantly positive helicity neutrino states and predominantly negative helicity antineutrino states that do not participate in the weak interactions:
\begin{equation}
\label{eq:diracmassterm}
-\mathcal{L}_D=m_D(\overline{\nu_L}\nu_R+\overline{\nu_R}\nu_L),
\end{equation}
where $m_D=y v/\sqrt{2}>0$, $y$ is a dimensionless Yukawa coupling coefficient and $v/\sqrt{2}$ is the vacuum expectation value of the neutral Higgs field after electroweak symmetry breaking. In eq.~(\ref{eq:diracmassterm}), $\nu_L$ and $\nu_R$ are, respectively, the negative and positive chirality components of the neutrino field $\nu$. The chiral spinors $\nu_L$ and $\nu_R$ have only two independent components each, leading to the four independent components in the spinor $\nu$. This is different from the the case of massless neutrinos, where only the 2-component spinor $\nu_L$ was needed.

The second way in which neutrino mass terms can be added to the Standard Model Lagrangian is unique to neutrinos. Majorana first realized \cite{Majorana:1937vz} that, for neutral particles, one can remove two of the four degrees of freedom in a massive Dirac spinor by imposing the \emph{Majorana condition}:
\begin{equation}
\nu^c = \nu
\label{eq:majoranacondition}
\end{equation}
where $\nu^c = C\bar{\nu}^T=C(\gamma^0)^T\nu^{\ast}$ is the CP conjugate of the field $\nu$, $C$ is the charge-conjugation operator, and $(\nu_L)^c$ ($(\nu_R)^c$) has positive (negative) chirality. This result can be obtained by decomposing both the left-hand and right-hand sides of eq.~(\ref{eq:majoranacondition}) into their chiral components, yielding:
\begin{equation}
\nu_R = (\nu_L)^c
\label{eq:nur}
\end{equation}
and therefore proving that the positive chirality component of the Majorana neutrino field $\nu_R$ is not independent of, but obtained from, its negative chirality counterpart $\nu_L$. By substituting eq.~(\ref{eq:nur}) into the mass term in eq.~(\ref{eq:diracmassterm}), we obtain a \emph{Majorana mass term}:
\begin{equation}
\label{eq:majoranamassterml}
-\mathcal{L}_L= \frac{1}{2}m_L(\overline{\nu_L}(\nu_L)^c+\overline{(\nu_L)^c}\nu_L)
\end{equation}
where $m_L$ is a free parameter with dimensions of mass. This lagrangian mass term implies the existence of a weak isospin triplet scalar (a Higgs triplet), with a neutral component acquiring a non-vanishing vaccum expectation value after electroweak symmetry breaking. Equation (\ref{eq:majoranamassterml}) represents a mass term constructed from negative chirality neutrino fields alone, and we therefore call it a \emph{negative chirality Majorana mass term}. If positive chirality fields also exist and are independent of negative chirality ones, this is not the only possibility. In this case, we may also construct a second Majorana mass term, a \emph{positive chirality Majorana mass term}:
\begin{equation}
\label{eq:majoranamasstermr}
-\mathcal{L}_R= \frac{1}{2}m_R(\overline{\nu_R}(\nu_R)^c+\overline{(\nu_R)^c}\nu_R)
\end{equation}

In the Standard Model, right-handed fermion fields such as $\nu_R$ are weak isosopin singlets. As a consequence, and in contrast with $m_D$ or $m_L$, the mass parameter $m_R$ is therefore not connected to a Higgs vaccum expectation value, and could be arbitrarily high. All three mass term in eqs.~(\ref{eq:diracmassterm}), (\ref{eq:majoranamassterml}) and (\ref{eq:majoranamasstermr}) convert negative chirality states into positive chirality ones\footnote{This is because the charge conjugate of a field with a given chirality, such as the ones appearing in eqs.~\ref{eq:majoranamassterml} and \ref{eq:majoranamasstermr}, always has the opposite chirality.}. Chirality is therefore not a conserved quantity, regardless of the Dirac/Majorana nature of neutrinos. Furthermore, the Majorana mass terms in eqs.~(\ref{eq:majoranamassterml}) and (\ref{eq:majoranamasstermr}) convert particles into their own antiparticles. As stated previously, they are therefore forbidden for all electrically charged fermions because of charge conservation. But not only: processes involving Majorana mass terms violate the Standard Model total lepton number $L\equiv L_e+L_{\mu}+L_{\tau}$ by two units ($|\Delta L|=2$), which is not a good quantum number anymore. 

\begin{figure}[t!b!]
\begin{minipage}[t]{0.56\textwidth}
\vspace{0pt}
\includegraphics[scale=0.36]{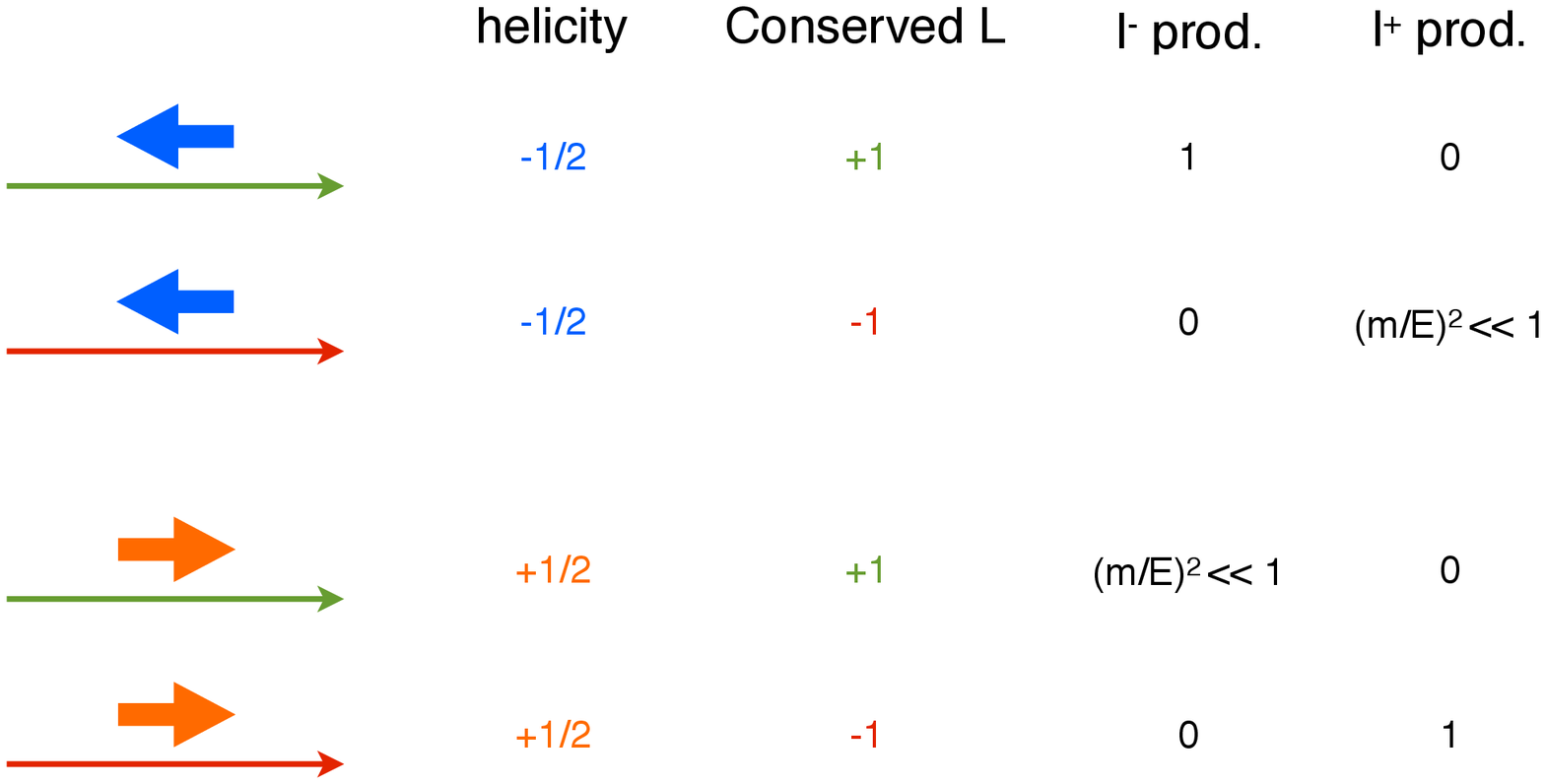}
\end{minipage}
\hfill
\begin{minipage}[t]{0.40\textwidth}
\vspace{0pt}
\includegraphics[scale=0.36]{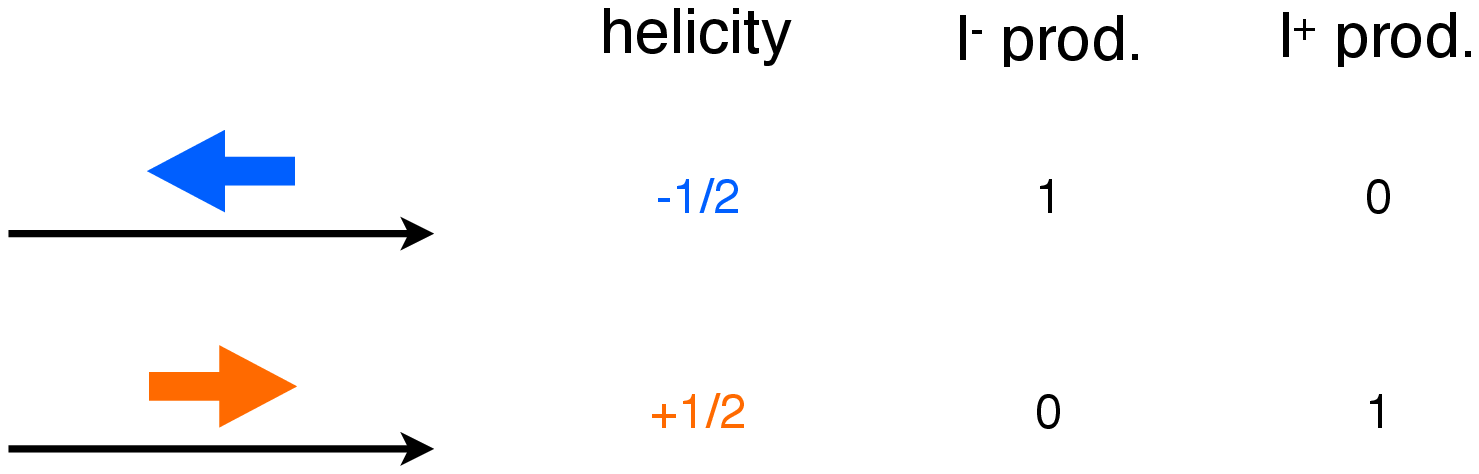}
\end{minipage} 
\caption{The difference between Dirac (left) and Majorana (right) massive neutrinos in a scattering experiment. See text for details. Adapted from \cite{Parke:2011zz}.}
\label{fig:DiracMajoranaNeutrinoInteractions}
\end{figure}

Which of the mass terms allowed in theory, among $\mathcal{L}_D$, $\mathcal{L}_L$ and $\mathcal{L}_R$ in eqs.~(\ref{eq:diracmassterm}), (\ref{eq:majoranamassterml}) and (\ref{eq:majoranamasstermr}) exist in nature? What are the numerical values of the corresponding coupling constants $m_D$, $m_L$, $m_R$? These questions can in principle be answered experimentally. Majorana and Dirac massive neutrinos will in fact have different Standard Model interactions. Let us consider for now an instructive, albeit unrealistic, scattering experiment, see fig.~\ref{fig:DiracMajoranaNeutrinoInteractions}.

In the Dirac case, Standard Model interactions conserve lepton number $L$, with $L(\nu)=L(l^-)=-L(\bar{\nu})=-L(l^+)=+1$, where $l^{\pm}$ indicates charged leptons. Particles are then identified as neutrinos or antineutrinos in accordance with the process through which they are produced. Charged-current interactions of Dirac neutrinos (as opposed to antineutrinos) produce only $l^-$ and carry a well-defined lepton number $L=-1$, and viceversa. As shown in fig.~\ref{fig:DiracMajoranaNeutrinoInteractions}, for Dirac neutrinos we would thus have four mass-degenerate states: for each of the two available helicity states\footnote{As mentioned above, the weak interaction is maximally parity violating, therefore the two helicity states are distinguishable.}, two distinct particle/antiparticle states characterized by a different $L$ value would be available. Standard Model interactions of neutrino (as opposed to antineutrino) states of positive helicity would have, however, much weaker $l^-$-producing interactions with matter compared to neutrino states of negative helicity, as indicated by the coefficients in fig.~\ref{fig:DiracMajoranaNeutrinoInteractions}. On the other hand, we have seen that in the Majorana case $L$ is not conserved. We would only have two mass-degenerate states, defined by the two available helicity states, see fig.~\ref{fig:DiracMajoranaNeutrinoInteractions}.

Given these differences between Dirac and Majorana massive neutrinos, can we establish which of the two possibilities is realized in Nature via a scattering experiment? In practice, no. The reason is that $l^-$ production from positive helicity Dirac neutrinos (and $l^+$ production from negative helicity Dirac antineutrinos) is expected to be highly suppressed in the ultra-relativisitc limit, and cannot be experimentally observed. Experimentally, all we know is that the neutral particle produced in association with a $l^+$ produces, when interacting, a $l^-$. In the Dirac case, lepton number conservation is assumed and such neutral particle is identified as the neutrino, with $L=-1$. In the Majorana case, such neutral particle is instead identified as the negative helicity state, interacting differently from its positive helicity counterpart. Both interpretations are viable, and what happens when a neutrino interacts can be understood without invoking a conserved lepton number \cite{Kayser:2002qs}.


\subsection{\label{subsec:massivenus_seesaw}The see-saw mechanism}
\begin{figure}[t!b!]
\begin{center}
\includegraphics[width=0.95\textwidth]{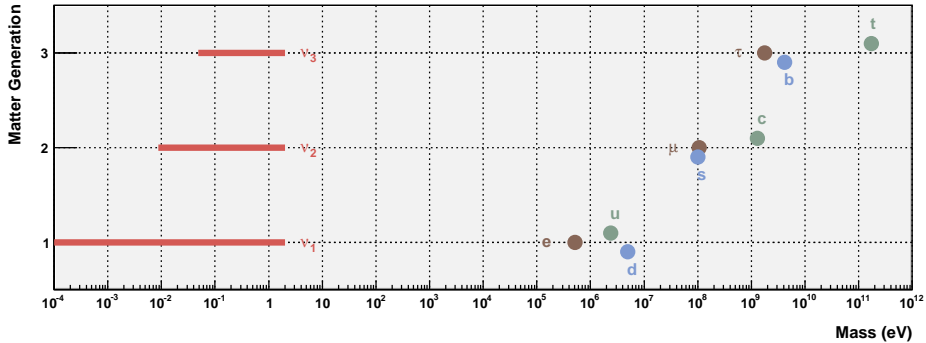}
\end{center}
\caption{ \label{fig:fermionmasses}Hierarchical structure of fermion masses. Only upper bounds for neutrino masses exist. The figure assumes a normal ordering for neutrino masses. Values taken from \cite{Nakamura:2010zzi}.}
\end{figure}

\indent Neutrino masses, although not measured yet, are known to be small,
 of the order of 1 eV or less, see sect.~\ref{subsec:massivenus_whereweare}. Such mass values are much smaller than the masses of all other elementary fermions, see fig.~\ref{fig:fermionmasses}. The explanation of neutrino masses via
 Dirac mass terms alone require neutrino Yukawa couplings of the
 order of $10^{-12}$ or less. The current theoretical prejudice is that
 neutrino Yukawa couplings with $y_{\nu}\ll 1$ and $y_{\nu}\ll y_l$
 are unnatural, if not unlikely. \\

%
\indent The so-called \emph{see-saw mechanism} provides a way to accommodate
 neutrino masses that is considered more natural. The simplest realization of the see-saw model is to add both a Dirac mass term and a positive chirality mass term to the Lagrangian, as given by eqs. (\ref{eq:diracmassterm}) and (\ref{eq:majoranamasstermr}), respectively, for each of the three neutrino flavors. This is sometimes called the \emph{type I see-saw mechanism}, where we take $m_L=0$, $m_D\neq 0$, and $m_R\neq 0$. In this case, the neutrino mass terms can be recast in the matrix form:
\begin{equation}
-\mathcal{L}_{\text{D+R}}
=
\frac{1}{2} \,
\overline{(\mathcal{N}_L)^c} \, M \, \mathcal{N}_L
+
\text{h.c.}
\,,
\label{eq:seesawmatrixform}
\end{equation}
\noindent where the matrix $M$ has the form:
\begin{equation}
M
=
\begin{pmatrix}
0 & m_{\text{D}}
\\
m_{\text{D}} & m_R
\end{pmatrix}
\end{equation}
\noindent and the negative chirality vector $\mathcal{N}_L$ is
\begin{equation}
\mathcal{N}_L
=
\begin{pmatrix}
\nu_L
\\
(\nu_R)^c
\end{pmatrix}
\,.
\label{eq:typeIseesaw_1family}
\end{equation}

The chiral fields $\nu_L$ and $\nu_R$ do not have a definite mass, since they are coupled by the Dirac mass term. In order to find the fields $\nu_{1L}$ and $N_{1L}$ with definite masses $m_1$ and $M_1$, respectively, it is necessary to diagonalize the mass matrix in eq.~(\ref{eq:seesawmatrixform}). In other words, it is necessary to find a unitary mixing matrix $\mathcal{U}$ such that:
\begin{equation}
\mathcal{U}^T \, M \, \mathcal{U} =
\begin{pmatrix}
m_1 & 0
\\
0 & M_1
\end{pmatrix}
\,,
\label{eq:diagonalization_1family}
\end{equation}
\noindent where:
\begin{equation}
\mathcal{N}_L = \mathcal{U}  \, n_L
\,,
\qquad
\text{with}
\qquad
n_L
=
\begin{pmatrix}
\nu_{1L}
\\
N_{1L}
\end{pmatrix}
\,,
\label{eq:diagonalization2_1family}
\end{equation}

 For each neutrino flavor, two fields of definite chirality and definite mass are therefore obtained, and the diagonalized mass terms can be written as:
\begin{equation}
-\mathcal{L}_{D+R}
=
\frac{1}{2}
\left(m_1 \, \overline{(\nu_{1L})^c} \, \nu_{1L} + M_1 \, \overline{(N_{1L})^c} \, N_{1L}\right) + \text{h.c.}
\,,
\label{eq:diagonalization3_1family}
\end{equation}

Both terms in eq.~(\ref{eq:diagonalization3_1family}) have the same form as the pure negative chirality Majorana mass term in eq.~(\ref{eq:majoranamassterml}). In other words, both mass eigenfields $\nu_{1L}$ and $N_{1L}$ are equal to their CP-conjugate fields, and thus both describe Majorana particles. The insertion of a Dirac mass term and a positive chirality Majorana mass term in the Lagrangian for massive neutrinos has resulted in Majorana particles.

Since the positive chirality fields are electroweak singlets in the Standard Model, the Majorana mass of the neutrino described by such field, $m_R$, may be orders of magnitude larger than the electroweak scale. In the so-called \emph{see-saw limit}, we assume that neutrino Yukawa couplings are of the order of the charged fermion couplings, and that $m_R\gg m_D$ is of the order of some high mass scale where new physics responsible for neutrino masses is supposed to reside. In this approximation, the see-saw mechanism naturally yields a small mass eigenvalue $m_1\simeq m_D^2/m_R\ll m_D$ for a predominantly negative helicity neutrino mass state, and a large mass eigenvalue $M_1\simeq m_R$ for a predominantly positive helicity (and therefore sterile) neutrino mass state. A very heavy $N_1$ corresponds to a very light $\nu_1$ and viceversa, as in a see-saw. 

The see-saw mechanism presented above can easily be generalized from the one-family case that we discussed to three neutrino species, yielding the three light neutrinos $\nu_i$ we are familiar with, and three heavy neutrinos $N_i$, with $i=1,2,3$ \cite{Giunti:2003qt}. In this case, the neutrino mass matrix in eq.~\ref{eq:seesawmatrixform} is a $6\times6$ mass matrix of the form:
\begin{equation}
M
=
\begin{pmatrix}
0 & (M^{\text{D}})^T
\\ \displaystyle
M^{\text{D}} & M^R
\end{pmatrix}
\,.
\label{eq:diagonalization_3families}
\end{equation}
\noindent where $M^D$ and $M^R$ are now $3\times3$ complex matrices, and the 6-component vector of negative chirality fields has the form:
\begin{equation}
N_L
=
\begin{pmatrix}
\nu_L
\\ \displaystyle
(\nu_R)^c
\end{pmatrix}
\,,
\qquad
\text{with}
\qquad
\nu_L
=
\begin{pmatrix}
\nu_{eL}
\\ \displaystyle
\nu_{{\mu}L}
\\ \displaystyle
\nu_{{\tau}L}
\end{pmatrix}
\qquad
\text{and}
\qquad
(\nu_R)^c
=
\begin{pmatrix}
(\nu_{s_1 R})^c
\\ \displaystyle
(\nu_{s_2 R})^c
\\ \displaystyle
(\nu_{s_3 R})^c
\end{pmatrix}
\,,
\label{eq:diagonalization_3families_2}
\end{equation}
In eq.~\ref{eq:diagonalization_3families_2}, the subscripts $e,\ \mu,\ \tau$ label the active neutrino flavors, while the subscripts $s_1,\ s_2,\ s_3$ indicate sterile states that do not participate in the weak interactions. The mass matrix is diagonalized via a $6\times6$ mixing matrix $\mathcal{V}$ analogous to $\mathcal{U}$ in eq.~\ref{eq:diagonalization_1family}, where the three negative chirality fields and the three positive chirality fields are now expressed in terms of the negative chirality components of 6 massive neutrino fields $\nu_{iL},\ i=1,\ldots ,6$. In the see-saw limit where the eigenvalues of $M^R$ are much larger than those of $M^D$, the $6\times6$ mass matrix in eq.~\ref{eq:diagonalization_3families} can be written in block-diagonal form $M\simeq \text{diag}(M_{\text{light}},M_{\text{heavy}})$, where the two $3\times3$ mass matrices of the light and heavy neutrino sectors are practically decoupled, and given by $M_{\text{light}}\simeq -(M^D)^T(M^R)^{-1}M^D$ and $M_{\text{heavy}}\simeq M^R$, respectively. 

For the low-energy phenomenology, it is sufficient to consider only $M_{\text{light}}$, sometimes called the \emph{neutrino mass matrix} $m_{\nu}$, that is the $3\times3$ matrix in the flavor basis which is diagonalized by the matrix $U$:
\begin{equation}
U^T \, M_{\text{light}} \, U
=
\text{diag}\!\left(m_1,m_2,m_3\right)
\,,
\label{eq:diagonalization3_3families}
\end{equation}
\noindent where the \emph{neutrino mixing matrix} $U$ appearing in eq.~(\ref{eq:diagonalization3_3families}) is the same matrix defined in eq.~(\ref{eq:neutrinomixing}), and $m_1,\ m_2,\ m_3$ are three light neutrino mass eigenvalues discussed in sect.~\ref{subsec:massivenus_whereweare}.

An important assumption in the simplest realization of the see-saw mechanism described above is that $m_L=0$. This assumption is not arbitrary, and directly follows from enforcing the gauge symmetries of the Standard Model, see for example \cite{Giunti:2003qt}. In models with a left-right symmetric particle content, this type I see-saw mechanism is often generalized to a \emph{type II see-saw}, where an additional direct mass term $m_L$ for the light neutrinos is present.

\subsection{\label{subsec:massivenus_leptogenesis}Leptogenesis}

Inflationary models of the Universe predict matter and antimatter to be equally abundant at the very hot beginning, given that any potential initial asymmetry would have been diluted away by inflation. However, the observable Universe today is almost entirely made of matter! This matter dominance today is consistent with the small level of baryon asymmetry that is inferred from BBN and CMB observations, given that annihilation of matter with antimatter would have left us in a matter-dominated Universe today. The baryon asymmetry has been precisely measured:
\begin{equation}
\eta \equiv \frac{n_B-n_{\bar{B}}}{n_{\gamma}} = 273.9\cdot 10^{-10}\Omega_bh^2 = (6.19\pm 0.15)10^{-10} \label{eq:measuredbaryonasymmetry}
\end{equation}
where $n_B$, $n_{\bar{B}}$, $n_{\gamma}$ are the number densities of baryons, antibaryons and photons, $\Omega_b=(0.0458\pm 0.0016)$ is the fraction of the critical energy density carried by baryons, and $h\equiv H_0/100\ \text{km}\cdot\text{s}^{-1}\cdot\text{Mpc}^{-1}=(0.742\pm 0.036)$ is the Hubble parameter, where $H_0$ is the Hubble constant today.

\begin{figure}[t!b!]
\begin{center}
\begin{tabular}{ccc}
\includegraphics[scale=0.45]{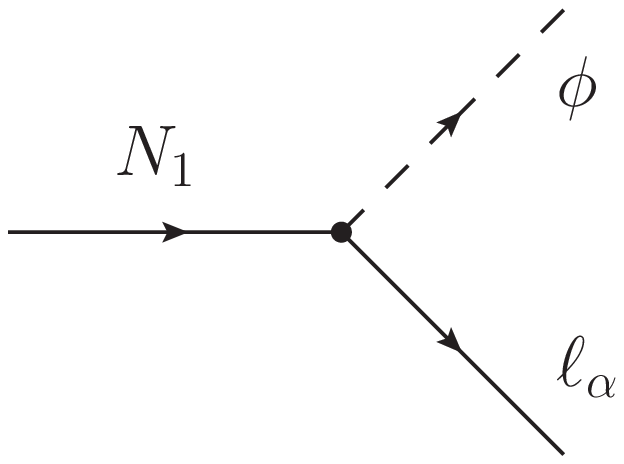} \hspace{0.03\textwidth} &
\includegraphics[scale=0.45]{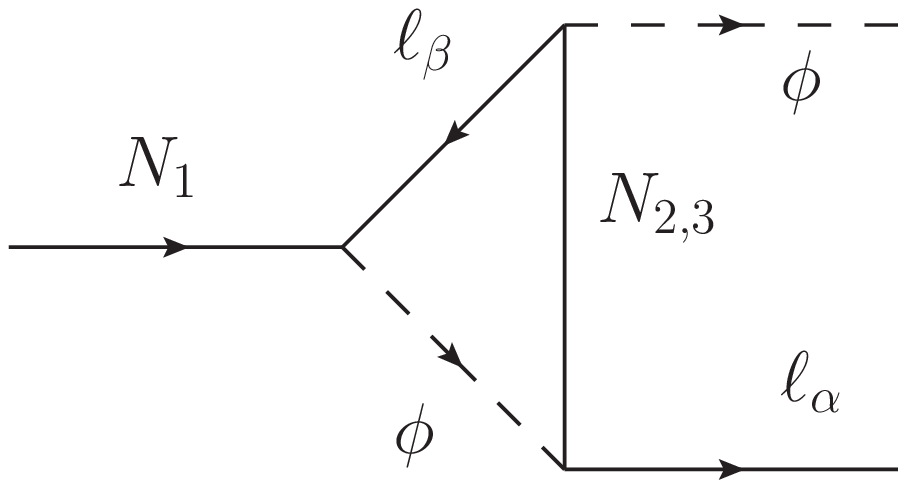} \hspace{0.03\textwidth} &
\includegraphics[scale=0.45]{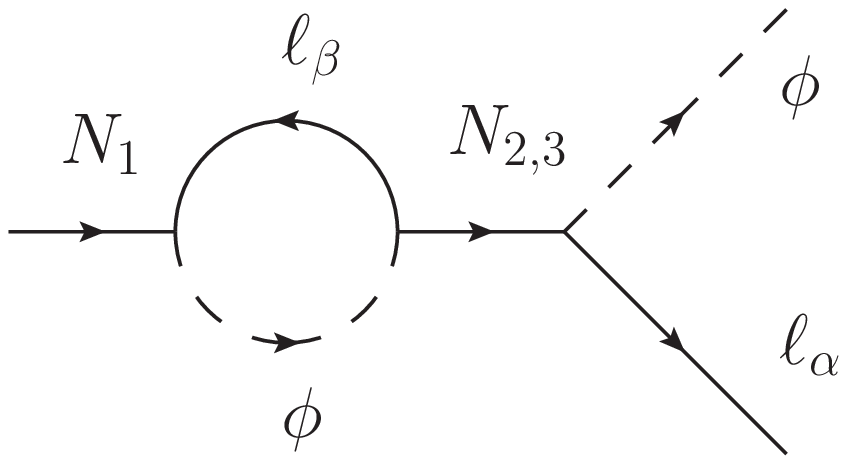} \\[10pt]
(a) & (b) & (c)
\end{tabular}
\end{center}
\caption{Feynman diagrams that contribute to the lepton number asymmetry through the decays of the heavy Majorana neutrino $N_1$ into the Higgs $\phi$ plus leptons $l_{\alpha}$. The asymmetry is generated via the interference of the tree-level diagram (a) with the one-loop vertex correction (b) and the self-energy (c) diagrams (adapted from \cite{Chen:2007fv}).} \label{fig:leptogenesis}
\end{figure}

What caused this matter-antimatter asymmetry in the early Universe? The baryon asymmetry could have been induced by a lepton asymmetry: \emph{leptogenesis} (see, for example, \cite{Chen:2007fv,Davidson:2008bu}). If neutrinos are Majorana particles, the decays of the heavy Majorana neutrinos into leptons $l_{\alpha}$ plus Higgs particles $\phi$ in the early Universe provides an ideal scenario for leptogenesis. Heavy Majorana neutrinos are their own anti-particles, so they can decay to both $l_{\alpha}\phi$ and $\bar{l_{\alpha}}\bar{\phi}$ final states. If there is an asymmetry in the two decay rates, a net lepton asymmetry will be produced. Figure \ref{fig:leptogenesis} shows the processes that would contribute to a net lepton asymmetry in the presence of heavy Majorana neutrino decays, in the simplest case where the asymmetry is dominated by the decay of the lightest among the three heavy neutrinos, $N_1$. Finally, this lepton asymmetry can be efficiently converted into a baryon asymmetry via the so-called \emph{sphaleron processes} (see \cite{Chen:2007fv,Davidson:2008bu} for details).

In more detail, for leptogenesis to occur, three conditions must be met. These conditions directly follow from the ingredients that are required to dynamically generate a baryon asymmetry (\emph{Sakharov's conditions} \cite{Sakharov:1967dj}): 
\begin{enumerate}
\item Presence of lepton number violating processes;
\item Beyond-SM sources of CP violation\footnote{CP violation is allowed by the Standard Model and has been measured; however, the magnitude of such CP-violating effects is far too small to provide the necessary amount of leptogenesis.};
\item Departure from thermal equilibrium.
\end{enumerate}
 The decay of heavy Majorana neutrinos can provide all of these conditions, namely
\begin{enumerate}
\item Total lepton number is violated in these decays;
\item CP can be violated in these decays, provided that there is more than one heavy Majorana field;
\item Departure from thermal equilibrium is obtained if the decay rate is slower than the expansion rate of the Universe at the time of decoupling, occurring for $T\sim M_1$, where $T$ is the temperature of the Universe's thermal bath, and $M_1$ is the mass of the lightest among the three heavy neutrinos. 
\end{enumerate}
In order to be fully successful, any theory of leptogenesis must be able to explain the observed magnitude of baryon asymmetry given in eq.~(\ref{eq:measuredbaryonasymmetry}). Leptogenesis via heavy Majorana neutrino decays is in principle able to do this. In this case, the asymmetry in lepton flavor $\alpha$ produced in the decay of $N_1$, defined as:
\begin{equation}
\varepsilon_{\alpha\alpha}\equiv \frac{\Gamma(N_1\to \phi l_{\alpha})-\Gamma(N_1\to\bar{\phi} \bar{l}_{\alpha})}{\Gamma(N_1\to \phi l)+\Gamma(N_1\to\bar{\phi} \bar{l})}
\label{eq:letptogenesis_asymmetry}
\end{equation}
\noindent should be of order $|\varepsilon_{\alpha\alpha}|>10^{-7}$ \cite{Davidson:2008bu}, where the factors $\Gamma$ in eq.~(\ref{eq:letptogenesis_asymmetry}) stand for the decay rates into the corresponding $N_1$ decay final states. It is at present unclear whether there is a direct connection between the high-energy CP-violating processes responsible for the asymmetry in the early Universe of eq.~(\ref{eq:letptogenesis_asymmetry}), and the low-energy CP-violating processes that may potentially affect laboratory-based experiments. Nonetheless, the discovery of CP violation in the lepton sector via neutrino oscillations on the one hand, and the discovery of the Majorana nature of neutrinos via neutrinoless double beta decay on the other, would undoubtedly strengthen the case for leptogenesis as a source of the baryon asymmetry of the Universe.


\subsection{\label{subsec:massivenus_lviolation}Lepton number violating processes}

We have seen that Majorana mass terms induce lepton number violating processes of the type $|\Delta L|=2$. The heavy neutrino decay needed for leptogenesis, discussed in sect.~\ref{subsec:massivenus_leptogenesis}, requires Majorana mass terms and is therefore an example of a lepton number violating process. However, heavy neutrino decay is unobservable in a laboratory-based experiment, given the tremendous energies needed for heavy neutrino production. A number of more promising lepton number violating processes have been proposed to probe the Majorana nature of neutrinos. The best known example is \bbonu, subject of this review and introduced in sect.~\ref{sec:bb0nu}. We anticipate that \bbonu\ is considered the most promising probe of the Majorana nature of neutrinos. However, and because of neutrino mixing, the phenomenology associated with $|\Delta L|=2$ processes is very rich. The basic process with $|\Delta L|=2$ is mediated by \cite{Atre:2005eb}:
\begin{equation}
W^-W^-\to l^-_{\alpha}l^-_{\beta}
\label{eq:dl2process}
\end{equation}
\noindent and we can categorize such processes according to the lepton flavors $(\alpha,\beta)$ involved. Assuming no lepton flavor violating contributions other than light Majorana neutrino exchange, the matrix element for the generic $|\Delta L|=2$ process in eq.~(\ref{eq:dl2process}) is proportional to the element $(\alpha,\beta)$ of the \emph{neutrino mass matrix}:
\begin{equation}
\left(m_{\nu}\right)_{\alpha\beta}\equiv \left(U^{\ast}\text{diag}(m_1,m_2,m_3)U^{\dagger}\right)_{\alpha\beta}=
 \sum_{i=1}^3 U^{\ast}_{\alpha i}U^{\ast}_{\beta i}m_i
\label{eq:mnu}
\end{equation}
\noindent where $m_{\nu}=M^{\text{light}}$ is the matrix appearing in eq.~(\ref{eq:diagonalization3_3families}), $U_{\alpha i}$ are the elements of the 3$\times$3 neutrino mixing matrix appearing in eq.~\ref{eq:neutrinomixing}, and $m_i$ are the three light neutrino masses. In a sense, this effective neutrino mass definition provides a metric to compare the sensitivity of various $|\Delta L|=2$ processes. The processes with the most competive constraints on $|\Delta L|=2$ processes involving the flavors $(\alpha,\beta)$ are reported in table \ref{tab:lviol_processes}. \\

\begin{table}[t!b!]
\caption{\label{tab:lviol_processes}Current bounds on effective neutrino masses from total lepton number violating processes, organized according to the flavors involved. Numbers taken (or derived) from \cite{Nakamura:2010zzi,Atre:2005eb}.}
\begin{tabular}{clll} \hline
Flavors & Exp.\ technique & Exp.\ bound & Mass bound (eV) \\ \hline
$(e,e)$ & \bbonu\ & $T_{1/2}(\GE \to {^{76}\mathrm{Se}}+2e^-)>1.9 \times 10^{25}\ \mathrm{yr}$ & $|m_{ee}| < 3.6 \times 10^{-1}$ \\ \hline
\multirow{2}{*}{$(e,\mu)$} & \multirow{2}{*}{$\mu^- \to e^+$ conversion} & $\Gamma(\mathrm{Ti}+\mu^- \to e^+ +\mathrm{Ca}_{\rm gs})\ /$ & \multirow{2}{*}{$|m_{e\mu}| < 1.7 \times 10^7$} \\
 & & $\Gamma(\text{Ti}+\mu^-\text{capture})<1.7 \times 10^{-12}$ & \\ \hline
$(e,\tau)$ & Rare $\tau$ decays & $\Gamma(\tau^-\to e^+\pi^-\pi^-)/\Gamma_{\text{tot}}<8.8 \times 10^{-8}$ & $|m_{e\tau}| < 2.6 \times 10^{12}$ \\ \hline
$(\mu,\mu)$ & Rare kaon decays & $\Gamma(K^+\to \pi^-\mu^+\mu^+)/\Gamma_{\text{tot}}<1.1 \times 10^{-9}$ & $|m_{\mu\mu}| < 2.9 \times 10^{8}$ \\ \hline
$(\mu,\tau)$ & Rare $\tau$ decays & $\Gamma(\tau^-\to \mu^+\pi^-\pi^-)/\Gamma_{\text{tot}}<3.7 \times 10^{-8}$ & $|m_{e\tau}| < 2.1 \times 10^{12}$ \\ \hline
$(\tau,\tau)$ & none & none & none \\ \hline 
\end{tabular}
\end{table}

As is apparent in table \ref{tab:lviol_processes}, indeed the constraint on the effective Majorana mass $m_{ee}$ coming from \bbonu\ searches outperforms by several orders of magnitude other searches involving a different flavor combination $(\alpha,\beta)$. The most important reason behind this is of statistical nature. While it is possible to amass macroscopic quantities of a \bb\ emitter to study \bbonu\ decay (as we will see, even a ton of isotope is in the cards of several experiments), this is not the case for the other experimental techniques listed in table \ref{tab:lviol_processes}. Nevertheless, it is important to keep exploring lepton flavor violating processes other than \bbonu\ for two reasons. First, it is in principle possible that phase cancellations are such that $m_{ee} \ll m_{\alpha\beta}$ with $(\alpha,\beta)\neq (e,e)$, making the search for \bbonu\ much less favorable than others because of Nature's choice of neutrino masses and mixings. Second, this effort may possibly lead to the identification of an even most promising experimental probe of lepton flavor violation in the future.

%% file: src/bb0nu.tex
\subsection{Double beta decay modes} \label{subsec:bbmodes}
\begin{figure}[t!b!]
\begin{center}
\includegraphics[width=0.8\textwidth]{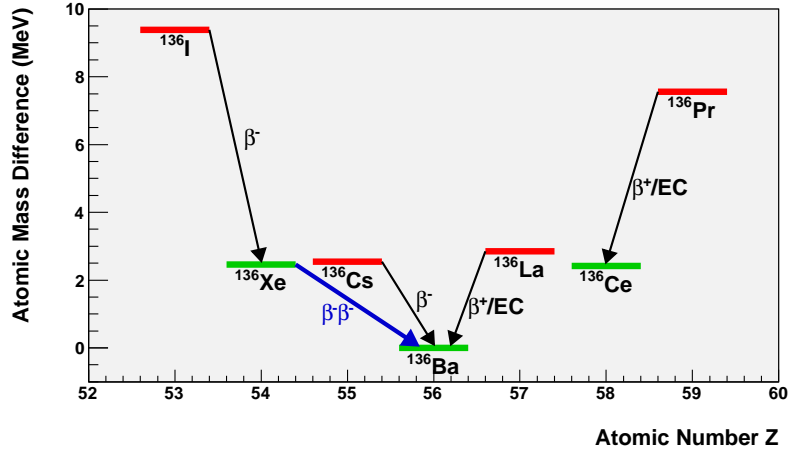}
\end{center}
\caption{Atomic masses of $A=136$ isotopes. Masses are given as differences with respect to the most bound isotope, $^{136}\text{Ba}$. The red (green) levels indicate odd-odd (even-even) nuclei. The arrows $\beta^-$, $\beta^+$, $\beta^-\beta^-$ indicate nuclear decays accompanied by electron, positron and double electron emission, respectively. The arrows EC indicate electron capture transitions.} \label{fig:atomicmasses_a136}
\end{figure}

Double beta decay is a rare nuclear transition in which a nucleus with $Z$ protons decays into a nucleus with $Z+2$ protons and the same mass number $A$. The decay can occur only if the initial nucleus is less bound than the final nucleus, and both more than the intermediate one, as shown in fig.~\ref{fig:atomicmasses_a136}. Such a condition is fulfilled by 35 nuclides in nature because of the nuclear pairing force (see sect.~\ref{subsubsec:nme_qrpa}), ensuring that nuclei with even $Z$ and $N$ are more bound than the odd-odd nuclei with the same $A=Z+N$.

The standard decay mode (\bbtnu), consisting in two simultaneous beta decays,
\begin{equation}
(Z,A) \rightarrow (Z+2,A) + 2\ e^{-} + 2\ \overline{\nu}_{e},
\label{eq:bb2nu}
\end{equation}
was first considered by Maria Goeppert-Mayer in 1935 \cite{GoeppertMayer:1935qp}. Total lepton number is conserved in this mode, and the process is allowed in the Standard Model of particle physics. This process was first detected in 1950 using geochemical techniques \cite{Inghram:1950qv}. The first direct observation of \bbtnu\ events, in \SE\ and using a time projection chamber, did not happen until 1987 \cite{Elliott:1987kp}. Since then, it has been repeatedly observed in several nuclides, such as \GE, \MO\ or \ND. Typical lifetimes are of the order of $10^{18}$--$10^{20}$ years, the longest ever observed among radioactive decay processes. For a list of \bbtnu\ half-lives measured in several isotopes, see table \ref{tab:bb2nu_exp} \cite{Barabash:2010ie}. The longest half-life in tab.~\ref{tab:bb2nu_exp} is the one for \XE , which has been measured for the first time only in 2011 \cite{Ackerman:2011gz}\footnote{The 10\% accuracy in the \XE\ \bbtnu\ decay measured half-life in \cite{Ackerman:2011gz} should be contrasted with a spread of more than one order of magnitude in the corresponding theoretical expectations from several nuclear structure calculations \cite{Staudt:1990qi,Engel:1988au,Caurier:1996zz}.}.

\begin{table}[t!b!]
\caption{Current best direct measurements of the half-life of \bbtnu\ processes.  The values reported are taken from the averaging procedure described in \cite{Barabash:2010ie}.} \label{tab:bb2nu_exp}
\begin{center}
\begin{tabular}{rcc}
\hline
Isotope & $T_{1/2}^{2\nu}\ \text{(year)}$ & Experiments \\ \hline
$^{48}\text{Ca}$ & $(4.4^{+0.6}_{-0.5}) \times 10^{19}$ & Irvine TPC \cite{Balysh:1996vr}, TGV \cite{Brudanin:2000in}, NEMO3 \cite{Flack:2008tf}  \\
$^{76}\text{Ge}$ & $(1.5\pm 0.1)\times 10^{21}$ & PNL-USC-ITEP-YPI \cite{Avignone:1994kn}, IGEX \cite{Morales:1998hu}, H-M \cite{Dorr:2003gf}  \\
$^{82}\text{Se}$ & $(0.92\pm 0.07)\times 10^{20}$ & NEMO3 \cite{Arnold:2005rz}, Irvine TPC \cite{Elliott:1992cf}, NEMO2 \cite{Arnold:1997ah}   \\
$^{96}\text{Zr}$ & $(2.3\pm 0.2)\times 10^{19}$ & NEMO2 \cite{Arnold:1999vg}, NEMO3 \cite{Argyriades:2009ph} \\
$^{100}\text{Mo}$ & $(7.1\pm 0.4)\times 10^{18}$ & NEMO3 \cite{Arnold:2005rz}, NEMO-2 \cite{Dassie:1994ru}, Irvine TPC \cite{DeSilva:1997cp} \\
$^{116}\text{Cd}$ & $(2.8\pm 0.2)\times 10^{19}$ & NEMO3 \cite{Flack:2008tf}, ELEGANT \cite{Ejiri:1995kd}, Solotvina \cite{Danevich:2003ef}, NEMO2 \cite{Arnold:1996wp}   \\
$^{130}\text{Te}$ & $(6.8^{+1.2}_{-1.1})\times 10^{20}$ & CUORICINO \cite{Arnaboldi:2002te}, NEMO3 \cite{Tretyak:2009zz}  \\
$^{136}\text{Xe}$ & $(2.11\pm 0.21)\times 10^{21}$ & EXO-200 \cite{Ackerman:2011gz} \\
$^{150}\text{Nd}$ & $(8.2\pm 0.9)\times 10^{18}$ & Irvine TPC \cite{DeSilva:1997cp}, NEMO3 \cite{Argyriades:2008pr}  \\
\hline
\end{tabular}
\end{center}
\end{table}

The neutrinoless mode (\bbonu),
\begin{equation}
(Z,A) \rightarrow (Z+2,A) + 2\ e^{-},
\label{eq:bb0nu}
\end{equation}
was first proposed by W.~H.~Furry in 1939 \cite{Furry:1939gr} as a method to test Majorana's theory \cite{Majorana:1937vz} applied to neutrinos. In contrast to the two-neutrino mode, the neutrinoless mode violates total lepton number conservation and is therefore forbidden in the Standard Model. Its existence is linked to that of Majorana neutrinos (see sect.~\ref{subsec:bb0nu_blackbox}). No convincing experimental evidence of the decay exists to date (see sect.~\ref{subsec:bb0nu_expstatus}).

The two modes of the \bb\ decay have some common and some distinct features \cite{Vogel:2008sx}. The common features are:
\begin{itemize}
\item The leptons carry essentially all the available energy, and the nuclear recoil is negligible;
\item The transition involves the $0^+$ ground state of the initial nucleus and, in almost all cases, the $0^+$ ground state of the final nucleus. For some isotopes, it is energetically possible to have a transition to an excited $0^+$ or $2^+$ final state\footnote{The transition to an excited $0^+$ final state has been observed for both $^{100}\text{Mo}$ \cite{Barabash:1995fn,Arnold:2006fk,Kidd:2009ai} and $^{150}\text{Nd}$ \cite{Barabash:2009wy}.}, even though they are suppressed because of the smaller phase space available;
\item Both processes are second-order weak processes, \emph{i.e.} their rate is proportional to $G_F^4$, where $G_F$ is the Fermi constant. They are therefore inherently slow. Phase space considerations alone would give preference to the \bb0nu\ mode which is, however, forbidden by total lepton number conservation. 
\end{itemize}

\begin{figure}[t]
\vspace{1cm}
\begin{center}
\includegraphics[angle=0,scale=0.4]{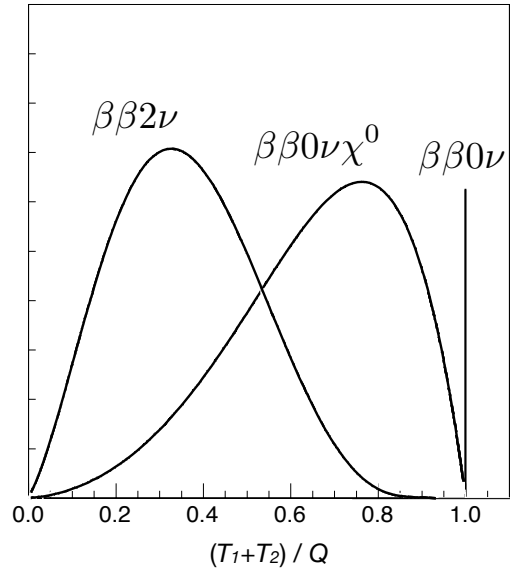}
\end{center}
\caption{Spectra for the sum kinetic energy $T_1+T_2$ of the two electrons, for different \bb\ modes: \bbtnu, \bbonu, and \bb\ decay with Majoron emission.} \label{fig:modes}
\end{figure}

The distinct features are:
\begin{itemize}
\item In the \bbtnu\ mode the two neutrons undergoing the transition are uncorrelated (but decay simultaneously), while in the \bbonu\ mode the two neutrons are correlated;
\item in the \bbtnu\ mode, the sum electron kinetic energy $T_1+T_2$ spectrum is continuous and peaked below $Q_{\beta\beta}/2$, where $Q_{\beta\beta}$ is the Q-value of the reaction. In the \bbonu\ mode, since no light particles other than the electrons are emitted and given that nuclear recoil is negligible, the $T_1+T_2$ spectrum is a mono-energetic line at $Q_{\beta\beta}$, smeared only by the detector resolution. This is illustrated in fig.~\ref{fig:modes}.
\end{itemize}

In addition to the the two basic decay modes described above, several decay modes involving the emission of a light neutral boson, the Majoron ($\chi^{0}$), have been proposed in extensions of the Standard Model, see sect.~\ref{subsec:bb0nu_alternativemechanisms}.

While in the following we will focus on \bbonu\ as defined in eq.~(\ref{eq:bb0nu}), there are three closely related lepton number violating processes that can be investigated:
\begin{eqnarray}
\beta^+\beta^+0\nu: & (Z,A) \to (Z-2,A) + 2\ e^+  \label{eq:b+b+}\\
\beta^+\text{EC}0\nu: & e^- + (Z,A) \to (Z-2,A) + e^+ \label{eq:b+EC}\\
\text{ECEC}0\nu: & 2\ e^- + (Z,A) \to (Z-2,A)^\ast \label{eq:ECEC}
\end{eqnarray}
Such processes are called \emph{double positron emission}, \emph{single positron emission plus single electron capture} (EC), and \emph{double electron capture}, respectively. All three involve transitions where the nuclear charge decreases (as opposed to increasing, as in \bbonu) by two units. From the the theoretical point of view, the physics probed by $\beta^+\beta^+0\nu$, $\beta^+\text{EC}0\nu$ and $\text{ECEC}0\nu$ is identical to the one probed by \bbonu. From the experimental point of view, however, $\beta^+\beta^+0\nu$ and $\beta^+\text{EC}0\nu$ are less favorable than \bbonu\ because of the smaller phase space available. On the other hand, the process $\text{ECEC}0\nu$ is gaining some attention recently as a promising (but still much less developed) alternative to \bbonu, since a resonant enhancement of its rate can in principle occur \cite{Eliseev:2011zza}.

In the following, the neutrinoless mode \bbonu\ is discussed in more detail, from both the theoretical and experimental point of views.

\subsection{The black box theorem } \label{subsec:bb0nu_blackbox}
\begin{figure}[t!b!]
\begin{center}
\includegraphics[scale=0.65]{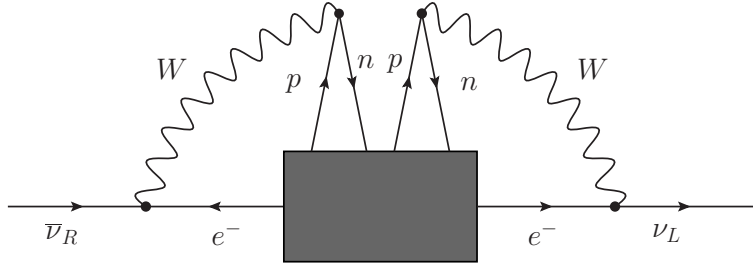}
\end{center}
\caption{Diagram showing how any neutrinoless double beta decay process induces a $\bar{\nu}$-to-$\nu$ transition, that is, an effective Majorana mass term. This is the so-called \emph{black box theorem} \cite{Schechter:1981bd}.} \label{fig:blackbox}
\end{figure}

In general, in theories beyond the Standard Model there may be several sources of total lepton number violation which can lead to \bbonu. Nevertheless, as it was first pointed out in reference \cite{Schechter:1981bd}, irrespective of the mechanism, \bbonu\ necessarily implies Majorana neutrinos. This is called the \emph{black box} (or Schechter-Valle) theorem. The reason is that any $\Delta L\neq 0$ diagram contributing to the decay would also contribute to the $(e,e)$ entry of the Majorana neutrino mass matrix, $(m_{\nu})_{ee}$. This is shown in fig.~\ref{fig:blackbox}, where a $\bar{\nu}_e-\nu_e$ transition, that is a non-zero $(m_{\nu})_{ee}$, is induced as a consequence of any $\Delta L\neq 0$ operator responsible for \bbonu.

From a quantitative point of view, however, the diagram in fig.~\ref{fig:blackbox} corresponds to a tiny mass generated at the four-loop level, and is far too small to explain the neutrino mass splittings observed in neutrino oscillation experiments \cite{Duerr:2011zd}. Other, unknown, Majorana and/or Dirac mass contributions must exist. As a consequence, therefore, the black box theorem says nothing about the physics mechanism dominating a \bbonu\ rate that is large enough to be observable. The dominant mechanism leading to \bbonu\ could then either be directly connected to neutrino oscillations phenomenology, or only indirectly connected or not connected at all to it \cite{Rodejohann:2011mu}. The former case is realized in the standard \bbonu\ mechanism of light neutrino exchange, discussed in sect.~\ref{subsec:bb0nu_lightmajoranaexchange}. The latter case involves alternative \bbonu\ mechanisms, briefly outlined in sect.~\ref{subsec:bb0nu_alternativemechanisms}.

\subsection{The standard \bbonu\ mechanism: light Majorana neutrino exchange} \label{subsec:bb0nu_lightmajoranaexchange}

\begin{figure}[t!b!]
\begin{center}
\includegraphics[scale=0.65]{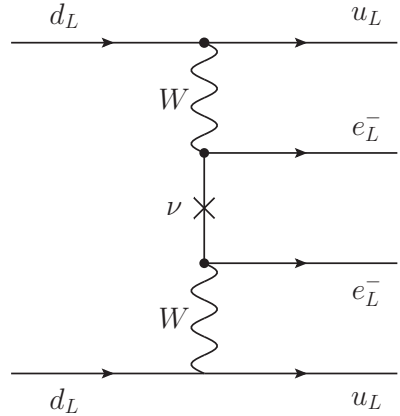}
\end{center}
\caption{\label{fig:bb0nu_standardmechanism}The standard mechanism for \bbonu\ decay, based on light Majorana neutrino exchange.}   
\end{figure}

Neutrinoless double beta decay can arise from a diagram (fig.~\ref{fig:bb0nu_standardmechanism}) in which the parent nucleus emits a pair of virtual $W$ bosons, and then these $W$ exchange a Majorana neutrino to produce the outgoing electrons. The rate is non-zero only for massive, Majorana neutrinos. The reason is that the exchanged neutrino in fig.~\ref{fig:bb0nu_standardmechanism} can be seen as emitted (in association with an electron) with almost total positive helicity. Only its small, $\mathcal{O}(m/E)$, negative helicity component is absorbed at the other vertex by the Standard Model electroweak current. Considering that the amplitude is in this case a sum over the contributions of the three light neutrino mass states $\nu_i$, and that is also proportional to $U_{ei}^2$, we conclude that the modulus of the amplitude for the \bbonu\ process must be proportional in this case to the \emph{effective neutrino Majorana mass}:
\begin{equation}
\mbb \equiv \Bigg| \sum_{i=1}^3 m_i U_{ei}^2 \Bigg|. \label{eq:mbb}
\end{equation}
In other words, the effective neutrino Majorana mass corresponds to the modulus of the $(e,e)$ element of the neutrino mass matrix of eq.~(\ref{eq:mnu}), $\mbb\equiv |\left(m_{\nu}\right)_{ee}|$.

In the case where light Majorana neutrino exchange is the dominant contribution to \bbonu, the inverse of the half-life for the process can be written as \cite{Doi:1985dx}:
\begin{equation}
\frac{1}{T^{0\nu}_{1/2}} = G^{0\nu}(Q,Z)\ |M^{0\nu}|^2\ \mbb^2,
\label{eq:Tonu}
\end{equation}
where \Gonu\ is a phase space factor that depends on the transition $Q$-value and on the nuclear charge $Z$, and $M^{0\nu}$ is the nuclear matrix element (NME). The phase space factor can be calculated analytically, in principle with reasonable accuracy\footnote{An accurate description of the effect of the nuclear Coulomb field on the decay electron wave-functions is, however, required.}. The  NME is evaluated using nuclear models, although with considerable uncertainty (see sect.~\ref{sec:nme}).
In other words, the value of the effective neutrino Majorana mass \mbb\ in eq.~(\ref{eq:mbb}) can be inferred from a non-zero \bbonu\ rate measurement, albeit with some nuclear physics uncertainties. Conversely, if a given experiment does not observe the \bbonu\ process, the result can be interpreted in terms of an upper bound on \mbb.  

If light Majorana neutrino exchange is the dominant mechanism for \bbonu, it is clear from eq.~(\ref{eq:mbb}) that \bbonu\ is in this case directly connected to neutrino oscillations phenomenology, and that it also provides direct information on the absolute neutrino mass scale, as cosmology and $\beta$ decay experiments do (see sect.~\ref{subsec:massivenus_whereweare}). The relationship between \mbb\ and the actual neutrino masses $m_i$ is affected by:
\begin{enumerate}
\item the uncertainties in the measured oscillation parameters;
\item the unknown neutrino mass ordering (normal or inverted);
\item the unknown phases in the neutrino mixing matrix (both Dirac and Majorana). 
\end{enumerate}

\begin{figure}[t!b!]
\begin{center}
\includegraphics[width=0.60\textwidth]{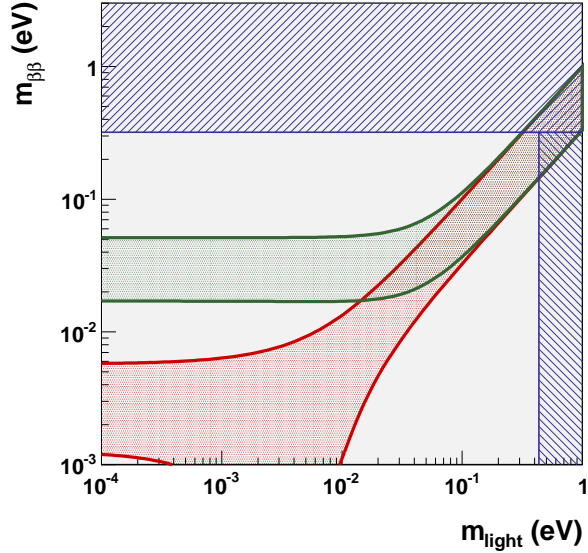}
\end{center}
\caption{\label{fig:mbetabetavsmlight}The effective neutrino Majorana mass \mbb\ as a function of the lightest neutrino mass, $m_{\text{light}}$. The red (green) band corresponds to the normal (inverted) ordering, respectively, in which case $m_{\text{light}}$ is equal to $m_1$ ($m_3$). The vertically-excluded region comes from cosmological bounds, the horizontally-excluded one from \bbonu\ constraints.}
\end{figure}

For example, the relationship between \mbb\ and the lightest neutrino mass $m_{\text{light}}$ (which is equal to $m_1$ or $m_3$ in the normal and inverted mass ordering cases, respectively) is illustrated in fig.~\ref{fig:mbetabetavsmlight}. This graphical representation was first proposed in \cite{Vissani:1999tu}. The width of the two bands is due to items 1 and 3 above, where the uncertainties in the measured oscillation parameters (item 1) are taken as $3\sigma$ ranges from a recent global oscillation fit \cite{Fogli:2011qn}. Figure \ref{fig:mbetabetavsmlight} also shows an upper bound on $m_{\text{light}}$ from cosmology ($m_{\text{light}}<0.43\ \text{eV}$), also shown in fig.~\ref{fig:mass_constraints_cosmo_beta}, and an upper bound on \mbb\ from current \bbonu\ data ($m_{\beta\beta}<0.32\ \text{eV}$), which we will discuss in sect.~\ref{subsec:bb0nu_expstatus}. As can be seen from fig.~\ref{fig:mbetabetavsmlight}, current \bbonu\ data provide a constraint on the absolute mass scale $m_{\text{light}}$ that is almost as competitive as the cosmological one.

In figs.~\ref{fig:mass_constraints_cosmo_beta} and \ref{fig:mbetabetavsmlight}, we have shown only upper bounds on various neutrino mass combinations, coming from current data. The detection of positive results for absolute neutrino mass scale observables would open up the possibility to further explore neutrino properties and lepton number violating processes. We give three examples in the following. First, the successful determination of both $m_{\beta}$ in eq.~(\ref{eq:mbeta}) and \mbb\ in eq.~(\ref{eq:mbb}) via $\beta$ and \bbonu\ decay experiments, respectively, can in principle be used to determine or constrain the phases $\alpha_{i}$ \cite{Avignone:2007fu}. Second, measurements of $m_{\beta}$ or $m_{\text{cosmo}}$ in eq.~(\ref{eq:mcosmo}) may yield a constraint on $m_{\text{light}}$ that is inconsistent with a \mbb\ upper limit. In this case, the non-observation of \bbonu\ would suggest that neutrinos are Dirac particles. Third, measurements of $m_{\beta}$ or $m_{\text{cosmo}}$ may yield a constraint on $m_{\text{light}}$ that is inconsistent with a measured non-zero \mbb. This scenario would demonstrate that additional lepton number violating physics, other than light Majorana neutrino exchange, is at play in the \bbonu\ process. We briefly describe some of these possible \bbonu\ alternative mechanisms in the following.

\subsection{Alternative \bbonu\ mechanisms} \label{subsec:bb0nu_alternativemechanisms}
\begin{figure}[t!b!]
\begin{center}
\begin{tabular}{ccc}
\includegraphics[width=0.30\textwidth]{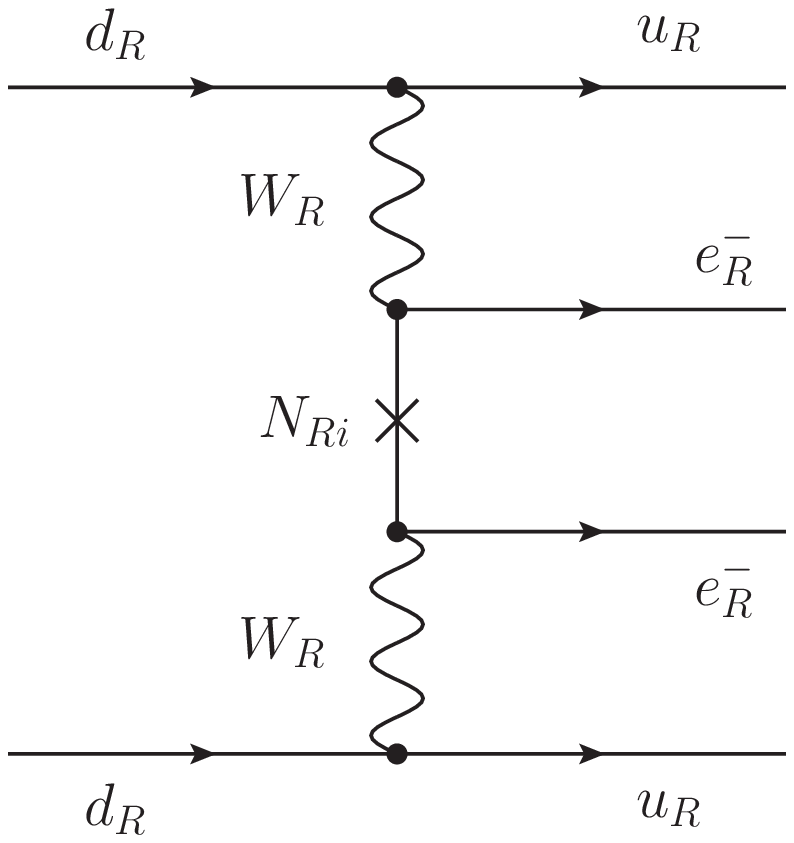} &
\includegraphics[width=0.30\textwidth]{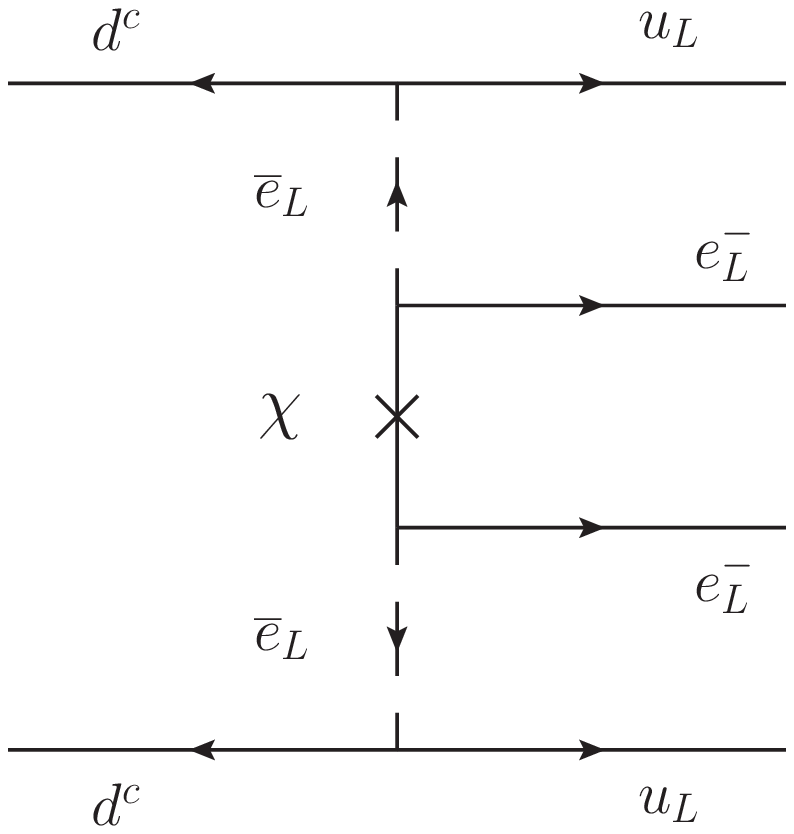} &
\includegraphics[width=0.30\textwidth]{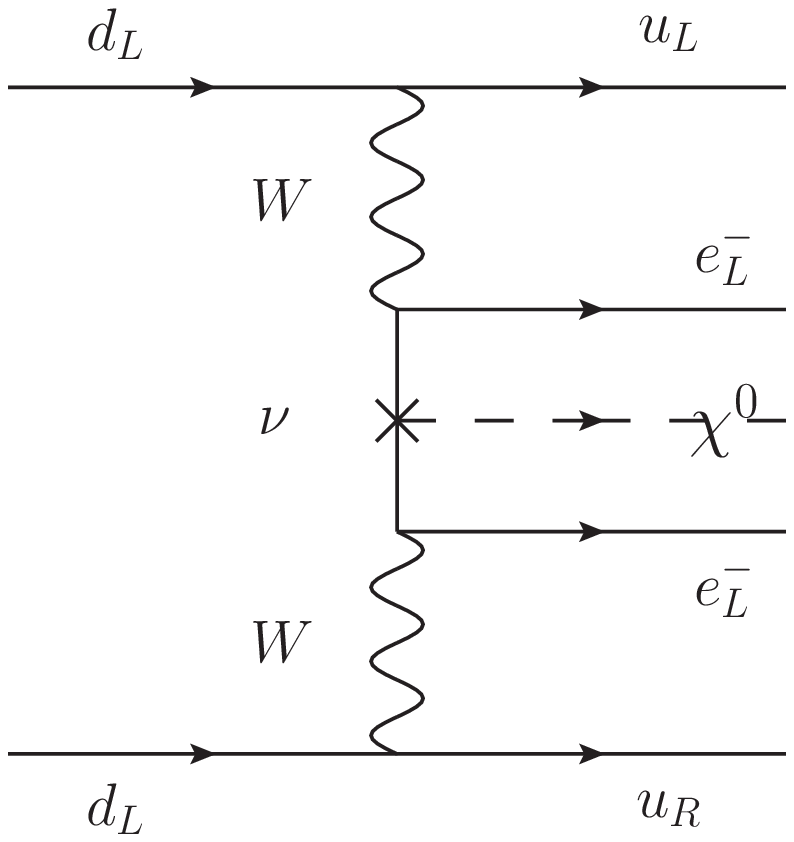} \\
(a) & (b) & (c) \\
\end{tabular}
\end{center}
\caption{\label{fig:nonstandardmechanisms}Examples of non-standard mechanism for \bbonu: (a) heavy neutrino exchange with positive chirality currents \cite{Mohapatra:1986pj}; (b) neutralino exchange in R-parity violating supersymmetry \cite{Mohapatra:1986su}; (c) Majoron emission \cite{Georgi:1981pg}.}   
\end{figure}

A number of alternative \bbonu\ mechanisms have been proposed. For an excellent and complete discussion of those, we refer the reader to \cite{Rodejohann:2011mu}. The realization of \bbonu\ can differ from the standard mechanism in one or several aspects:
\begin{itemize}
\item The Lorentz structure of the currents. Positive chirality currents mediated by a $W_R$ boson can arise, for example, in left-right symmetric theories. A possible diagram involving positive chirality current interactions of heavy Majorana neutrinos $N_i$ is shown in fig.~\ref{fig:nonstandardmechanisms}(a).
\item The mass scale of the exchanged virtual particles. One example would be the presence of ``sterile'' (that is, described by positive chirality fields) neutrinos, either light or heavy, in the neutrino propagator of fig.~\ref{fig:bb0nu_standardmechanism}, in addition to the three light, active, neutrinos we are familiar with. Another example would be the exchange of heavy supersymmetric particles, as in fig.~\ref{fig:nonstandardmechanisms}(b).
\item The number of particles in the final state. A popular example involves decay modes where additional Majorons, that is very light or massless particles which can couple to neutrinos, are produced in association with the two electrons (see fig.~\ref{fig:nonstandardmechanisms}(c)).
\end{itemize}
In non-standard \bbonu\ mechanisms, the scale of the lepton number violating physics is often larger than the momentum transfer, in which case one speaks of \emph{short-range} processes. This is in contrast to the standard \bbonu\ mechanism of light Majorana neutrino exchange, in which case the neutrino is very light compared to the energy scale, resulting in a \emph{long-range} process. Non-standard and long-range \bbonu\ processes are, however, also possible.

In general, several contributions to the total \bbonu\ amplitude can add coherently, allowing for interference effects. Neutrinoless double beta decay observables alone may be able to identify the dominant mechanism responsible for \bbonu. We give two examples. First, if Majorons are also emitted in association with the two electrons, energy conservation alone requires the electron kinetic energy sum $T_1+T_2$ to be a continuous spectrum with $Q_{\beta\beta}$ as endpoint. This spectrum is potentially distinguishable from the \bbtnu\ one (see fig.~\ref{fig:modes}), provided that the Majoron-neutrino coupling constant is large enough. Second, if positive chirality current contributions dominate the \bbonu\ rate, electrons will be emitted predominantly as positive helicity states. As a consequence, both the energy and angular correlation of the two emitted electrons will be different from the ones of the standard \bbonu\ mechanism. A detector capable of reconstructing individual electron tracks may therefore be able to distinguish this type of non-standard \bbonu\ mechanism from light Majorana neutrino exchange (see, for example, \cite{Arnold:2010tu}).

\subsection{Existing experimental results} \label{subsec:bb0nu_expstatus}

Neutrinoless double beta decay searches have been carried out over more than half a century, exploiting the same experimental techniques used for measuring the two-neutrino mode rate (see also sect.~\ref{subsec:past}). Several \bb\ emitting isotopes have been investigated, as shown in table \ref{tab:bb0nu_exp}. 

\begin{table}[t!b!]
\begin{center}
\begin{narrowtabular}{1cm}{rcc}
\hline
Isotope & $T_{1/2}^{0\nu}\ \text{(years)}$ & Experiment \\ \hline
$^{48}\text{Ca}$ & $>5.8 \times 10^{22}$ & ELEGANT \cite{Umehara:2008ru} \\
$^{76}\text{Ge}$ & $>1.9 \times 10^{25}$ & Heidelberg-Moscow \cite{KlapdorKleingrothaus:2000sn} \\
$^{82}\text{Se}$ & $>3.6 \times 10^{23}$ & NEMO3 \cite{Barabash:2010bd} \\
$^{96}\text{Zr}$ & $>9.2 \times 10^{21}$ & NEMO3 \cite{Argyriades:2009ph}  \\
$^{100}\text{Mo}$ & $>1.1 \times 10^{24}$ & NEMO3 \cite{Barabash:2010bd}  \\
$^{116}\text{Cd}$ & $>1.7 \times 10^{23}$ & Solotvina \cite{Danevich:2003ef}   \\
$^{130}\text{Te}$ & $>2.8 \times 10^{24}$ & CUORICINO \cite{Andreotti:2010vj} \\
$^{136}\text{Xe}$ & $>4.5 \times 10^{23}$ & DAMA \cite{Bernabei:2002bn} \\
$^{150}\text{Nd}$ & $>1.8 \times 10^{22}$ & NEMO3 \cite{Argyriades:2008pr} \\
\hline
\end{narrowtabular}
\end{center}
\caption{\label{tab:bb0nu_exp}Current best limits on the half-life of \bbonu\ processes for the most interesting isotopes. All values are at 90\% CL.}
\end{table}

The most sensitive limit to date was set by the Heidelberg-Moscow (HM) experiment \cite{KlapdorKleingrothaus:2000sn}: \mbox{$T^{0\nu}_{1/2}(^{76}{\rm Ge}) > 1.9 \times 10^{25}$} years (90\% CL), corresponding to an effective Majorana mass bound of $\mbb\ < 0.32\ \text{eV}$. A subset of this collaboration claimed  to observe evidence for a \bbonu\ signal, with a best value for the half-life of $1.5\times10^{25}$ years \cite{KlapdorKleingrothaus:2001ke}. The claim is very controversial \cite{Aalseth:2002dt}. Also, a subsequent re-analysis by the same group updated this result to $(2.23^{+0.44}_{-0.31}) \times 10^{25}\ \text{years}$ \cite{KlapdorKleingrothaus:2006ff}, resulting in an effective Majorana mass of about 0.30 eV. The authors claim a statistical significance for the evidence of 6$\sigma$, and do not present any systematic uncertainty analysis. It should be mentioned that a different germanium detector, IGEX, did not observe any evidence for \bbonu, and reported a lower bound on the half-life of:  \mbox{$T^{0\nu}_{1/2}(^{76}{\rm Ge}) > 1.57 \times 10^{25}$} years (90\% CL) \cite{Aalseth:2002rf}.

%% file: src/nme.tex
All nuclear structure effects in $\beta\beta0\nu$ are included in the nuclear matrix element (NME). Its knowledge is essential in order to relate the measured half-life to the neutrino masses, and therefore to compare the sensitivity and results of different experiments, as well as to predict which are the most favorable nuclides for $\beta\beta0\nu$ searches. Unfortunately, NMEs cannot be separately measured, and must be evaluated theoretically.

In the last few years the reliability of the calculations has greatly improved, with several techniques being used, namely: the Interacting Shell Model (ISM) \cite{Caurier:2007wq}; the Quasiparticle Random Phase Approximation (QRPA) \cite{Vogel:2008sx}; the Interacting Boson Model (IBM) \cite{Barea:2009zza}; and the Generating Coordinate Method (GCM) \cite{Rodriguez:2010mn}. Before briefly reviewing the different approaches, we discuss the ingredients that are common to all. 

It is beyond the scope of this review, and beyond our expertise, to provide a complete derivation of NME calculations. Here, we limit ourselves to outline the theoretical framework used to carry out the calculations, the approximations used, and the most significant differences among the different approaches. For more details, the reader should refer to \cite{Elliott:2002xe,Avignone:2007fu,Vogel:2008sx}, where most of the material covered below was taken from.

\subsection{\label{subsec:nme_commoningredients}Common elements in calculations}

The rate for the \bbonu\ process can be written as:
\begin{equation}
\label{eq:rate_firstprinciples}
 [T^{0\nu}_{1/2}]^{-1}=\sum_{\textrm{spins}} \int |Z_{0\nu}|^2
\delta(E_{e1}+E_{e2}+E_f-M_i) \frac{d^3p_1}{2\pi^3}\frac{d^3p_2}
{2\pi^3}~.
\end{equation}
In this formula, $E_{1(2)}$ and $\vec{p}_{1(2)}$ are the total energies and momenta of the electrons, $E_f(M_i)$ is the energy of the final (mass of the initial) nuclear state, and $Z_{0\nu}$ is the reaction amplitude, to be evaluated in time-dependent perturbation theory to second order in the weak interaction (that is, to second order in the Fermi constant $G_F$). The reaction amplitude can be factorized into the product of a leptonic and a hadronic part. Assuming that the decay is mediated solely by the exchange of light neutrinos (standard \bbonu\ mechanism, see sect.~\ref{subsec:bb0nu_lightmajoranaexchange}), the leptonic part can be written as the product of two negative chirality currents. After substitution for the neutrino propagator, the lepton amplitude acquires the form:
\begin{equation}
\label{eq:lep} 
-\frac{i}{4} \int \sum_k \frac{d^4q}{(2\pi)^4}e^{-i
q\cdot(x-y)} \overline{e}(x) \gamma_{\mu}(1-\gamma_5)
\frac{q^{\rho}\gamma_{\rho}+m_k}{q^2-m_k^2}(1-\gamma_5) \gamma_{\nu}
 e^c(y)~U_{ek}^2~,
\end{equation}
\noindent where the integral is over the 4-momentum transfer $q$ (that is, the momentum of the virtual neutrino), the sum $k$ is over the three neutrino mass eigenstates of mass $m_k$, the mixing matrix elements $U_{ek}$ specify the electron flavor content of the mass states, and $\overline{e}(x)$ and $e^c(y)$ are the electron creation operators. This lepton part implies a contraction over the two neutrino operators, which is allowed only if neutrinos are Majorana particles. Furthermore, from the commutation properties of the gamma matrices it follows that
\begin{equation}
\frac{1}{4}(1-\gamma_5)(q^{\rho}\gamma_{\rho}+m_k)(1-\gamma_5)=\frac{m_k}{2}(1-\gamma_5)
\label{eq:lep2}
\end{equation}
From eqs.~(\ref{eq:lep}) and (\ref{eq:lep2}), we obtain that the decay amplitude for purely negative chirality lepton currents is proportional to the neutrino Majorana mass $\sum_k U_{ek}^2 m_k$, as discussed in the previous section.

Integration over the virtual neutrino energy leads to the replacement of the propagator $(q^2-m_k^2)^{-1}$ by the residue $\pi/\omega_k$ with $\omega_k=\sqrt{\vec{q}^2+m_k^2}$. Integration over the space part $d\vec{q}$ 
leads to an expression representing the effect of the
neutrino propagation between the two nucleons. This expression has the form of a  \emph{neutrino potential}, $H(r)$, where $r <R$, and $R$ is the \emph{nuclear radius},  
$R = 1.2\ A^{1/3}$~ fm. $H(r)$ appears in the corresponding nuclear matrix elements, introducing a
dependence of the transition operator on the coordinates of the two nucleons, as well as
a weak dependence on the excitation energy $E_m -E_i$~ of the virtual state in the odd-odd
intermediate nucleus.

The momentum of the virtual neutrino is determined by the uncertainty relation $q \sim 1/r$. Here $r$~is a typical spacing between two nucleons, $r \sim 2-3$~ fm. Therefore the momentum transfer is $q \sim 100-200$~ MeV. For light neutrinos the neutrino mass $m_j$ can then be safely neglected
in the potential $H(r)$. Also, given the large value of $q$, the dependence on the
difference of nuclear energies $E_m -E_i$~ is weak.  One can then neglect the variation of the energy from state to state when integrating over the virtual neutrino energies.  In this \emph{closure approximation} the contributions of the two electrons are  added coherently, and the neutrino potential is of the form: 
\begin{equation}
H(r)=\frac{R}{r}\Phi(\omega r)
\label{eq:nup}
\end{equation}
The nuclear radius $R$ in eq.~(\ref{eq:nup}) is introduced in order to make the potential H dimensionless. A corresponding $1/R^2$ factor compensates for this auxiliary quantity in the phase space formula (see eq.~(\ref{eq:g0nu}) below). Also, in eq.~(\ref{eq:nup}), $\Phi(\omega r) \le 1$~ is a relatively slowly varying function of $r$. A typical value of $H(r)$ is larger than unity, but less than 5-10. 

To go any further, we need an expression for the hadronic current. In the \emph{impulse approximation}, the hadronic current is obtained from that of free nucleons. The latter can be written as
\begin{equation}
J^{\rho \dagger}
=  \overline{\Psi} \tau^+ \left[ g_V(q^2) \gamma^\rho
 - g_A(q^2) \gamma^\rho\gamma_5 - g_P(q^2) q^\rho \gamma_5 \right] \Psi~,
\label{eq:hadronic_current}
\end{equation}
\noindent where $m_p$ is the nucleon mass, $\Psi$ is a nucleon field, $\tau^+=(\tau_1+i\tau_2)/2$, $\tau_{1(2)}$ are the isospin Pauli matrices, and $g_V$, $g_A$ and $g_P$ are the so-called \emph{vector}, \emph{axial-vector} and \emph{induced pseudoscalar} form factors, parametrizing the composite structure of nucleons. Since in $\beta\beta0\nu$ decay we have $\vec{q}^{~2}\gg q_0^2$, we take $q^2\simeq -\vec{q}^{~2}$.

The $q^2$ dependence of the vector and axial-vector form factors is parametrized via the usual dipole approximation
\begin{equation}
\label{eq:dipole}
g_V({\vec q}^{~2}) = {g_V}/{(1+{\vec q}^{~2}/{M_V^2})^2},~~~
g_A({\vec q}^{~2}) = {g_A}/{(1+{\vec q}^{~2}/{M_A^2})^2}~,
\end{equation}
\noindent with $g_V\simeq 1$, $g_A\simeq 1.25$, $M_V\simeq 850\ \text{MeV}$, and $M_A\simeq 1090\ \text{MeV}$. It is also customary to use the \emph{Goldberger-Treiman relation} for the induced pseudoscalar term
\begin{equation}
g_P({\vec q}^{~2}) = {2 m_p g_A({\vec q}^{~2})}/({{\vec q}^{~2} + m^2_\pi}) ~.
\end{equation}
\noindent where $m_{\pi}$ is the pion mass. For the ground state to ground state transitions, \emph{i.e.} $0_i^+\to 0_f^+$, it is sufficient to consider $s$-wave outgoing electrons (\emph{long-wave approximation}), and the nonrelativisitic approximation for the nucleons. The rate then takes the form given in eq.~(\ref{eq:Tonu}), and repeated here for convenience:
\begin{equation}
\label{eq:rate}
[T^{0\nu}_{1/2}]^{-1}=G^{0\nu}(Q,Z)\ \left|M^{0\nu}\right|^2\ m_{\beta\beta}^2~,
\end{equation}
\noindent where $Q=M_i-E_f$, $G^{0\nu}(Q,Z)$ comes from the
phase-space integral, and the nuclear matrix element turns into a sum of the Gamow-Teller and Fermi nuclear matrix elements, where :
\begin{equation}
\label{eq:m0nu}
M^{0\nu}\simeq \left(\frac{g_A}{1.25} \right)^2 \left( M^{0\nu}_{GT} - \frac{g_V^2}{g_A^2} M^{0\nu}_F \right)
\end{equation}
\noindent with, to first order:
\begin{equation}
M_{F}^{0\nu} = \langle f |\sum_{a,b} H(r) \tau^+_a \tau^+_b
|i\rangle~ 
\label{eq:ferminme}
\end{equation}
and
\begin{equation}
M_{GT}^{0\nu} = \langle f |\sum_{a,b} H(r) \vec{\sigma}_a
\cdot \vec{\sigma}_b \tau^+_a \tau^+_b |i\rangle ~.
\label{eq:gamowtellernme}
\end{equation}
In these equations the neutrino potential $H(r)$ is of the form defined in eq.~(\ref{eq:nup}) (for explicit realizations of the neutrino potential see, for example, \cite{Vogel:2008sx}), $\vec{\sigma}_{a(b)}$ are spin Pauli matrices, and $|f\rangle$ ($|i\rangle$) are the final (initial) nuclear states. In contrast to $\beta\beta2\nu$, which involves only Gamov-Teller transitions through intermediate $1^+$ states (because of low momentum transfer), the nuclear matrix element for $\beta\beta0\nu$ involves all multipolarities in the intermediate odd-odd $(A,Z+1)$ nucleus, and contains both a Fermi (F) and a Gamov-Teller (GT) part. 

In eq.~(\ref{eq:rate}) the explicit form of the phase-space integral is:
\begin{eqnarray}
\label{eq:g0nu}
 G^{0\nu}(Q,Z)  =  (G_F V_{ud} g_A )^4 \left( \frac{1}{R} \right)^2 
\frac{1}{\ln(2) 32 \pi^5} \cdot \nonumber \\
  \left( \int F(Z,E_{e1})F(Z,E_{e2}) p_{e1}p_{e2}E_{e1}E_{e2} \delta(E_0 - E_{e1} - E_{e2}) dE_{e1}dE_{e2} \right)~.
\end{eqnarray}
\noindent where $E_0=Q+2m_e=M_i-M_f$ is the available energy, $p_{e1}$ ($p_{e2}$) are the electron 3-momenta, $F(Z,E)$ is the Fermi function that describes the nuclear Coulomb effect on the outgoing electrons, and $Z$ is the charge of the daughter nucleus. 

\begin{figure}[t!b!]
\begin{center}
\includegraphics[width=0.90\textwidth]{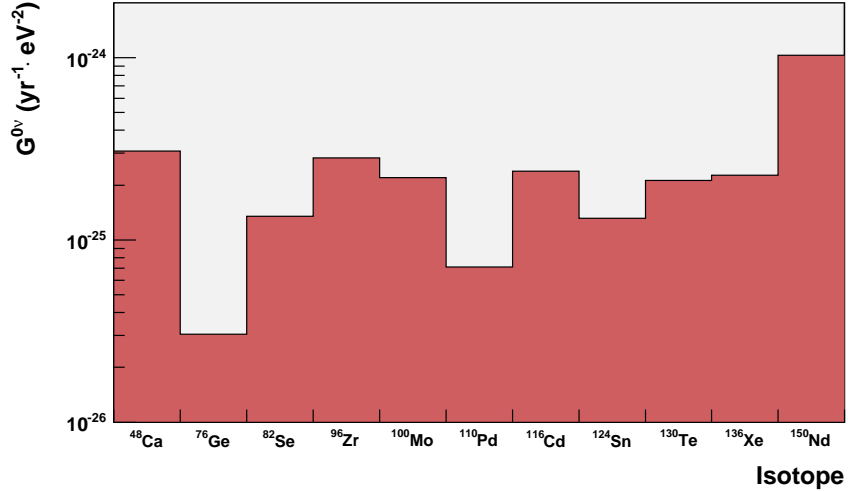}
\end{center}
\caption{ \label{fig:g0nu}The phase space factor for all ${\beta\beta}$ emitters with $Q>2$ MeV. Values taken from \cite{Pantis:1996py,Simkovic:1999re}}
\end{figure}

If an accurate result
is required, the relativistic form of the function $F(Z,E)$ must be used and a numerical
evaluation is necessary \cite{Doi:1985dx}. The phase space factor for all $\beta\beta$~ emitters with $Q > 2$~MeV are given in fig.~\ref{fig:g0nu}.

 For a qualitative picture, one can use the simplified nonrelativistic
Coulomb expression, the so-called \emph{Primakoff-Rosen approximation} \cite{Primakoff:1970jy}:

\begin{equation}
\label{eq:g0nu_pr1}
F(Z,E) = \frac{E}{p}\frac{2\pi Z \alpha}{1- e^{2\pi Z \alpha}}
\end{equation}

In this approximation, $G^{0\nu}$ is independent of Z:
\begin{equation}
\label{eq:g0nu_pr2}
G^{0\nu} \sim (\frac{E_0^5}{30}+ \frac{2 E_0^2}{3} + E_0 - \frac{2}{5})
\end{equation}
where $E_0$ is expressed in units of electron mass. Notice that the phase space dependence of the $\beta\beta0\nu$~mode goes with $E_0^5$, while the phase space
of the corresponding two-neutrino mode goes with $E_0^{11}$. That is, based on phase space considerations alone, the $\bb0\nu$ mode would be much faster than the $\beta\beta2\nu$ mode, if the neutrino mass were of the order of the electron mass. 

\begin{figure}[t!b!]
\begin{center}
\includegraphics[angle=270,width=0.70\textwidth]{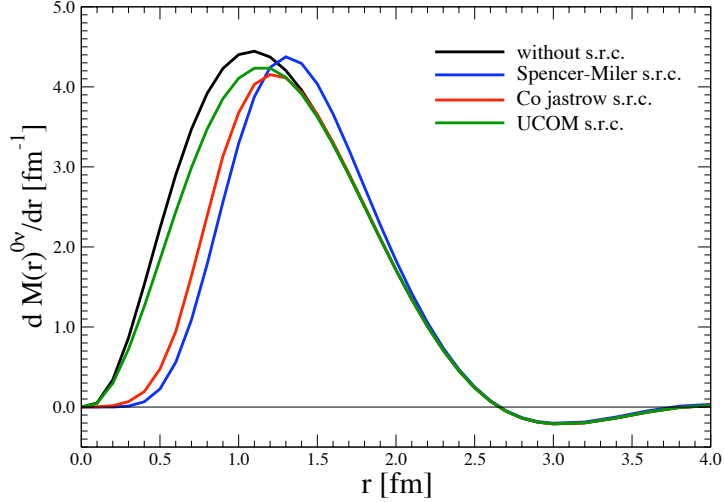}
\end{center}
\caption{ \label{fig:SRC}The dependence of $M^{0\nu}$ as a function of the distance $r$ among the two neutrons participating in \bbonu, in
$^{76}$Ge. The four curves show the effects of different treatments
of nucleon-nucleon short-range correlations \cite{Vogel:2008sx}.}
\end{figure}

In addition to the total NME in eq.~(\ref{eq:m0nu}), the different nuclear structure approaches discussed in sect.~\ref{subsec:nme_nuclear physicsmodels} allow to predict also what are the typical distances among the two decaying nucleons that contribute the most to $M^{0\nu}$. The result is shown in fig.~\ref{fig:SRC}. One can see that only relatively short distances, $r<2-3\ \text{fm}$, contribute significantly. In other words, essentially only the nearest neighbour neutrons undergo \bbonu\ transitions. Although these short distances justify the above-mentioned closure approximation, the fact that two nucleons strongly repel each other for distances $r<0.5-1.0\ \text{fm}$ should be taken into account. The nuclear wavefunctions computed according to the methods described in sect.~\ref{subsec:nme_nuclear physicsmodels} do not take such effect into account. The usual and simplest way to include this effect is by introducing a phenomenological function in eq.~(\ref{eq:m0nu}). A popular procedure to obtain such function is based on the \emph{Unitary Correlation Operator Method} (UCOM) \cite{Feldmeier:1997zh}. This procedure reduces the value of  $M^{0\nu}$ in fig.~\ref{fig:SRC} by only about 5\%. However, other prescriptions for short-range correlations introduce a much more significant (of order 20--25\%) reduction in $M^{0\nu}$.


\subsection{\label{subsec:nme_nuclear physicsmodels}The different nuclear structure approaches}

In order to compute the $\beta\beta0\nu$ decay rates for a given neutrino mass, we need to evaluate the initial and final state wavefunctions $|i\rangle$ and $|f\rangle$, and the nuclear matrix elements connecting the two in eq.~(\ref{eq:m0nu}). Given the complicated nuclear many-body nature of the problem, this calculation cannot be done exactly, and some approximations need to be introduced. Different nuclear physics approaches have been used to this end. Only a very schematic description of those will follow.

\subsubsection{\label{subsubsec:nme_ism}The Interacting Shell Model}
\begin{figure}[t!b!]
\begin{center}
\includegraphics[width=0.50\textwidth]{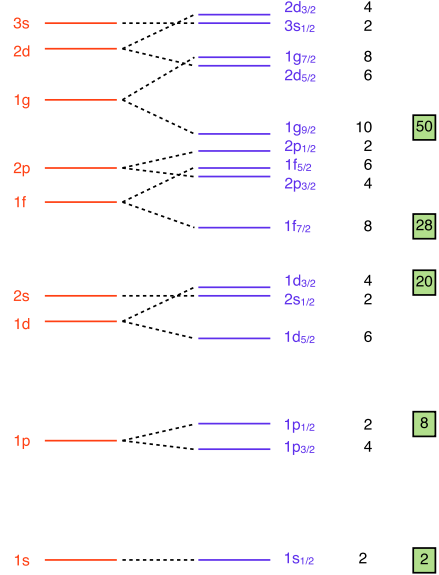}
\end{center}
\caption{ \label{fig:shells}Low-lying energy levels in a single-particle shell model with an oscillator potential (with a small negative $\vec{l}^2$ term) without spin-orbit (left) and with spin-orbit (right) interaction. The number to the right of a level indicates its degeneracy, (2j+1). The boxed integers indicate the magic numbers.}
\end{figure}

In the \emph{Interacting Shell Model} (ISM) \cite{Caurier:2004gf}, all microscopic calculations are based on the \emph{Independent Particle (Shell) Model} (IPM). The basic premise of such a model is that the nucleons are moving independently in a mean field with a strongly attractive spin-orbit term:
\begin{equation}
U(r) = \frac{1}{2} \; \hbar\omega \;r^2 + D \; \vec{l} \;^2
+ C \; \vec{l} \cdot \vec{s}.
\label{eq:ho+ls} 
\end{equation}
\noindent where the harmonic oscillator (plus the surface correction $D \; \vec{l} \;^2$) part describes the bound nucleon nature of the problem, and the spin-orbit part $C \; \vec{l} \cdot \vec{s}$ is added to give the proper separation of the subshells and explain the nuclear magic numbers, \emph{i.e.} specific values of the number of protons $Z$ and neutrons $N$ ($N$ or $Z$ = 2, 8, 20, 28, 50, 82, 126) accounting for the existence of shell closures at those occupation numbers. In Eq.~\ref{eq:ho+ls}, $\vec{l}$ is the orbital angular momentum, and $\vec{s}$ the spin of single nucleons. The effect of the spin-orbit potential on the nuclear energy levels is schematically shown in Fig.~\ref{fig:shells}.

As the number of protons and neutrons depart from the magic numbers, it becomes indispensable to include the ``residual'' two-body nucleon interaction among nucleons. This point marks the passage from the IPM to the ISM model. This residual interaction contains both a kinetic (${\cal K}$) and a potential (${\cal V}$) term:
\begin{equation}
  \label{Hec}
  {\cal H}=\sum_{ij} {\cal K}_{ij}\, a^{\dagger}_ia_j- \sum_{i\le j\, k\le
    l}{\cal V}_{ijkl}\,a^{\dagger}_ia^{\dagger}_ja_ka_l
\end{equation}
\noindent that adds one or two particles in orbits of total angular momentum $i,\, j$ and removes one or
two from orbits $k,\, l$, subject to the Pauli principle
($\{a_i^{\dagger}a_j\}=\delta_{ij}$). While complicated in practice, this approach is conceptually simple: given a good enough residual interaction ${\cal V}_{ijkl}$, the problem is reduced to diagonalizing a matrix in a sufficiently large basis (``valence space''). In this framework, a limited valence space is used but all configurations of valence nucleons are included. The ISM describes well properties of low-lying nuclear states.

\subsubsection{\label{subsubsec:nme_qrpa}The Quasiparticle Random Phase Approximation}
\begin{figure}[t!b!]
\begin{center}
\vspace{0.8cm}
\includegraphics[width=0.65\textwidth]{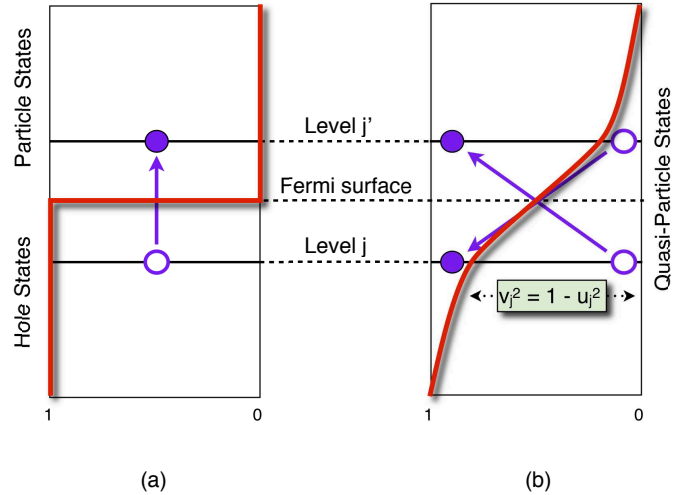}
\end{center}
\caption{ \label{fig:quasiparticles}Occupation probabilities of single particle orbitals in (a) the Independent Particle Model, and (b) with the inclusion of pairing forces. The arrows indicate possible excitations of the nucleus induced by transfer of a particle from a (partially) occupied orbital to a (partially) unoccupied orbital. Adapted from \cite{rowe2010nuclear}.}
\end{figure}

The basic idea behind the \emph{proton-neutron Quasiparticle Random Phase Approximation} (QRPA) \cite{Vogel:2008sx} is that the most important part of the residual interaction among nucleons is the \emph{pairing force}. The pairing force accounts for the tendency of nucleons to couple pairwise to especially stable configurations, \emph{i.e.} into nuclei with even $N$, even $Z$. This force favors the coupling of neutrons with neutrons, and protons with protons, so that the orbital angular momentum and spin of each couple adds to zero. As the result of the pairing force, the nuclear ground state is mainly composed of Cooper-like pairs of neutrons and protons coupled to $J^{\pi}=0^+$ total angular momentum. In QRPA, the nucleon pairing is introduced via the \emph{BCS theory} of superconductivity. A unitary (Bogoliubov) transformation is first performed to change from a particle to a \emph{quasiparticle} basis. Quasiparticles are generalized fermions which are partly particles (with probability $u_j^2$, where $j$ is the single-particle orbital the quasiparticle belongs to, see fig.~\ref{fig:quasiparticles}) and partly holes (with probability $v_j^2$). Quasiparticles are just a mathematical construct to account for pairing between like nucleons in a simple fashion while retaining the simplicity of the independent particle model, since the quasiparticles are kept, to first order, independent. This transformation smears out the nuclear Fermi surface over several orbitals, for both protons and neutrons, as shown schematically in fig.~\ref{fig:quasiparticles}.

 Once the problem has been transformed into the simpler quasiparticle basis, the QRPA goal is to evaluate the transition amplitudes associated with charge changing one-body operator $T^{JM}$ connecting the $0^+$ vacuum of quasiparticles in the even-even nucleus with any of the $J^{\pi}$ excited states in the neighboring odd-odd nuclei. Such states are described as harmonic oscillations above this vacuum. The creation of such particle-hole pairs (or \emph{phonons}) from a BCS-only vacuum would, however, overestimate this transition amplitude. The QRPA is the simplest theory which admits the possibility that the ground state is not of purely independent quasiparticle character, but may contain correlations. As a consequence, two-particle, two-hole excitations are included in the QRPA vacuum state, as opposed to the BCS vacuum. The transition amplitude is then modified as needed, since the creation of a particle-hole pair from the BCS vacuum (the so-called \emph{forward-going amplitude} $X$) can lead to the same final state $J^{\pi}$ as the destruction of a particle-hole pair from a two-particle, two-hole excitation (the \emph{backward-going amplitude} $Y$). The amplitudes $X$ and $Y$ as well as the corresponding energy eigenvalues $\omega_m$ are determined by solving the \emph{QRPA equations of motion} for each $J^{\pi}$:
\begin{equation}
\left( \begin{array}{cc}
A & B \\ -B & - A 
\end{array} \right)
\left( \begin{array}{c}
X \\ Y \end{array} \right)
= \omega \left(
 \begin{array}{c}
X \\ Y \end{array} \right) ~.
\label{eq:rpa}
\end{equation}
In eq.~(\ref{eq:rpa}) the terms $A$ and $B$ depend on the interaction matrix elements between quasi-particle configurations. They can be written in terms of particle-hole (p-h) and particle-particle (p-p) matrix elements. Customarily, these interaction matrix elements are multiplied by adjustable coupling constants $g_{ph}$ and $g_{pp}$, respectively. If a realistic nucleon-nucleon interaction is used, then the values of these constants are $g_{ph}\simeq g_{pp}\simeq 1$. These particle-particle interactions enhance the backward-going amplitude $Y$, thereby reducing the transition amplitude.

\subsubsection{\label{subsubsec:nme_gcm}The Generating Coordinate Method}
A nucleon coupling scheme that competes with nucleon pairing (see sect.~\ref{subsubsec:nme_qrpa}) to fix the equilibrium shape of a nucleus, and its collective motion, is the so-called \emph{aligned coupling scheme}. In this scheme, each nucleon has the tendency to align its orbit with the average field produced by all other nucleons. This preferentially gives rise to nuclei with deformed equilibrium shapes and collective rotational motion. A common representation of the shape of these nuclei is that of an ellipsoid. The \emph{quadrupole deformation parameter} $\beta$ is related to the eccentricity of the ellipse: $\beta\neq 0$ represents a non-spherical nucleus, with $\beta>0$ ($\beta<0$) corresponding to a prolate (oblate) ellipsoid.

Nuclear collective rotors are associated with ``intrinsic states'' very well approximated by deformed mean field determinants, that is antisymmetrized products of independent particle wavefunctions. Nuclear wavefunctions of this type represent the basic assumption of the \emph{Hartree-Fock self-consistent field theory} (see, for example, \cite{rowe2010nuclear}), an approximation for reducing the problem of many interacting particles to one of non-interacting particles in a mean field. 

In the Generating Coordinate Method with Particle Number and Angular Momentum Projected product-type wave functions (GCM-PNAMP, or GCM in short) \cite{Rodriguez:2010eu}, one starts by building a set of Hartree-Fock-Bogoliubov (HFB) intrinsic axial symmetric wave functions ${|\phi_{\beta}\rangle}$ along the quadrupole deformation parameter $\beta$. These HFB intrinsic states are found by solving the so-called \emph{constrained Particle Number Variation After Projection} equations (PN-VAP), $\delta(E^{N,Z}(|\phi_{\beta}\rangle))=0$. This is a variational equation constrained to a fixed value of the quadrupole deformation $\beta$. The Gogny D1S interaction is used as the underlying nucleon-nucleon interaction. Exact eigenstates can be found by projecting from the HFB wavefunctions the components of well-defined angular momentum, proton number and neutron number. The initial and final ground states $|0^+\rangle$ can therefore be written as \emph{GCM wavefunctions}:
\begin{equation}
|0^+\rangle = \sum_{\beta}g_{\beta}P^{I=0}P^NP^Z|\phi_{\beta}\rangle
\label{eq:gcm}
\end{equation}
\noindent where $P^{I=0}$, $P^N$ and $P^Z$ are the angular momentum ($I=0$ for axial symmetric wavefunctions), neutron number and proton number projection operators. In this method, one starts with the projected HFB approach, but allows for admixtures of different deformations $\beta$, as described by eq.~(\ref{eq:gcm}). The coefficients $g_{\beta}$ of this admixture are found by solving the so-called \emph{Hill-Wheeler-Griffin} (HWG) equation of generator coordinates (see, for example, \cite{rowe2010nuclear}).

\subsubsection{\label{subsubsec:nme_ibm}The Interacting Boson Model}
A somewhat intermediate path between the ``microscopic'' view of nuclear structure (ISM) and the ``collective'' views (QRPA, GCM) mentioned above was opened by the \emph{Interacting Boson Model}, IBM \cite{iachello1987interacting}. In the interacting boson model, collective excitations of nuclei are described by bosons, and the microscopic foundation of such collective nuclear states is rooted in the shell model. As the number of valence nucleons increases, the direct application of the shell model becomes prohibitively difficult, and some approximation is needed. First, one usually assumes that the closed shells are inert. Second, one assumes that the important particle configurations in even-even nuclei are those in which identical particles are paired together in states with total angular momentum and parity $J^P=0^+$ or $J^P=2^+$. Third, one treats the pairs as bosons, much in the same way as Cooper pairs in a gas of electrons. 

If one retains all three approximations, one is led to consider a system of interacting bosons of two types, proton bosons and neutron bosons. The proton (neutron) bosons with $J^P=0^+$ are denoted by $s_{\pi}$ ($s_{\nu}$), the ones with $J^P=2^+$ are denoted by $d_{\pi}$ ($d_{\nu}$). The multitude of shells which appears in the shell model is then reduced to the simple $s$-shell ($J=0$) and the $d$-shell ($J=2$). The number of proton ($N_{\pi}$) and neutron ($N_{\nu}$) bosons is counted from the nearest closed shell, \emph{i.e.} if less (more) than half of the shell is full, $N_{\pi(\nu)}$ is taken as the number of particle (hole) pairs. 

All fermionic operators, for example the operators yielding the Gamow-Teller and Fermi nuclear matrix elements in eqs.~\ref{eq:ferminme} and \ref{eq:gamowtellernme}, are similarly mapped into bosonic operators by the \emph{Otsuka, Arima and Iachello (OAI) method} \cite{Otsuka:1978zz}. Using this method one is assured that the matrix elements between fermionic states in the collective subspace are identical to the matrix elements in the bosonic space \cite{Barea:2009zza}. 

A realistic set of wavefunctions of even-even nuclei with mass $A\gtrsim 60$ is provided by the proton-neutron IBM-2 \cite{iachello1987interacting}. The wavefunctions are generated by diagonalizing the IBM-2 Hamiltonian. These IBM-2 wavefunctions provide an accurate description of many properties (energies, electromagnetic transition rates, quadrupole and magnetic moments, etc.) of the final and initial nuclei. Using these wavefunctions, and the bosonic operators of the OAI method, it is possible to calculate \bbonu\ NMEs (for details, see \cite{Barea:2009zza}).


\subsection{\label{subsec:nme_pmr}Quantifying uncertainties in NME calculations}

\begin{figure}[t!b!]
\begin{center}
\includegraphics[angle=90,width=0.80\textwidth]{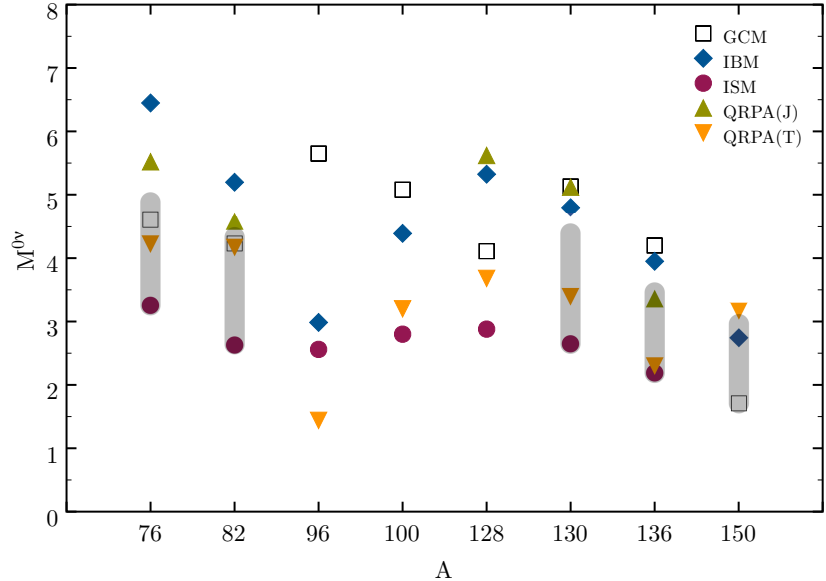}
\end{center}
\caption{ \label{fig:nme}Recent NME calculations from different techniques (GCM \cite{Rodriguez:2010mn}, IBM \cite{Barea:2009zza}, ISM \cite{Menendez:2008jp,Menendez:2009xa}, QRPA(J) \cite{Suhonen:2010zzc}, QRPA(T) \cite{Simkovic:2009pp,Simkovic:2008cu,Fang:2010qh}) with UCOM short range correlations. All the calculations use $g_A$ = 1.25; the IBM-2 results are multiplied by 1.18 to account for the difference between Jastrow and UCOM, and the RQRPA are multiplied by 1.1/1.2 so as to line them up with the others in their choice of $r_0$ = 1.2 fm. The shaded intervals correspond to the proposed physics-motivated ranges (see text for discussion).}
\end{figure}

Figure \ref{fig:nme} shows the results of the most recent NME calculations with the methods described in sect.~\ref{subsec:nme_nuclear physicsmodels}. We can see that in most cases the results of the ISM calculations are the smallest ones, while the largest ones may come from the IBM, QRPA or GCM. For a detailed study quantifying the spread of NME results resulting from different calculations, we refer the reader to ref.~\cite{Dueck:2011hu}. 
Shall the differences between the different methods in sect.~\ref{subsec:nme_nuclear physicsmodels} be treated as an uncertainty in sensitivity calculations? Should we assign an error bar to the distance between the maximum and the minimum values? This approach, we argue, does not reflect the recent progress in the theoretical understanding of the treatment of nuclear matrix elements. In quantifying the uncertainties in NME calculations, we follow \cite{GomezCadenas:2010gs}.

Each one of the major methods has some advantages and drawbacks, whose effect in the values of the NME can be sometimes explored. The clear advantage of the ISM calculations is their full treatment of the nuclear correlations, while their drawback is that they may underestimate the NMEs due to the limited number of orbits in the affordable valence spaces. It has been estimated \cite{Blennow:2010th} that the effect can be of the order of 25\%. On the contrary,  the QRPA variants, the GCM in its present form, and the IBM are bound to  underestimate the multipole correlations in one or another way.  As it is well established that these  correlations tend to diminish the NMEs, these methods should tend to overestimate them \cite{Caurier:2007wq, Menendez:2010id}.

With these considerations in mind, \emph{physics-motivated ranges} (PMR) of theoretical values for \GE, \SE, \TE , \XE\ and \ND\ NMEs have been proposed in \cite{GomezCadenas:2010gs}. In quantifying the uncertainties, the results of the major nuclear structure approaches which share the following common ingredients were considered: (a) nucleon form factors of dipole shape, see eq.~(\ref{eq:dipole}); (b) soft short-range correlations computed with the UCOM method; (c) unquenched axial coupling constant g$_{\text A}=1.25$; (d) higher order corrections to the nuclear current \cite{Simkovic:1999re} accounted for; and (e) nuclear radius $R=r_0\ A^{1/3}$, with $r_0=1.2$~fm \cite{Smolnikov:2010zz}. Therefore, the remaining discrepancies between the diverse approaches are solely due to the different nuclear wavefunctions that they employ. The uncertainties in NME calculations for \GE, \SE, \TE , \XE\ and \ND\ are shown as grey bands in fig.~\ref{fig:nme}, and are in the 20--30\% range.


%% file: src/ingredients.tex
The discovery of \bbonu\ would represent a substantial breakthrough in particle physics. A single, unequivocal observation of the decay would prove the Majorana nature of neutrinos and the violation of lepton number. Alas, that is not, by any means, an easy task. The design of a detector capable of identifying efficiently and unambiguously such a rare signal represents a major experimental problem.

To start with, one needs a large mass of the scarce \bb\ isotope in order to probe in a reasonable time the extremely long lifetimes expected. For instance, for a Majorana neutrino mass of 50 meV, it can be estimated using eq.~(\ref{eq:Tonu}) and a sound assumption for the NMEs that half-lives in the range of $10^{26}$--$10^{27}$ years must be explored (\textit{i.e.}, 17 orders of magnitude longer than the age of the universe!). A better sense of what such extremely long half-lives mean can be grasped with a simple calculation. Consider the radioactive decay law in the approximation $T_{1/2}\gg t$, where $t$ is the exposure time; in that case, the expected number of \bbonu\ events is given by
\begin{equation}
N_{\bbonu} = \log2 \cdot \frac{\Mbb\cdot N_{A}}{W_{\beta\beta}} \cdot \varepsilon \cdot \frac{t}{T_{1/2}^{0\nu}}, 
\label{eq:Nbb}
\end{equation}
where \Mbb\ is the mass of the \bb\ emitting isotope, $N_{A}$ is the Avogadro constant, $W_{\beta\beta}$ is the molar mass of the \bb\ isotope, and $\varepsilon$ is the signal detection efficiency.

It follows from eq.~(\ref{eq:Nbb}) that, in order to observe (assuming perfect detection efficiency and no disturbing background) as little as one decay per year and assuming a Majorana neutrino mass of 50 meV ($T_{1/2}^{0\nu}\sim 10^{26}$--$10^{27}$ years), ``macroscopic'' masses of \bb\ isotope of the order of 100 kg are needed.

The situation becomes even more desperate when considering real experimental conditions. The background processes that can mimic a \bbonu\ signal in a detector are copious. In the first place, the experiments have to deal with an intrinsic background, the \bbtnu, that can only be distinguished by measuring the energy of the emitted electrons, since the neutrinos escape the detector undetected (see fig.~\ref{fig:modes}). Good energy resolution is therefore essential to prevent the \bbtnu\ spectrum tail from spreading over the \bbonu\ peak. Nevertheless, this \emph{energy signature} could not be enough \textit{per se}: a continuous spectrum arising from natural radioactivity can easily overwhelm the signal peak. Other signatures, like particle identification or the observation of the daughter nucleus, are a bonus to provide a robust result.

Several other factors such as detection efficiency or the scalability to large masses must be taken into account as well when choosing the experimental technique. The simultaneous optimization of all these parameters is most of the time conflicting, and consequently, many different experimental approaches have been proposed and are under development. In order to compare their merits, a figure of merit, the experimental sensitivity to \mbb, is normally used. We describe it below, followed by a discussion on the main parameters entering this figure.

Before doing so, we note that our notation in eq.~(\ref{eq:Nbb}) --- and in the rest of this review --- differs from the usually adopted one, derived from source-equals-detector experimental configurations. In the source-equals-detector notation, one refers to the total active mass $M$ of the detector, which is related to the mass \Mbb\ in the \bb\ isotope via the following relationship:
\begin{equation}
\Mbb\ = W_{\beta\beta}\cdot \frac{M}{W}\cdot a\cdot \eta ,
\label{eq:mbbversusm}
\end{equation}
where $W$ is the molecular weight of the molecule of the active material, $a$ is the isotopic abundance of the candidate \bbonu\ nuclide, and $\eta$ is the number of \bbonu\ element nuclei per molecule of the active mass. For example, TeO$_2$ bolometric detectors with a natural isotopic abundance in \TE\ are characterized by $W_{\beta\beta}=129.9\ \mathrm{g/mol}$, $W=159.6\ \mathrm{g/mol}$, $a=0.34167$ and $\eta=1$, such that $\Mbb = 0.278 M$.\footnote{To stress this somewhat unconventional mass notation and to avoid any confusion, we will make use in the following of \kgbb\ as the mass unit to indicate one kilogram of \bb\ emitter mass.} \footnote{As pointed out to us, in principle the best quantity to express the \bbonu\ exposure, and the background rate per unit exposure and unit energy discussed below, is neither kg$\cdot$year nor \kgbb $\cdot$year, but rather $n_{\beta\beta}\cdot$year, where $n_{\beta\beta}=\Mbb \cdot N_A/W_{\beta\beta}$ is the number of moles of the \bb\ nuclide. The reason is that 1 \kgbb\ of, say, \GE\ contains almost twice as many \bb\ nuclides as 1 \kgbb\ of \ND . To avoid an even more ``radical'' departure from commonly employed units, we stick to \kgbb$\cdot$year units in the following.}


\subsection{Sensitivity of a $\beta\beta0\nu$ experiment} \label{subsec:sensitivitydefinition}

All \bbonu\ experiments have to deal with non-negligible backgrounds, an only partially efficient \bbonu\ event selection, and more or less difficulties to extrapolate their detection technique to large masses. It is instructive, however, to imagine an ideal, background-free, experiment. If such an experiment, after running for an exposure $\Mbb\cdot t$, observes no events, it would report an upper limit in the \bbonu\ decay rate $(\Tonu)^{-1}$, or possibly in the more relevant physical parameter \mbb:
\begin{equation}
\mbb = K_{1} \sqrt{\frac{1}{\varepsilon\cdot \Mbb \cdot t}}, \label{eq:mbbx4}
\end{equation}
where $K_{1}$ is a constant that depends only on the isotope type, and on the details of the statistical method (and the confidence level) chosen to report such limit. Equation~(\ref{eq:mbbx4}) follows directly from eqs.~(\ref{eq:Tonu}) and (\ref{eq:Nbb}), see \cite{GomezCadenas:2010gs} for details.

Let us now consider the sensitivity in the case of an experiment with background. In the large background approximation, the sensitivity as a function of the background rate $b$ follows the classical limit: $\mathcal{S}(b) \propto \sqrt{b}$, where $b$ is the mean predicted background level. In this limit, the \mbb\ sensitivity can be written as
\begin{equation}
\mbb =K_{2} \sqrt{\frac{b^{1/2}}{\varepsilon\cdot \Mbb \cdot t}} 
\label{eq:mbbx1}
\end{equation}
where $K_{2}$ is a constant depending on the isotope. If the background $b$ is proportional to the exposure $\Mbb \cdot t$ and to an energy window $\Delta E$ around \Qbb :
\begin{equation}
b = c\cdot \Mbb \cdot t\cdot \Delta E
\label{eq:mbbx2}
\end{equation}
with the background rate $c$ expressed in \ckky, then:
\begin{equation}
\mbb = K_2  \ \sqrt{1/\varepsilon} \ \Big(\frac{c\cdot \Delta E}{\Mbb \cdot t}\Big)^{1/4} \label{eq:mbbx3}
\end{equation}

In short, the background limits dramatically the sensitivity of a double beta decay experiment, improving only as $(\Mbb \cdot t)^{-1/4}$ instead of the $(\Mbb \cdot t)^{-1/2}$ expected in the background-free case.

Two aspects of eq.~(\ref{eq:mbbx2}), and in particular of our definition of the background rate $c$, deserve further clarification. First, for a given background level $b$, the background rate $c$ will in general depend on the choice of the energy window $\Delta E$. This is the case if the background energy spectrum around \Qbb\ is not flat. In the following, all background rate values refer to a $\Delta E$ choice of 1 FWHM energy resolution total width, that is computed for background events whose reconstructed energy falls in the $[\Qbb-0.5\cdot\text{FWHM},\Qbb+0.5\cdot\text{FWHM}]$ range. Similarly, the background rate $c$ will in general depend on the mass \Mbb\ of the \bb\ emitting material considered. This is the case for backgrounds that are not uniformly distributed within the active mass, such as surface contaminations of materials or backgrounds that are of external origin. As already assumed in deriving eq.~(\ref{eq:mbbx3}), all background rate values are relative to the total mass \Mbb\ appearing in the signal count rate computation of eq.~(\ref{eq:Nbb}).


\subsection{Choice of the \bb\ isotope} \label{subsec:isotope}
In nature, 35 naturally-occurring isotopes are \bb\ emitters. Which ones are the most favorable in terms of \bbonu\ searches? 

\begin{figure}[t!b!]
\begin{center}
\includegraphics[width=0.65\textwidth]{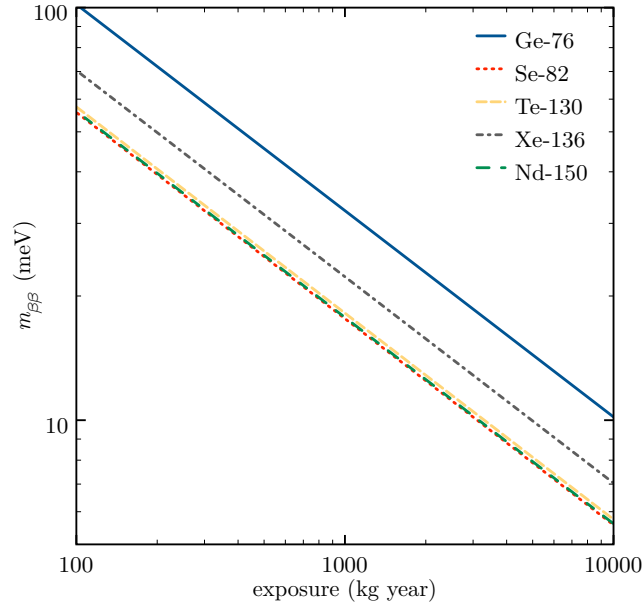}
\end{center}
\caption{Sensitivity of ideal experiments at 90\% CL for different \bb\ isotopes. Since the yields are very similar, the sensitivities of \SE, \TE\ and \ND\ overlap . From reference \cite{GomezCadenas:2010gs}.} \label{fig:SensiIdeal}	
\end{figure}

Let us start with considerations about the most favorable \bbonu\ phase space factors and nuclear matrix elements. We are interested in the isotopes that provide the highest \bbonu\ rate for the same \mbb\ mass, or, in other words, those that minimize the constants $K_1$ and/or $K_2$ appearing in eqs.~(\ref{eq:mbbx1}) and (\ref{eq:mbbx3}), respectively. To a first approximation, the phase space factor $G^{0\nu}(Q,Z)$ appearing in eq.~(\ref{eq:g0nu}) varies as $Q_{\beta\beta}^5$, see eq.~(\ref{eq:g0nu_pr2}). Isotopes with large $Q$-values are therefore favored. For this reason, only isotopes with $Q_{\beta\beta}>2$ MeV are usually considered for \bbonu\ searches. The 11 isotopes satisfying this criterion are shown in fig.~\ref{fig:g0nu}. The isotopes with the most favorable phase space factors are ${\rm ^{150}Nd}$, ${\rm ^{48}Ca}$ and ${\rm ^{96}Zr}$. As far as the nuclear matrix elements are concerned, variations from one isotope to another are significantly smaller than $G^{0\nu}(Q,Z)$ variations, as can be seen from fig.~\ref{fig:nme}. Considering the relevant product of the phase space factor times the nuclear matrix element squared for the most promising \bb\ isotopes, we find variations of about a factor of 2 in \mbb\ sensitivity depending on the isotope and for an ideal experiment, as can be seen in fig.~\ref{fig:SensiIdeal}. From this figure, and from phase space factor and nuclear matrix element considerations alone, we would conclude that \SE, \TE\ and \ND\ would be preferable than \GE. However, other factors enter in the isotope choice, as discussed in this section.
 
Another advantage in choosing a \bb\ isotope with a high $Q_{\beta\beta}$ value relies in background control. As we will see later, backgrounds to \bbonu\ searches from natural radioactivity populate the energy region below $\sim$3 MeV. The possibility to use an isotope with $Q_{\beta\beta}$ above these background energies is therefore desirable.

\begin{figure}[t!b!]
\begin{center}
\includegraphics[width=0.65\textwidth]{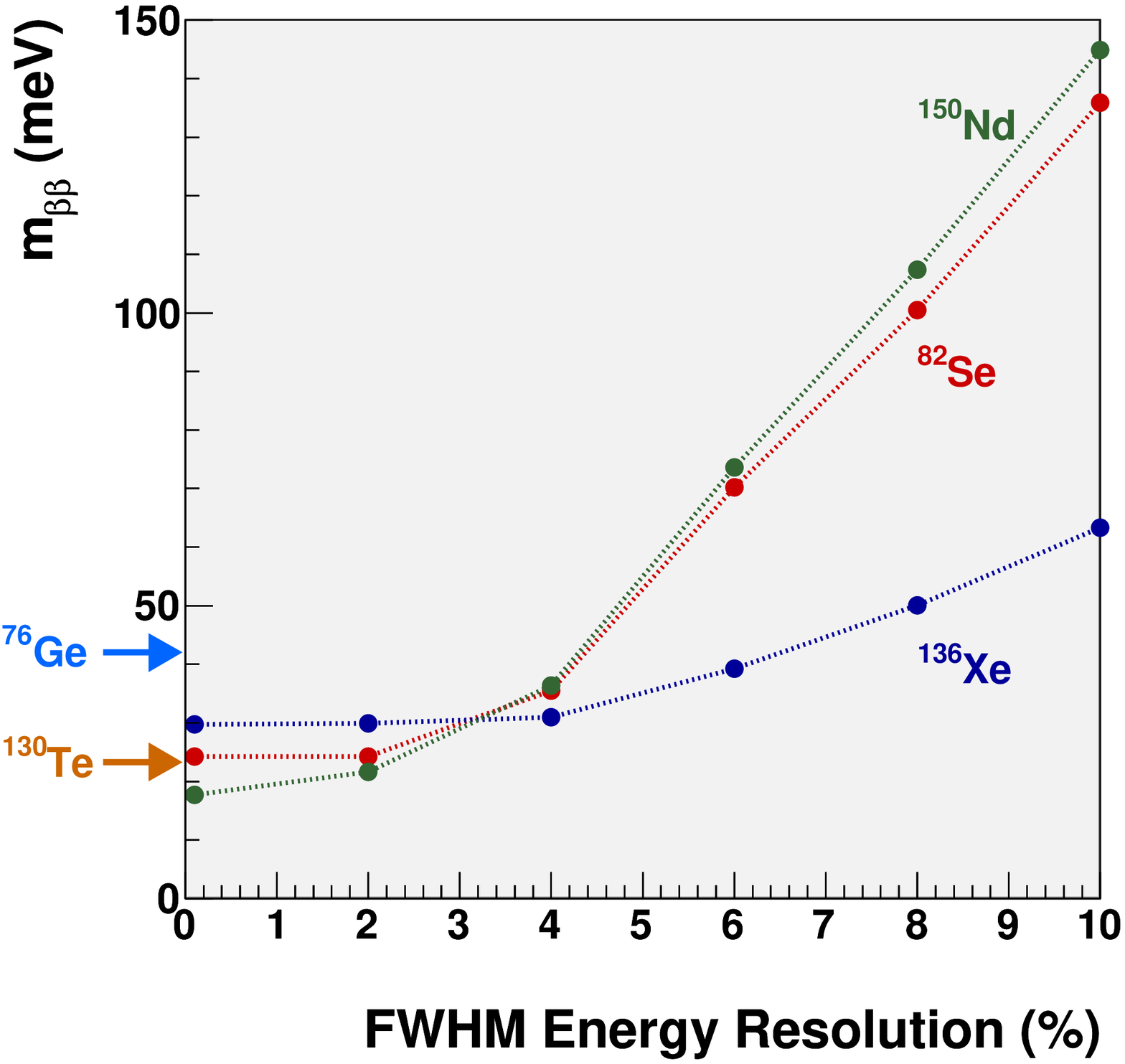}
\end{center}
\caption{\label{fig:twonubgr}Sensitivity to \mbb\ at 90\% CL as a function of FWHM energy resolution, for ideal experiments using five different isotopes, each with 100 \kgbb\ of \bb\ emitter mass and 5 years of data-taking. The experiments are assumed to have perfect efficiency and to be affected only by \bbtnu\ backgrounds. In practice, experiments using \GE\ and \TE\ always feature an excellent energy resolution and are therefore not affected by \bbtnu\ backgrounds, hence only the background-free sensitivity limit is shown in those cases, with an arrow.}  
\end{figure}

One is also typically interested in choosing an isotope with a relatively slow \bbtnu\ mode. As the energy resolution degrades, the experiments are affected by \bbtnu\ backgrounds in a more or less pronounced way, depending on the isotope.  This is true unless the energy resolution of the experiment is truly excellent, in which case even relatively fast \bbtnu\ modes do not constitute a serious background to \bbonu\ searches. This is illustrated in fig.~\ref{fig:twonubgr}. In this figure, the \mbb\ sensitivity (computed according to the prescription described later, in sect.~\ref{subsec:sensi}) at 90\% CL is shown for ideal experiments using five different isotopes as a function of FWHM energy resolution. The experiments, each assumed to use 100 \kgbb\ of \bb\ emitter mass and to run for five years, are ideal in the sense of having perfect \bbonu\ efficiency and of being affected only by \bbtnu\ backgrounds. As fig.~\ref{fig:twonubgr} illustrates, and as far as the \bbtnu\ background is concerned and for the same moderate energy resolution (say, 5-10\% FWHM), \XE\ is to be preferred over \SE\ and \ND , thanks to its much longer \bbtnu\ half-life (see tab.~\ref{tab:bb2nu_exp}). For experiments featuring excellent energy resolution, say $<2\%$ FWHM, all experiments would operate in a essentially background-free regime, for the assumed 500 kg$\cdot$ yr exposure. As we have seen above, in this regime an isotope such as \ND\ is to be preferred over, say, \GE \footnote{The cases of \GE\ and \TE\ are shown in fig.~\ref{fig:twonubgr} only in this background-free regime, given that in practice experiments using such isotopes always feature excellent energy resolution, see sect.~\ref{subsec:energyresolution}.}. In practice, however, other backgrounds are always present, typically creating a continuum through the region of interest, and a better resolution improves the experimental sensitivity even in the $<2\%$ FWHM energy resolution range.

Another factor entering in the \bb\ isotope choice has to do with how well understood the nuclear physics for that isotope is. As we have seen in sect.~\ref{sec:nme}, the calculation of nuclear matrix elements is a very complicated task. In sect.~\ref{subsec:nme_pmr}, we have made an attempt at quantifying the uncertainties in the NMEs for various isotopes. Our conclusion is that no \emph{magic isotope} exists, and uncertainties in the 20--30\% range (according to our evaluation) exist for the five isotopes we have considered, \GE, \SE, \TE, \XE\ and \ND.


\subsection{Isotope mass} \label{subsec:isotope_mass}

As explained above, large masses of \bb\ isotope are needed to explore the expected half-lives. The previous generation of double beta decay experiments used masses of the order of 10 kg. New-generation experiments will range from tens of kilograms to several hundreds, depending on the proposal. 

Unfortunately, the \bb\ isotopes are not always abundant in nature, requiring enrichment in order to obtain large, concentrated masses. It was argued in \cite{GomezCadenas:2010gs} that ${\rm ^{136}Xe}$ would be a particularly favorable isotope to use, since it permits target masses of 1~ton or more and low-background experimental techniques. Isotope enrichment appears relatively easier from a technical point of view (and therefore, cheaper) for ${\rm ^{136}Xe}$. Experimental proposals involving both liquid scintillator detectors (see KamLAND-Zen, sect.~\ref{subsec:kamland}), as well as liquid-phase or gaseous-phase TPCs (see EXO and NEXT, sects.~\ref{subsec:exo} and \ref{subsec:next}, respectively), are using or planning to use ${\rm ^{136}Xe}$. 

Liquid scintillator proposals permit in principle to reach large \bbonu\ isotope masses with other isotopes as well. Given its very favorable phase space, the SNO+ Collaboration (see sect.~\ref{subsec:sno+}) will dissolve a neodymium salt in the liquid scintillator. Scalability to large ${\rm ^{150}Nd}$ masses will ultimately depend on the feasibility to enrich neodymium, most likely a difficult enterprise. 

High-resolution calorimeters, that is germanium diodes and bolometers, are compact detectors and might be therefore, in principle, scalable to large masses. In germanium experiments such as GERDA (see sect.~\ref{subsec:gerda}), \bb\ isotope masses of the order of 100 kg appear feasible, but it might be difficult, in particular for economical reasons, to go much beyond that. The CUORE bolometers (see sect.~\ref{subsec:cuore}) plan to use \TE. This isotope has the highest natural isotopic abundance among the commonly-considered \bb\ emitters (34\%). The need for isotope enrichment is therefore less important in this case, and \bb\ masses of the order of hundreds of kilograms appear within reach of new-generation experiments.


\subsection{Energy resolution} \label{subsec:energyresolution}
Together with a large isotope mass, good energy resolution is a necessary (but not sufficient!) requirement for the \emph{ultimate} \bbonu\ experiment. It is the only protection against the intrinsic \bbtnu\ background, and improves the signal-to-noise ratio in the region of interest around \Qbb. 

The detectors for \bb\ searches that have achieved the best energy resolution so far are the \emph{germanium diodes} and the \emph{bolometers}. In germanium detectors the energy is measured via ionization (creation of electron-hole pairs in the semiconductor). An energy resolution as low as 0.1\% FWHM at \Qbb\ has been obtained \cite{Agostini:2010ke}. Partly thanks to their superior energy resolution, germanium diodes have dominated the \bbonu\ searches so far (see sect.~\ref{subsec:past}). The GERDA and MAJORANA proposals are based on germanium diodes, see sect.~\ref{subsec:gerda}. In bolometers the energy is measured by detecting a temperature rise in crystals with very small specific heat. Several bolometric crystals have been proposed and tested for \bb\ searches, with tellurite (TeO$_{2}$) being the favored one due to its reasonable mechanical and thermal properties, and the natural high content (28\% in mass) of the \TE\ \bb\ isotope. An energy resolution of about 0.2\% FWHM at \Qbb\ has been reached using TeO$_{2}$ crystals \cite{Andreotti:2010vj}. The CUORE proposal uses this technique (see sect.~\ref{subsec:cuore}).


\subsection{Backgrounds} \label{subsec:backgrounds}

Double beta decay experiments are mostly about suppressing backgrounds. As we have seen already, the mere presence of background in the region of interest around \Qbb\ changes the regime of the \mbb\ sensitivity from a $(\Mbb \cdot t)^{-1/2}$ dependence to $(\Mbb \cdot t)^{-1/4}$.

\begin{figure}[t!b!]
\begin{center}
\includegraphics[scale=0.3]{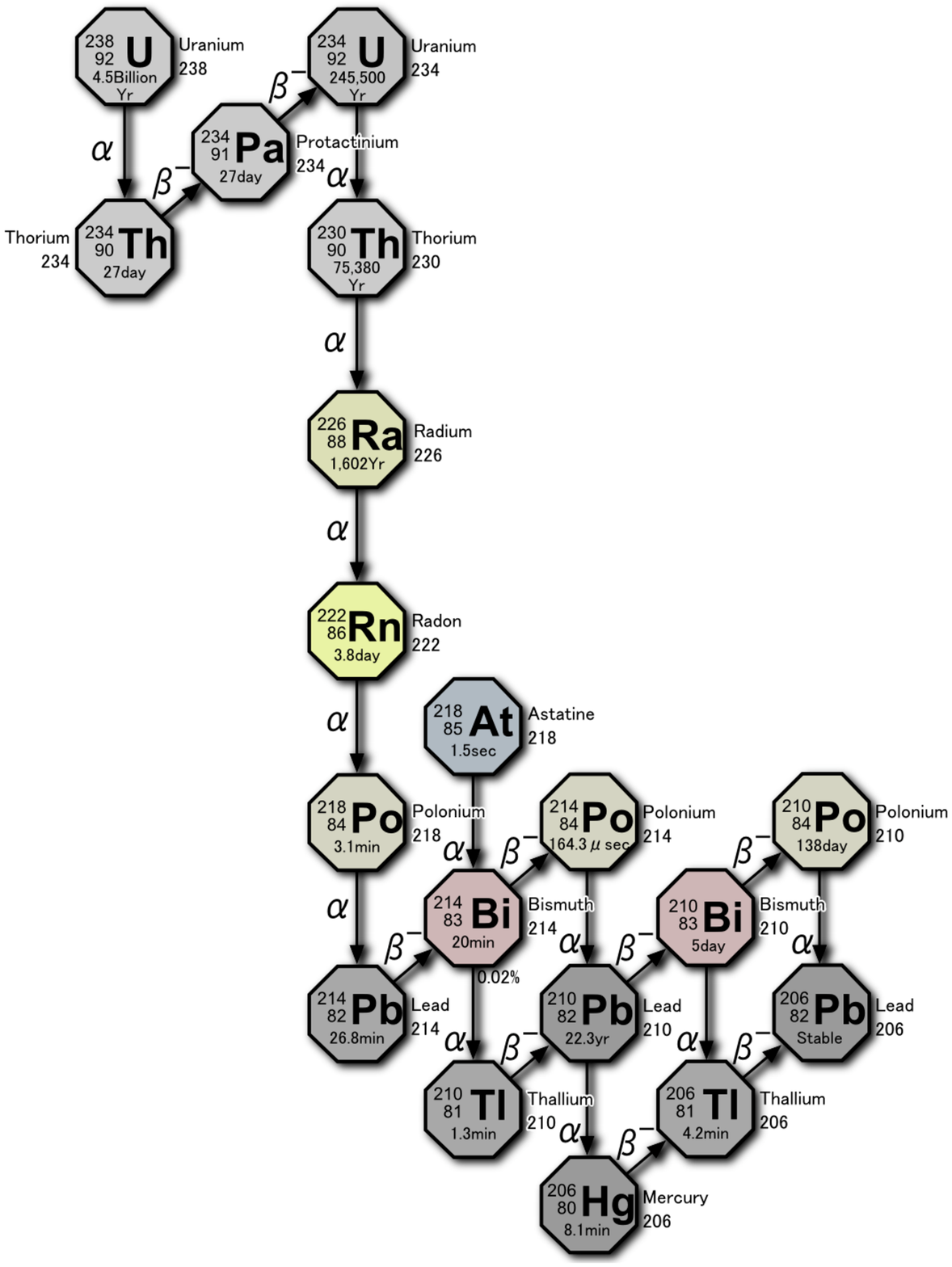} \hspace{0.09\textwidth}
\includegraphics[scale=0.40]{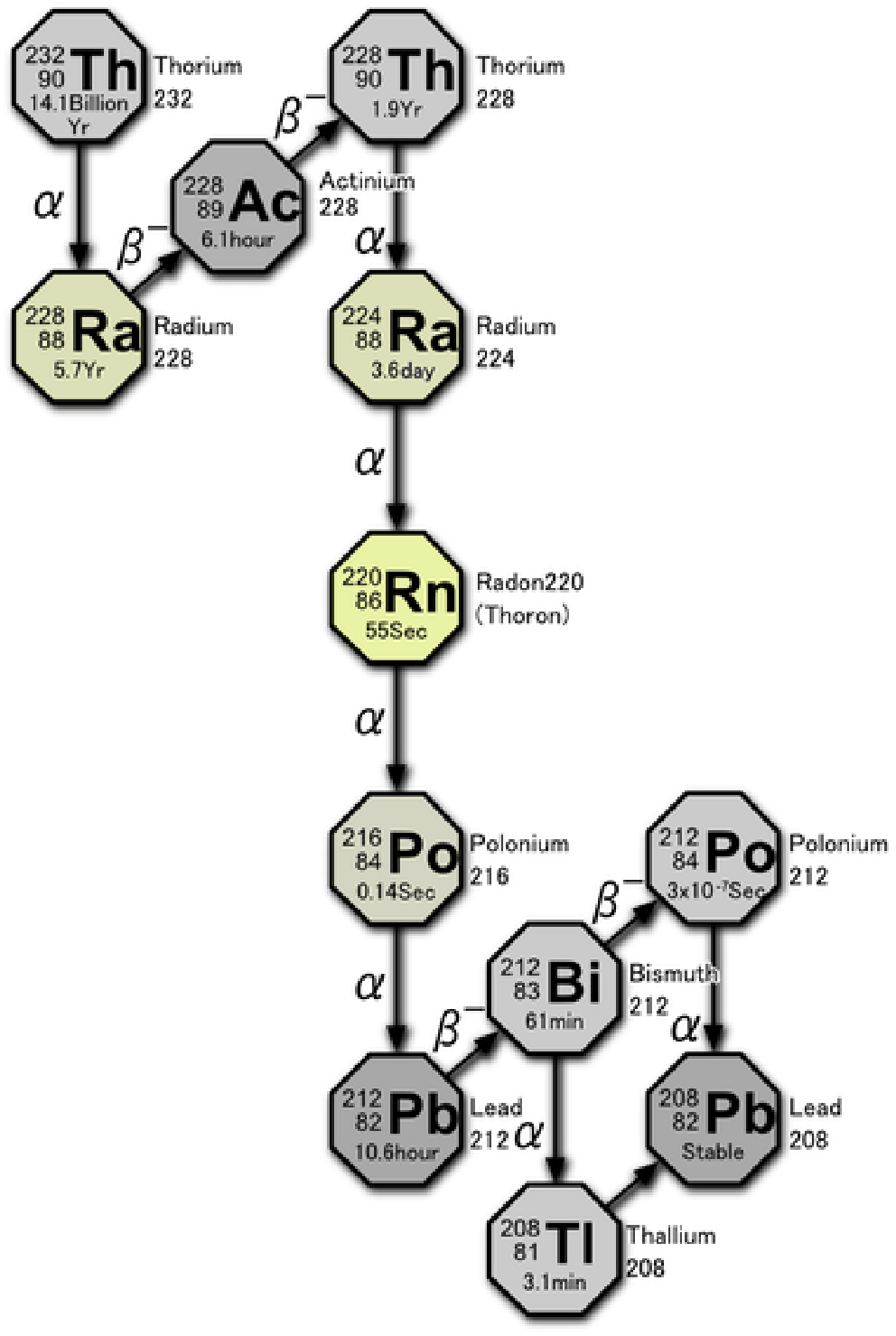} 
\caption{Decay chains of uranium (left) and thorium (right) \cite{wiki:decaychain}.} \label{fig:decaychain}
\end{center}
\end{figure}

The natural radioactivity of detector components is often the main background in \bbonu\ experiments. Even though the half-lives of the natural decay chains are comparable to the age of the Universe, they are very short compared to the half-life sensitivity of the new-generation \bbonu\ experiments. Therefore, even traces of these nuclides can become a significant background. The decays of \TL\ and \BI\ are particularly pernicious, given the high Q-values of these reactions, therefore polluting the energy region of interest of most \bb\ emitters. These isotopes are produced as by-products of the natural thorium and uranium decay chains (see fig.~\ref{fig:decaychain}), and they are present at some level in all materials. Careful selection of material and purification is mandatory for all \bbonu\ experiments. The new-generation experiments are being fabricated from amazingly radiopure components, some with activities as low as 1 $\mu$Bq/kg or less.

Radon gas, either $^{222}$Rn or $^{220}$Rn, is also a worry for most experiments. These isotopes, present in the natural decay chains, diffuse easily through many materials, infiltrating the detectors sensitive region. Their daughters tend to be charged and stick to surfaces. Many experiments eliminate radon from the detector surroundings by flushing pure nitrogen gas. Also, some laboratories have installed radon-traps in the air circulation system.

In addition to internal backgrounds coming from radioactive impurities in detector components themselves, there are external backgrounds originated outside the detector. Those backgrounds can be in principle suppressed by placing the detector at an underground location and by enclosing it into a shielding system. 

\begin{figure}[t!b!]
\begin{center}
\includegraphics[width=0.65\textwidth]{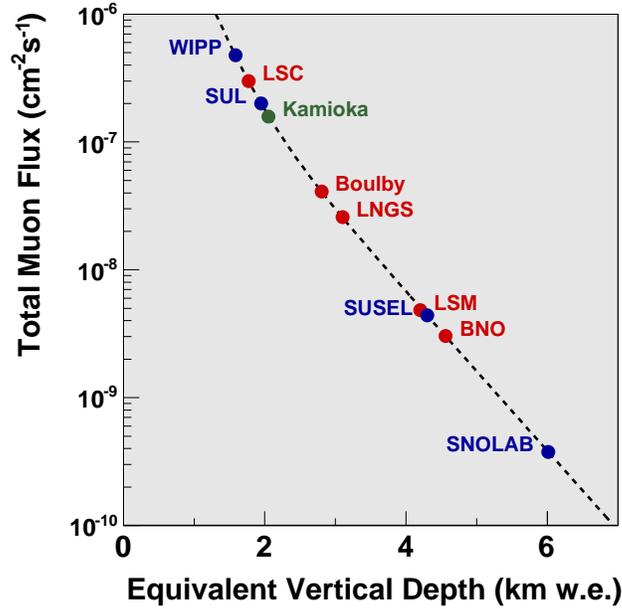}
\caption{\label{fig:underground_labs}Total muon flux as a function of the equivalent vertical depth for a flat overburden. The empirical parametrization, shown as a dashed line, is taken from \cite{Mei:2005gm}. The fluxes measured at various underground sites currently available to host physics experiments are taken from \cite{Mei:2005gm,Bettini:2011zza}. Facilities shown in brown, blue and red are located in Europe, America and Asia, respectively. The full names of the facilities shown in the figure are: Waste Isolation Pilot Plant (WIPP), Laboratorio Subterraneo de Canfranc (LSC), Soudan Underground Laboratory (SUL), Kamioka Observatory (Kamioka), Boulby Palmer Laboratory (Boulby), Laboratorio Nazionale del Gran Sasso (LNGS), Laboratoire Souterrain de Modane (LSM), Sanford Underground Science and Engineering Laboratory (SUSEL), Baksan Neutrino Observatory (BNO), SNOLAB.}
\end{center}
\end{figure}

In general, the depth requirement for a \bbonu\ experiment varies according to the detector technology. A very efficient shielding and additional detection signatures such as topological information can compensate the benefits of a very deep location. Figure \ref{fig:underground_labs} shows the depth (and corresponding cosmic ray muon flux) of several underground facilities currently available to host physics experiments around the world. The deepest laboratory is SNOLAB (Canada), an expansion of the existing facilities constructed for the Sudbury Neutrino Observatory (SNO). The SNO+ experiment (sect.~\ref{subsec:sno+}), and perhaps also EXO (sect.~\ref{subsec:exo}), will be located there. Other deep laboratories include SUSEL (USA) and LSM (France), which will host the demonstrators for the MAJORANA (sect.~\ref{subsec:majorana}) and SuperNEMO (sect.~\ref{subsec:snemo}) experiments, respectively. In addition to depth, other important factors characterizing the underground sites include the size of the excavated halls and the services provided to the experiments. The size is an important factor to take into account especially for experimental proposals at the ton-scale, given that some of them (\emph{e.g.}, the SuperNEMO experiment) need large volumes. For a recent review of the currently available underground facilities around the world, see reference \cite{Bettini:2011zza}. 

At the depths of underground laboratories, muons (and neutrinos) are the only surviving radiation from cosmic rays. However, their interactions can produce high-energy secondaries such as neutrons or electromagnetic showers. 

Charged backgrounds (such as muons) can be easily eliminated using a veto system. Neutrons, on the other hand are often a more serious problem. They can have sizable penetrating power, impinging on the detector materials and \emph{activating} them through large $\Delta A$ transitions in nuclei, ultimately resulting in radioactive nuclides. Cosmogenic activation is, of course, more severe on surface. Therefore, for experiments using materials that can get activated (like germanium-based experiments), underground fabrication and storage of the detector components is essential. The detectors can be shielded against neutrons with layers of hydrogenous material. 

Radioactive decays in the rock of the underground cavern result in a $\gamma$-ray flux that can interact in the detector producing background. Dense (high $Z$), radiopure materials such as lead and copper are used as shielding to suppress this background. Water, being inexpensive and easy to purify, is also a good alternative for shielding against $\gamma$-rays.

Finally, very massive detectors such as liquid-scintillator calorimeters suffer from an irreducible external background: the solar neutrino flux.

In the design of a shielding system against external backgrounds, a \emph{graded shielding} principle is followed: the thickness of a shield component does not need to reduce the flux below the contribution of the next inner component, with the innermost shield component selected to be the radiopurest.

\begin{figure}[t!b!]
\begin{center}
\includegraphics[width=0.60\textwidth]{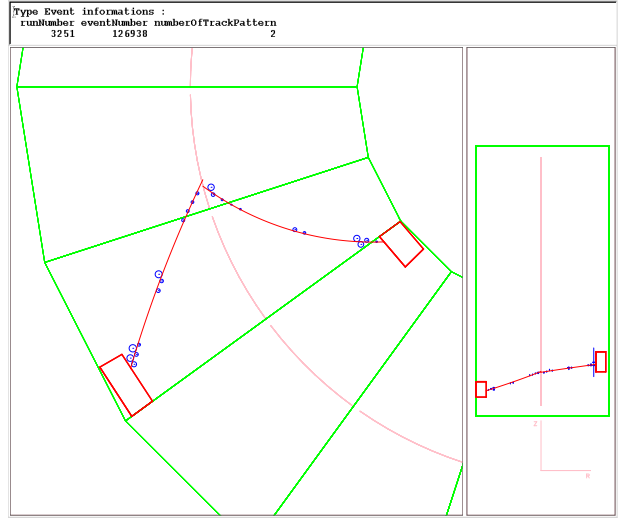}
\caption{\label{fig:nemo3eventdisplay}Top (left) and side (right) view of a reconstructed \bb\ event selected from NEMO-3 data with a two electron energy sum of 2812 keV \cite{Arnold:2005rz}.}
\end{center}
\end{figure}

The lowest background rate (expressed in terms of background events per unit energy, \bb\ isotope mass and exposure time) in a \bbonu\ experiment so far was achieved by a \emph{tracker-calo experiment}, NEMO-3 (see sect.~\ref{subsec:past}). In this experimental approach,  foils of the \bb\ source are surrounded by a tracking detector that provides a direct detection of the two electron tracks emitted in the decay. The topological reconstruction of the events provides a powerful active handle to reject backgrounds, together with relatively radiopure detectors (see fig.~\ref{fig:nemo3eventdisplay}). The NEMO-3 experiment measured a background rate of a few times $10^{-3}$ \ckkbby\ \cite{Argyriades:2009vq}. Time projection chambers using xenon in gas phase, as proposed by the NEXT experiment (see sect.~\ref{subsec:next}), provide also some topological information that can be used to reject backgrounds.  High-resolution calorimeters, such as germanium diodes and bolometers, have so far achieved somewhat worse background rates, in the neighborhood of $10^{-1}$ \ckkbby\ \cite{KlapdorKleingrothaus:2000sn,Gonzalez:2003pr,Andreotti:2010vj}. The goal of the new-generation experiments is typically to reach $10^{-3}$ \ckkbby\ background levels, and sometimes significantly better than that. This may require, in some cases, activities for detector components of $1\ \mu$Bq/kg or less! A recent example witnessing the challenges to build and characterize such extremely radiopure detectors is provided by the EXO Collaboration. In this case, a thorough and systematic study of trace radioactive impurities in a large variety of parts and materials required to construct the EXO-200 detector has been carried out and documented in detail \cite{Leonard:2007uv}.

\begin{figure}[t!b!]
\begin{center}
\includegraphics[angle=270,width=0.50\textwidth]{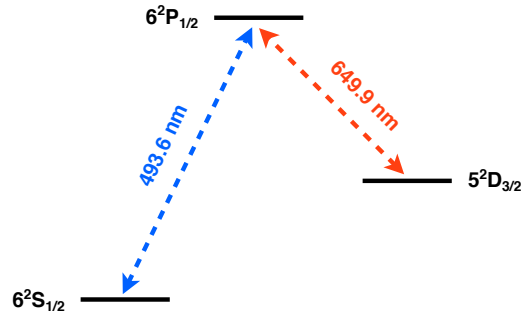}
\caption{Simplified energy level diagram of the barium ion, and the wavelength in vacuum for the transitions.} \label{fig:bariumenergylevels}
\end{center}
\end{figure}

Another handle to suppress non-\bbtnu\ backgrounds is \emph{daughter ion tagging}. This has been proposed, and is actively being pursued, for the EXO xenon-based detector. In this case, the \bbonu\ decay is ${\rm ^{136}Xe}\to {\rm ^{136}Ba^{++}}+2e^-$. The ${\rm ^{136}Ba^{++}}$ ion rapidly captures an electron, resulting in ${\rm ^{136}Ba^{+}}$ which is stable in xenon. The ${\rm ^{136}Ba^{+}}$ ions can be identified via atomic spectroscopy, exciting them with a blue laser and observing the resulting red light (see fig.~\ref{fig:bariumenergylevels}). Daughter ion tagging is undoubtedly very challenging from the technical point of view, but the payoff would be huge if the R\&D were to be successful.

\subsection{\label{subsec:efficiency}Detection efficiency}

Neutrinoless double beta decay events are extremely rare, if present at all, thus a high detection efficiency is an important requirement for a \bb\ experiment. Equation~(\ref{eq:mbbx3}) clearly indicates that the detector design should prioritize a high detection efficiency. To obtain the same increase in \mbb\ sensitivity obtained by doubling the efficiency, the mass would have to be increased by a factor of 4, assuming the same background.

In general, the simpler the detection scheme, the higher the detection efficiency. For instance, pure calorimetric approaches such as germanium diodes or bolometers have detection efficiencies in excess of 80\%. This is to be contrasted with experiments performing, for example, particle tracking, which will typically cause a significant efficiency loss. Also, homogeneous detectors, where the source material is the detection medium, provide in principle higher efficiency than the separate-source approach. This is due to a number of reasons, including geometric acceptance, absorption in the \bb\ source, backscattering of electrons, and the tracking requirement. On the other hand, some relatively dense homogeneous detectors use some of the \bb\ mass close to the detector borders effectively for self-shielding, paying it with some efficiency loss.


%% file: src/exps.tex
During decades the search for double beta decay was a rather marginal activity carried out with geochemical techniques. It was not until 1987 that the \bbtnu\ was directly observed in the laboratory. In the 1990's, the field was dominated by germanium detectors, devices characterized by superb energy resolution and a high efficiency. 

After the positive results of neutrino oscillation experiments, the field has gone through a revolution. The community is preparing a rich and varied new generation of experiments that should explore ultimately the inverted-hierarchy region of neutrino masses (see fig.~\ref{fig:mbetabetavsmlight}). This will require a multi-ton experiment. It seems prudent to build, as a first step, an experiment containing about 100 kg of isotope that can be expanded at a later time. However, the scalability will not be possible for all experimental techniques.

In this section, after summarizing the results of past experiments, some proposals for the new generation are described. This discussion does not pretend to be exhaustive, but focused on the pros and cons of the different techniques. Finally, the sensitivity to \mbb\ of the proposals is evaluated.

\subsection{Past experiments} \label{subsec:past} 
\input{src/past.tex}

\subsection{CUORE} \label{subsec:cuore}
\input{src/cuore.tex}

\subsection{EXO} \label{subsec:exo}
\input{src/exo.tex}

\subsection{GERDA} \label{subsec:gerda}
\input{src/gerda.tex}

\subsection{MAJORANA} \label{subsec:majorana}
\input{src/majorana.tex}

\subsection{KamLAND-Zen} \label{subsec:kamland}
\input{src/kamland.tex}

\subsection{NEXT} \label{subsec:next}
\input{src/next.tex}

\subsection{SNO+} \label{subsec:sno+}
\input{src/sno+.tex}

\subsection{SuperNEMO} \label{subsec:snemo}
\input{src/snemo.tex}

\subsection{Other proposals} \label{subsec:other}
\input{src/other_exps.tex}

\subsection{Sensitivity of new-generation experiments} \label{subsec:sensi}
\input{src/sensi.tex}

\subsection{Validity of sensitivity assumptions} \label{subsec:sensi_assumptions}
\input{src/sensi_assumptions.tex}

%% file: src/past.tex
For almost half a century the only evidence of the existence of double beta decay came from geochemical methods consisting in measuring the concentrations of the stable daughter isotopes $(Z+2,A)$, produced over geologic times ($\sim 10^9$ years). An excess of the daughter isotope over its natural concentration is interpreted as evidence for \bb\ decay (either \bbtnu\ or \bbonu, since the method cannot distinguish between them).

The first direct measurement of \bbtnu, in \SE, did not happen until 1987 \cite{Elliott:1987kp}. It was done using a fairly large ($\sim1$ m$^{3}$) time projection chamber, the well-known Irvine TPC. The source, 14 g of 97\% enriched \SE, was deposited on a thin Mylar foil forming the central electrode of the chamber. The trajectories of the electrons emitted from the source foil were recorded by the TPC and analyzed to infer their energy and kinematic characteristics. Since this initial detection, the two-neutrino mode has been directly observed for 8 isotopes in several experiments (see table \ref{tab:bb2nu_exp} and ref.~\cite{Barabash:2010ie} for further details).

The most restricting limits to date in the search for \bbonu\ were obtained with germanium detectors. The Heidelberg-Moscow (HM) experiment \cite{KlapdorKleingrothaus:2000sn} searched for the \bbonu\ decay of \GE\ using five high-purity Ge semiconductor detectors enriched to 86\% in \GE. The experiment ran in the Laboratori Nazionali del Gran Sasso (LNGS), Italy, from 1990 to 2003, totaling an exposure of 71.7 kg$\cdot$year. The background rate reached by the experiment in the \Qbb\ region was ($0.19\pm 0.01$) \ckky, or, in units of \bb\ emitter mass, $0.22\pm 0.01$ \ckkbby. Pulse shape discrimination (PSD) was used in a subset of the data (35.5 kg$\cdot$year) to separate single-site events, like \bbonu\ decays, from multi-site events, like $\gamma$ interactions, resulting in a background rate of ($0.06\pm 0.01$) \ckky, or ($0.07\pm 0.01$) \ckkbby. A lower limit on the \bbonu\ half-life of $T^{0\nu}_{1/2}(\GE) \geq 1.9 \times 10^{25}$ years (90\% CL) was obtained \cite{KlapdorKleingrothaus:2000sn}.

A subset of the collaboration re-analyzed the data claiming evidence for \GE\ \bbonu\ decay \cite{KlapdorKleingrothaus:2001ke}. The latest publication by this group reports a $6\sigma$ evidence for \bbonu\ and a half-life measurement of $T_{1/2}^{0\nu}=(2.23^{+0.44}_{-0.31})\times 10^{25}$ years \cite{KlapdorKleingrothaus:2006ff}, corresponding to $\mbb\ = (0.30^{+0.02}_{-0.03})\ \text{eV}$ according to the central value of the PMR nuclear matrix element for \GE\ given in sect.~\ref{subsec:nme_pmr}. This claim sparked an intense debate in the community, and at the moment no consensus exists about its validity (see, for example, ref.~\cite{Aalseth:2002dt}).

The International Germanium Experiment (IGEX) \cite{Aalseth:2002rf} also searched for \bbonu\ using enriched germanium crystals. It ran in the Homestake gold mine (USA), the Canfranc Underground Laboratory (Spain) and the Baksan Neutrino Observatory (Russia) from 1991 to 2000, accumulating a total exposure of 8.87 kg$\cdot$year. It reached a sensitivity similar to that of Heidelberg-Moscow, but not enough to disprove the claim. The lowest background rate reached by the IGEX experiment was 0.26 (0.10) \ckky\ without (with) pulse shape discrimination for a 8.87 (4.65) kg$\cdot$year total exposure \cite{Gonzalez:2003pr}, corresponding to 0.30 (0.12) \ckkbby\ per unit \bb\ emitter mass.

The Cuoricino experiment, an array of 62 TeO$_{2}$ bolometric crystals, ran for five years in Gran Sasso searching for \bbonu\ in \TE. It reached a sensitivity to \mbb\ comparable to that of the HM experiment, but it cannot disprove the claim due to the uncertainties in the nuclear matrix elements. The average background rate for the 5$\times$5$\times$5 cm$^3$ Cuoricino crystals, computed in a 60 keV wide region centered around \Qbb , was $0.161\pm 0.006$ \ckky\ \cite{Alessandria:2011rc}, corresponding to $0.58\pm 0.02$ \ckkbby\ per unit \bb\ emitter mass. The average FWHM energy resolution in all crystals was $6.3\pm 2.5$ keV at 2615 keV \cite{Alessandria:2011rc}.

The lowest levels of background so far were achieved by the NEMO3 experiment \cite{Argyriades:2009vq}: a few times $10^{-3}$ \ckky . This detector represents the state of the art of separate-source \bb\ experiments. Reconstruction of the electron tracks emerging from the source provided a powerful signature to discriminate signal from background. The NEMO3 experiment ran from 2003 to 2010 at the Modane Underground Laboratory (LSM), in France. The detector, of cylindrical shape, had 20 segments of thin source planes, with a total area of 20 m$^{2}$, supporting about 10 kg of source material. The sources were within a drift chamber, for tracking, surrounded by plastic scintillator blocks, for calorimetry. A solenoid generated  a magnetic field of 25 Gauss which allowed the measurement of the tracks electric charge sign. The detector was shielded against external gammas by 18 cm of low-background iron. Fast neutrons from the laboratory environment were suppressed by an external shield of water, and by wood and polyethylene plates. The air in the experimental area was constantly flushed, and processed through a radon-free purification system embedding the detector volume. In addition to searching for \bbonu , NEMO3 very successfully served as a ``\bbtnu\ factory'', providing precise \bbtnu\ half-life measurements for seven \bb\ isotopes, see table \ref{tab:bb2nu_exp}. Apart from representing the ultimate background for \bbonu\ searches , an accurate measurement of \bbtnu\ in several nuclides is also important as input to NME calculations. 

%% file: src/cuore.tex
The Cryogenic Underground Observatory for Rare Events (CUORE) \cite{Ardito:2005ar}
has been designed following the successful experience of the  MiDBD \cite{Arnaboldi:2002te} and Cuoricino \cite{Andreotti:2010vj} $^{130}$Te experiments, where for the first time arrays of bolometers were used to search for $\beta\beta$ decay.

\begin{figure}[t!]
\begin{center}
\includegraphics[scale=.8]{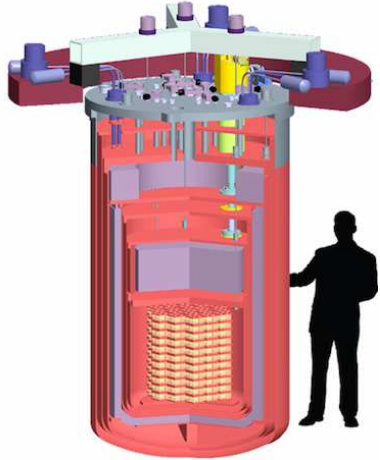}
\end{center}
\caption{The CUORE detector and cryostat. In yellow, the 19 towers of bolometers; the lavender volumes are the lead shielding.}\label{fig:cuore}
\end{figure}

CUORE will be placed in the hall A of the Gran Sasso Underground Laboratory and will consist of a system of 988 bolometers, each being a crystal of TeO$_2$ of $5\times5\times5$ cm$^3$, arranged in 19 vertical towers consisting of 13 layers of 4 crystals each. The four crystals are held between two copper frames joined by copper columns. PTFE pieces are inserted between the copper and TeO$_2$, as a heat impedance and to clamp the crystals. There is a few mm gap between crystals with no material between them.
A system of lead shields will be hosted inside the cryostat close to the detectors, to shield them from environmental radioactivity and from radioactive contaminations originating in the dilution unit located above the detector and in the dewar structure \cite{Bellini:2007zz}. A 6 cm thick roman lead shield surrounds the detector array on its sides, while a 30 cm thick layer of low-activity lead is placed above the detector. The $^{210}$Pb activity of the roman lead was measured to be less than 4 mBq/kg. A sketch of the detector is shown in fig.~\ref{fig:cuore}.

The total mass of the detector will be 741 kg for a $^{130}$Te mass of 206 kg. The energy released in a single particle interaction within the crystal is measurable as a change in temperature by  Neutron Transmutation Doped (NTD) germanium thermistors. The measured energy resolution is $\sim 5$ keV FWHM at the $\beta \beta \; (0\nu)$ transition energy ($\sim 2.53$ MeV).

The CUORE bolometers will operate at temperatures between 10 and 15 mK. A challenging $^3$He/$^4$He dilution refrigerator, with a cooling power of 3 mW at 120 mK, has been designed on purpose and is under construction.

A single tower of CUORE, CUORE-0, is presently under construction and will begin  operations within 2011. It will be hosted in the old Cuoricino dilution refrigerator, placed in the hall A of the LNGS. CUORE-0 is a real test of the CUORE assembly chain and procedure, will directly test the level of backgrounds of the CUORE setup and improve the Cuoricino sensitivity on \bbonu.

The CUORE setup will possibly allow in the long term for powerful upgrades. An obvious, though expensive, possibility is to substitute the natural tellurium bolometers with enriched $^{130}$Te units (provided that the enrichment procedure can keep the internal backgrounds very low). A more sophisticated option is to use scintillating crystals containing interesting double beta emitters \cite{Pirro:2005ar}. The contemporary read-out of scintillation light and thermal signal could indeed allow for a dramatic reduction of the background rate and a better characterization of the signal. An array of ZnSe scintillating bolometers, LUCIFER, has been recently proposed as a prototype experiment exploring the performances of such an approach \cite{Ferroni:2011zz} (see sect.~\ref{subsec:other}).

%% file: src/exo.tex
The Enriched Xenon Observatory \cite{Hall:2010zz} will search for \bbonu\ in \XE. The ultimate goal of the Collaboration is the development of the barium tagging for a multi-ton xenon-based detector, which would lead to a virtually background-free experiment. Prior to that, the Collaboration has built the EXO-200 detector, a $\sim$200-kg liquid xenon (enriched to 80.6\% in \XE) time projection chamber that detects both scintillation and ionization.

\begin{figure}[t!]
\begin{center}
\includegraphics[width=0.475\textwidth]{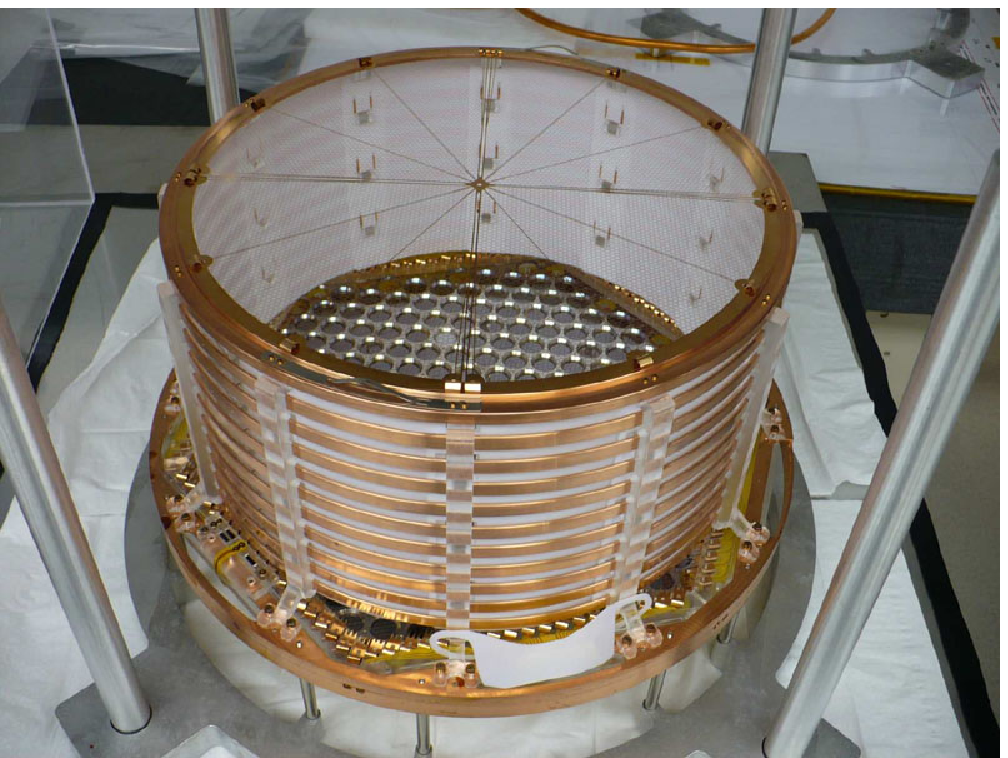}
\includegraphics[width=0.475\textwidth]{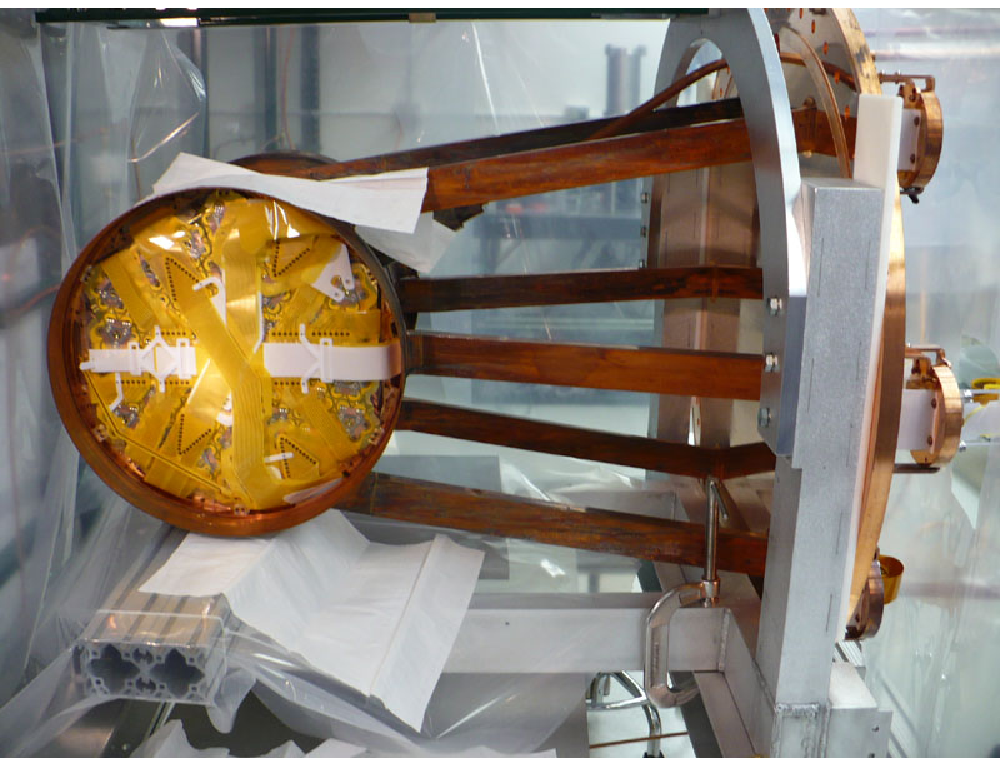}
\end{center}
\caption{Left: one half of the EXO chamber, viewed from the cathode plane. Right: the chamber attached to the cryostat door, as viewed from the bottom of the APD plane. The legs contain the readout cabling and are also the conduits for xenon circulation.} \label{fig:exo}
\end{figure}

The fiducial volume of the chamber, 44 cm in length, is divided in two halves by a central cathode (see fig.~\ref{fig:exo}, left). Ionization charges created in the xenon by charged particles drift under the influence of an electric field towards the two ends of the chamber. There, the charge is collected by a pair of crossed wire planes which measure its amplitude and transverse coordinates. Each end of the chamber includes also an array of avalanche photodiodes (APDs) to detect the 178-nm scintillation light produced by primary interactions. The sides of the chamber are covered with teflon sheets that act as VUV reflectors, improving the light collection. The simultaneous measurement of both the ionization charge and scintillation light of the event may in principle allow to reach a detector energy resolution as low as 3.3\% FWHM at the \XE\ Q-value, for a sufficiently intense drift electric field \cite{Conti:2003av}. 

The xenon is held inside a thin copper vessel immersed in a cryofluid that also shields the experiment from external radioactive backgrounds. The HFE heat-transfer fluid is stored in a vacuum-insulated low-activity copper cryostat. The cryostat is surrounded on all sides by 25 cm of low-activity lead. The entire assembly is surrounded by a radon-free tent and housed in a class 100 clean room, the exterior of which is instrumented on five sides with plastic scintillator panels for vetoing cosmic rays with 95.9\% efficiency. The detector is located 2150 feet underground for an overburden of 1585 meters water equivalent, at WIPP (Waste Isolation Pilot Plant), in the United States.

The EXO-200 TPC was installed in its cryostat in early 2010. The detector was first filled with natural (unenriched) xenon in late 2010. The data collected during these engineering runs were used to make a first assessment of the performance of the detector, and to perform a first round of calibrations. Low-background running with enriched xenon started in the spring of 2011. The EXO Collaboration announced the observation of the \bbtnu\ mode of \XE\ in August 2011. Prior to this measurement, reported in table~\ref{tab:bb2nu_exp}, the \bbtnu\ had been observed in all other important \bbonu\ candidate nuclei except \XE . The measured \bbtnu\ half-life is significantly lower than previously reported lower limits \cite{Bernabei:2002bn,Gavriljuk:2005xc}.

To identify the daughter barium, several methods are under study, including single-ion fluorescence, resonant ionization spectroscopy (RIS), and mass spectroscopy. Single ion fluorescence is a highly sensitive and highly selective method to observe a barium ion while held under vacuum in a RF trap. In this technique, the Ba$^+$ ion is rapidly cycled from its $6^2\text{S}_{1/2}$ ground state to its $6^2\text{P}_{1/2}$ excited state by illuminating it with lasers of the appropriate wavelength (493 nm and 650 nm), see fig.~\ref{fig:bariumenergylevels}. As the electronic state changes, the laser photons are scattered in all directions, and the scattered light can be easily detected by a photo-multiplier tube. EXO has achieved good single barium ion identification with this technique, even in the presence of low pressure xenon and helium gas mixtures. However, this technique also requires that the barium ion be retrieved from the TPC volume, transported to the RF trap, released, and trapped, while not altering its chemical or ionization state. Resonant ionization spectroscopy, on the other hand, is a technique which allows single barium ions to be observed without requiring a vacuum ion trap. In RIS, barium ions are desorbed from the surface of a transport probe, and subsequently resonantly ionized under illumination by 554 nm and 390 nm lasers. The ionized barium can then be observed with a Channeltron electron multiplier. Initial tests with the RIS technique have successfully identified barium being desorbed from the probe tip, so this technique is promising. Other avenues of research include barium identification within xenon ice, and barium extraction from a high pressure gas TPC using gas nozzles.

%% file: src/gerda.tex
The GERmanium Detector Array (GERDA) experiment \cite{Abt:2004yk}, located in Hall A of the Laboratori Nazionali del Gran Sasso (LNGS), will make use of naked Ge detectors immersed in a large cryostat of ultra-pure LAr.

\begin{figure}[t!]
\begin{center}
\includegraphics[]{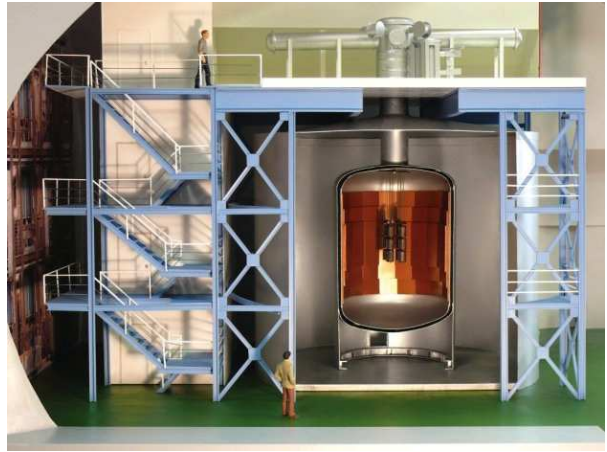}
\end{center}
\caption{Sketch of the GERDA experiment. The germanium arrays can be seen inside the copper cryostat, and this one placed inside the cylindrical water tank.} \label{fig:gerda}
\end{figure}

The Ge detectors are organized in strings (2--5 detectors) and mounted in special low-mass ($\sim 80$ g) holders made of ultra-pure copper and PTFE. The array of strings is contained in a vacuum insulated stainless steel cryostat of 4.2 m diameter and 8.9 m height. A copper shield covers the inner cylindrical shell of the cryostat with a maximum thickness of 6 cm. The cryostat is placed in a water tank, of 10 m diameter and 9.4 m height, serving as a gamma and neutron shield. It will be also used as a veto against cosmic rays thanks to its instrumentation with 66 photomultipliers, with good efficiency in detecting the Cherenkov light. The cosmic muon veto is reinforced by plastic scintillator panels on top of the detector, for a surface of about 20 m$^2$. A drawing of the detector and shielding is shown in fig.~\ref{fig:gerda}.

In its first phase, GERDA-1, eight fully refurbished germanium diodes (17.7 kg total active mass, 86\% isotopic enrichment in \GE ) from the previous Heidelberg-Moscow and IGEX experiments will be used. In the subsequent step, GERDA-2, new diodes will be used for a total active mass of 35.4 kg. These new diodes will be p-type Broad Energy (BEGe) detectors \cite{Budjas:2008wb,Agostini:2010ke}, allowing for a better discrimination of backgrounds thanks to a sophisticated pulse shape discrimination.

The experiment started commissioning runs in June 2010 using natural Ge, low-background, detectors, refurbished from the Genius-TF experiment. These commissioning runs focused on the investigation of the background sources (most notably on the unexpectedly large contribution from $^{42}$Ar-$^{42}$K in the LAr) and on $^{42}$Ar background mitigation strategies, see sect.~\ref{subsec:sensi_assumptions}. Data taking for the GERDA-1 physics run will start in the next months. The background level of the natural Ge setup was measured to be $0.06\pm 0.02$ \ckky\ (corresponding to $0.07\pm 0.02$ \ckkbby ), consistent with early indications from the first string of enriched Ge detectors deployed \cite{Cattadori:taup2011}. This rate, obtained without using pulse shape information, is a factor of 3--4 lower than the HM and IGEX measured ones (see sect.~\ref{subsec:past}), but still about a factor of 6 higher than the GERDA-1 goal. The reason for this higher than expected background rate is at present not fully understood. The goals of GERDA-2 are to start data taking in about two years, with about twice the isotope mass of GERDA-1, and with a background level of 0.001 \ckky\ (or 0.0012 \ckkbby ).

In the very long term a third phase of the experiment, GERDA-3, is foreseen to make use of about 1 ton of $^{76}$Ge target material together with a further reduction of background. Such an effort, common with the MAJORANA project (see below), would be feasible only in a word-wide collaboration, and provided that the GERDA approach could demonstrate to be the best candidate technology to push the double beta decay sensitivity below the inverted hierarchy mass threshold (about 30 meV).

%% file: src/majorana.tex
The MAJORANA Collaboration is following a more classic approach than GERDA in the design of a germanium-based experiment \cite{Guiseppe:2011me}. The Ge detectors will be mounted in a string-like arrangement in ultra-pure vacuum cryostat made from radiopure copper. The cryostat will be surrounded by a passive shielding of Cu and Pb, and an active muon veto.

The Collaboration is building a demonstrator module, to be placed at the Deep Underground Science and Engineering Laboratory (DUSEL) in the United States, with about 20 kg of natural BEGe detectors. The goal is to demonstrate a background rate of about 4 counts per tonne and per year in the 4-keV wide region of interest \cite{Schubert:2011nm}. The demonstrator is expected to operate with enriched detectors in 2013.

%% file: src/kamland.tex
The KamLAND-Zen experiment \cite{Efremenko:2011} will search for \bbonu\ in \XE\ using enriched xenon dissolved in liquid scintillator. This will allow a calorimetric measurement of the \bb\ electrons, as first proposed in \cite{Raghavan:1994qw}. Xenon is relatively easy to dissolve (with a mass fraction of more than 3\% being possible) and also easy to extract from the scintillator. 

\begin{figure}[t!]
\begin{center}
\includegraphics[scale=0.25]{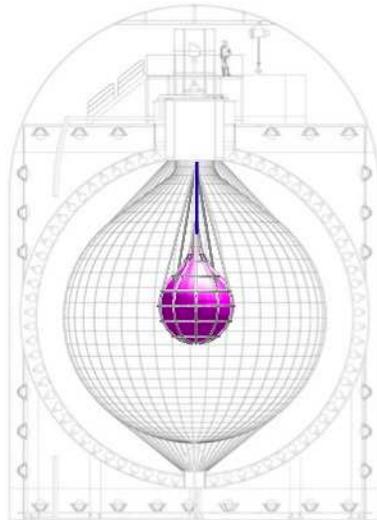}
\end{center}
\caption{Sketch of the KamLAND-Zen detector. The ball-on containing the dissolved xenon (purple) hangs in the center of the active volume.} \label{fig:kamlandzen}
\end{figure}

The major modification to the existing KamLAND detector \cite{Eguchi:2002dm} was the construction of an inner, very radiopure (of order $3\times 10^{-12}$ g/g of \URANIUM\ and \THORIUM) and very transparent balloon to hold the dissolved xenon. This balloon, 1.58 m in radius, is placed at the center of the KamLAND active volume as shown in fig.~\ref{fig:kamlandzen}.

The KamLAND-Zen experiment plans to dissolve 389 kg of \XE\ in the liquid scintillator of KamLAND in the first phase of the experiment, and up to 1 ton in a projected second phase. 

The proven resolution (from the previous operation of the KamLAND experiment) is 16\% FWHM at 1 MeV. The main sources of expected background are the \bbtnu\ tail, \BI\ impurities in the scintillator or in the balloon, $^{10}$C generated in the scintillator by cosmic rays, and $^{8}$B solar neutrinos. The expected background rate in the region of interest is $2\times10^{-4}$ \ckky\ \cite{Kozlov:taup2011}, corresponding to $2.2\times10^{-4}$ \ckkbby .

At the time of writing this report, the mini-balloon installation into the KamLAND detector has been completed, and detector commissioning is ongoing. Physics data-taking with the xenon-loaded liquid scintillator is expected to start in the fall of 2011.

%% file: src/next.tex
The Neutrino Experiment with a Xenon TPC (NEXT) \cite{Alvarez:2011my} will search for \bbonu\ in \XE\ using a 100-kg high-pressure gaseous xenon (HPXe) time projection chamber. 
Such a detector can provide both good energy resolution and event topological information for background rejection \cite{Nygren:2009zz}.

\begin{figure}[t!]
\vspace{0.75cm}
\begin{center}
\includegraphics[angle=-90,scale=0.45]{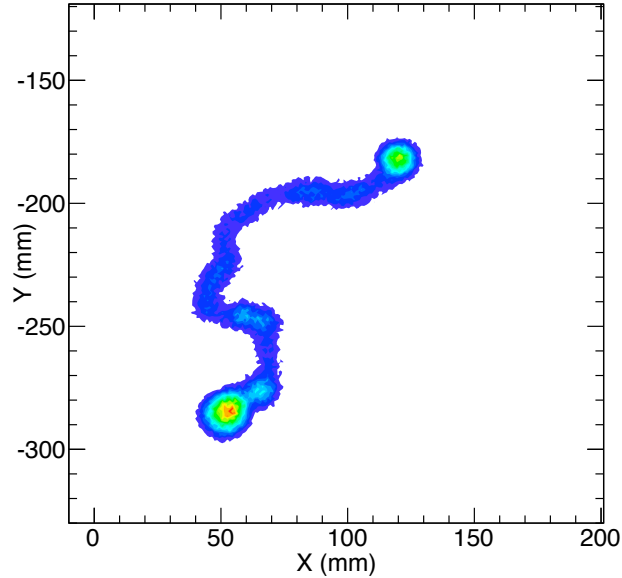}
\end{center}
\caption{Simulation of a \bbonu\ track in gaseous xenon at 10 bar \cite{Alvarez:2011my}.} \label{fig:next_track}
\end{figure}

Double beta decay events leave a distinctive topological signature in HPXe: a ionization track, of about 30 cm long at 10 bar, tortuous due to multiple scattering, and with larger energy depositions at both ends (see fig.~\ref{fig:next_track}). The Gotthard experiment \cite{Luscher:1998sd}, consisting in a small xenon TPC (5.3 kg of xenon, 68\% enrichment in \XE ) operated at 5 bar, proved the effectiveness of such a signature to reject background, achieving a background rate of only $\sim0.01$ \ckky.

The design of NEXT is optimized for energy resolution (better than 1\% FWHM at $Q_{\beta \beta}$) by using proportional electroluminescent (EL) amplification of the ionization signal. The detection process is as follows. Particles interacting in the HPXe transfer their energy to the medium through ionization and excitation. The excitation energy is manifested in the prompt emission of VUV ($\sim$178 nm) scintillation light. The ionization tracks (positive ions and free electrons) left behind by the particle are prevented from recombination by a strong electric field (0.5--1.0 kV/cm). Negative charge carriers drift toward the TPC anode, entering a region, defined by two highly-transparent meshes, with an even more intense electric field (3.5 kV/cm/bar). There, further VUV photons are generated isotropically by electroluminescence. Therefore, both scintillation and ionization produce an optical signal, to be detected with a sparse plane of PMTs located behind the cathode. The detection of the primary scintillation light constitutes the start-of-event ($t_0$), whereas the detection of EL light provides an energy measurement. Electroluminescent light provides tracking as well, since it is detected also a few mm away from production at the anode plane, via a dense array (1 cm pitch) of 1-mm$^{2}$ SiPMs.

The NEXT detector will operate at 10 bar, with xenon enriched at 90\% in the \XE\ isotope. At that pressure the 100 kg mass of xenon results in a volume of $\sim$2.5 m$^3$.

The major benefits of the NEXT 100 proposal are its high background rejection factor, resulting in an expected background rate of $2\times 10^{-4}$ \ckkbby , and the fact that xenon is relatively easy (cheap) to enrich and obtain in large quantities.

The NEXT Collaboration expects to commission the detector at the end of 2013. The experiment plans to start its physics run in the second half of 2014.

%% file: src/sno+.tex
SNO+ \cite{Kraus:2010zz} is the follow-up of the successful SNO experiment \cite{Boger:1999bb}, located at SNOLAB, in Canada. It re-uses the existing equipment of the detector (acrylic vessel, photomultipliers and their support structure, electronics and the light water shield) replacing the heavy water by $\sim$780 tonnes of liquid scintillator (linear alkylbenzene, LAB).

The physics program of the SNO+ detector includes measurements of low energy solar neutrinos and \bbonu\ searches using \ND. In order to do that, the liquid scintillator will be loaded with a neodymium salt, resulting in about 50 kg of \ND. This isotope has the second highest endpoint, 3.37 MeV, and the fastest predicted neutrinoless double beta decay rate due to its large phase space factor, see fig.~\ref{fig:g0nu}. The high endpoint is above most radioactive backgrounds, such as radon, and this is a significant advantage. However, enrichment of this isotope seems difficult.

The energy resolution of the SNO+ detector is estimated to be 6.5\% FWHM at 3.4 MeV. External backgrounds can be rejected with a relatively tight fiducial volume selection, cutting however about 50\% of the signal. The most important sources of background are expected to be \TL\ impurities in the scintillator, the irreducible background from $^{8}$B solar neutrinos and \bbtnuº events from \ND . Assuming the radiopurity levels for the liquid scintillator achieved by BOREXINO ($\sim10^{-17}$ g/g of \TL) \cite{Franco:2009wg}, simulations predict a background rate of $\sim10^{-2}$ \ckkbby\ \cite{Wright_PhDthesis}. 

\begin{figure}[t!b!]
\begin{center}
\includegraphics[width=0.55\textwidth]{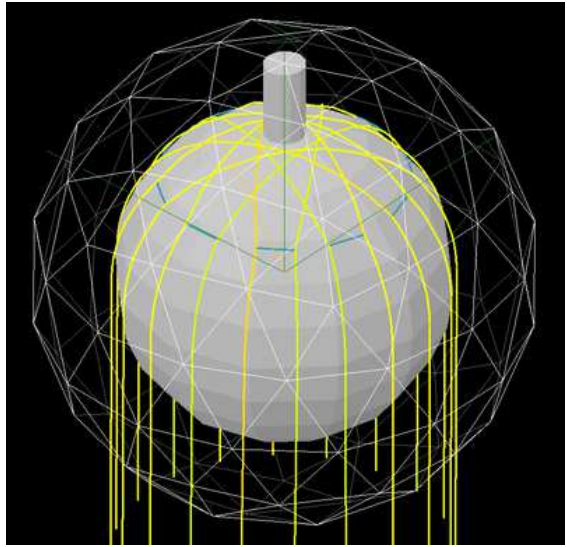}
\end{center}
\caption{\label{fig:snoplus_anchor_possibility}One of the candidate configurations for the SNO+ acrylic vessel anchor system. The acrylic vessel is shown in grey, and the anchor system in yellow. The outer sphere made of triangles is the PMT support structure.} 
\end{figure}

The SNO+ experiment is expected to start commissioning in the spring of 2013 with pure liquid scintillator, to be followed by the Nd-loaded liquid scintillator phase. Given that the LAB liquid scintillator is about 15\%  less dense than the surrounding light water, one of the major technical challenges of the SNO+ upgrade is the design of a hold-down system for the acrylic vessel using a net of radiopure ropes, see fig.~\ref{fig:snoplus_anchor_possibility}.  

%% file: src/snemo.tex
This proposed new installment of the NEMO detectors series consists of up to 20 tracker-calo modules, each one containing a thin foil of about 5 kg of \bb-decaying material, probably \SE, although other isotopes such as \ND\ or \CA\ are also under consideration. 

\begin{figure}[t!b!]
\begin{center}
\includegraphics[angle=270,width=0.45\textwidth]{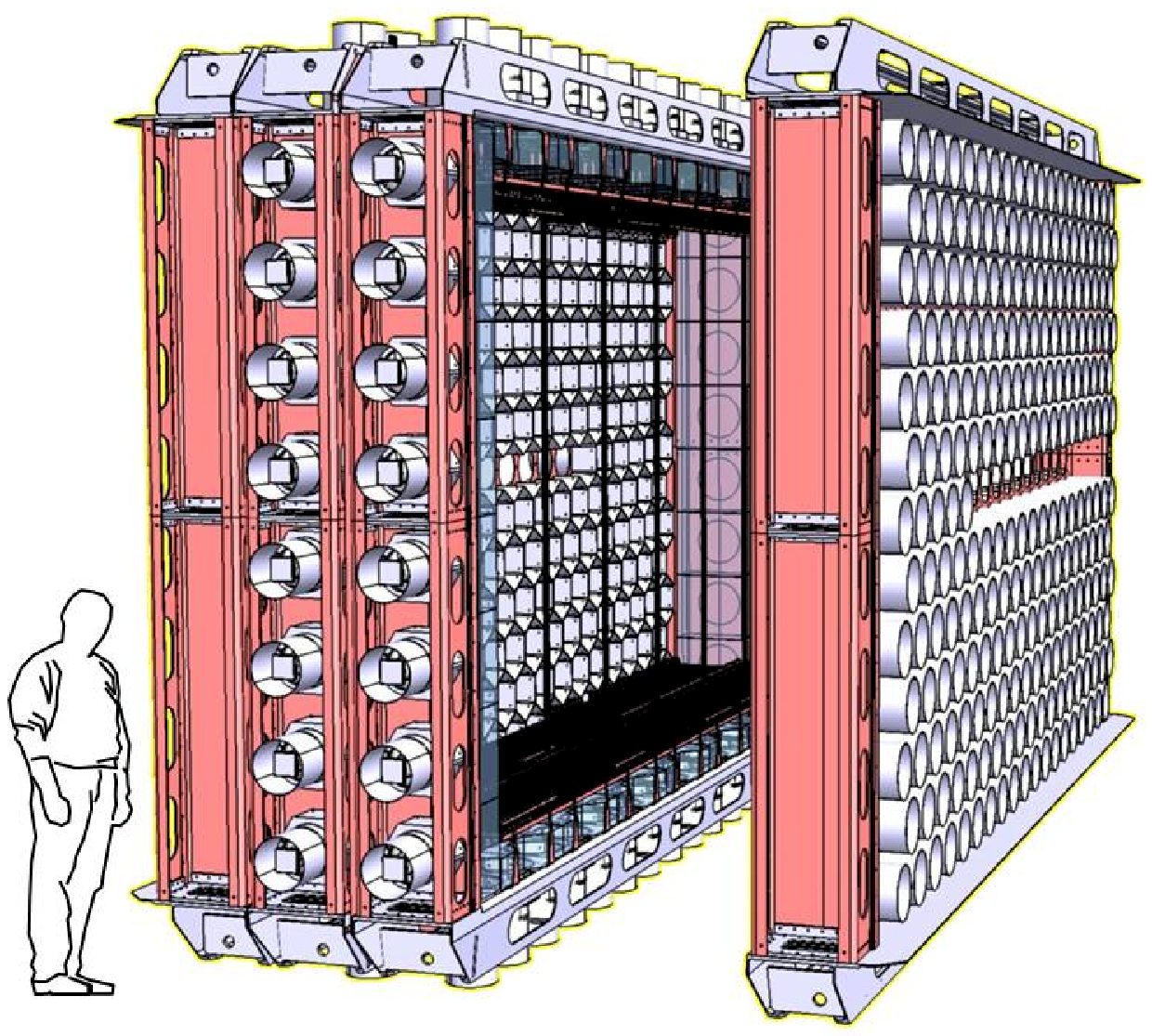} \hspace{0.04\textwidth}
\includegraphics[angle=270,width=0.45\textwidth]{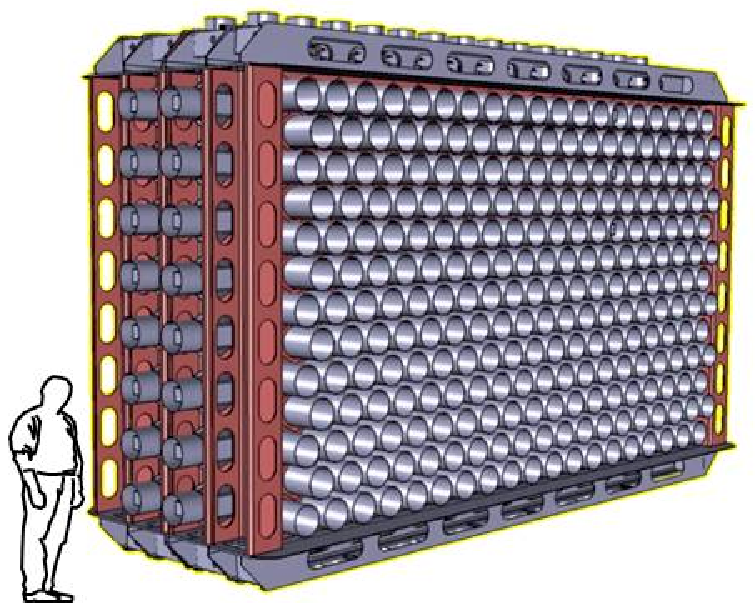}
\end{center}
\caption{A SuperNEMO module. The source foil (not shown) is placed in the center of a tracking volume consisting of drift cells operating in Geiger mode. The tracking volume is surrounded by calorimetry consisting of scintillator blocks connected to PMTs (grey). The support frame is shown in red.} \label{fig:snemo}
\end{figure}

A sketch of a SuperNEMO module can be seen in fig.~\ref{fig:snemo}. The source foil, 3 meters high and 4.5 meters long, with a surface density of about 40 mg/cm$^{2}$, is placed in the center of a tracking chamber with overall dimensions of 4 m height, 5 m length and 1 m width. Nine planes of drift cells operating in Geiger mode and a magnetic field of 25 Gauss allow to reconstruct the trajectory and charge of particles crossing the chamber. A calorimeter consisting of blocks of plastic scintillator coupled to low-activity PMTs surrounds the tracking chamber on four sides. Its granularity allows the energy of individual particles to be measured.

The physics case of SuperNEMO relies on several significant improvements over the NEMO-3 detector performance \cite{Shitov:2010nt}. The energy resolution is expected to be 7\% FWHM at 1 MeV, a factor of 2 better than in NEMO-3. Such a resolution has been attained with a 28 cm hexagonal PVT scintillator directly coupled to a 8-inch PMT \cite{Freshville:2011zz}. The detection efficiency of SuperNEMO is estimated by means of simulation to be about 30\%, almost a factor of 2 better than in NEMO-3. As far as the backgrounds are concerned, SuperNEMO goals require an impressive improvement in the purification (both chemical and via distillation methods) of the source foils. In particular, \BI\ and \TL\ contamination in \SE\ foils are to be reduced by factors of 50 and 170, respectively. A dedicated setup, the BiPo detector, installed in the Laboratorio Subterr\'aneo de Canfranc (LSC), will measure the radiopurity of the foils in order to make sure that the required levels are achieved. Finally, in order to decrease radon gas levels in the tracking chamber down to negligible levels ($<$0.15 mBq/m$^3$ ), a reduction of at least a factor of 40 with respect to NEMO-3 is needed. 

The first SuperNEMO module, called the demonstrator, will be the first step from R\&D to construction with the aims to demonstrate the feasibility of large scale mass production, to measure the backgrounds (especially from radon emanation), and to finalize the detector design. The demonstrator will be installed in the space previously occupied by the NEMO-3 detector at the Modane Underground Laboratory.

The current plans of the SuperNEMO Collaboration for the following: (a) demonstrator construction, 2010--2012; (b) demonstrator physics run start-up, 2013; and (c) full detector construction start-up, 2014. 

%% file: src/other_exps.tex
\subsubsection*{CANDLES} This project \cite{Umehara:2010zz} proposes the use of CaF$_{2}$ scintillating crystals to search for \bbonu\ in \CA. The crystals would be immersed in liquid scintillator providing shielding and an active veto against external backgrounds. Among the \bb\ isotopes, \CA\ has the highest $Q$-value, 4.27 MeV. This places the signal well above the energy region of the natural radioactive processes. Unfortunately, the natural abundance of the isotope is only 0.187\% and enrichment seems complicated. Therefore, many tons of crystals are needed for a competitive new-generation experiment.  
\subsubsection*{COBRA} The COBRA experiment \cite{Zuber:2001vm, Zuber:2010zz} is exploring the potentials of Cadmium Zinc Telluride (CdZnTe) room-temperature semiconductor detectors for \bbonu\ searches. Out of the several \bb\ candidate isotopes in CdZnTe, COBRA is focusing on \TE, because of its natural abundance, and \CD, because of its high $Q$-value of 2.8 MeV. Activities are split in two main directions: (a) the identification of the main background components in a setup of 64 commercial 1-cm$^{3}$ CdZnTe diodes located at LNGS; and (b) the development of pixelized devices that would allow to reduce the background by particle identification.
\subsubsection*{DCBA} The Drift Chamber Beta-ray Analyzer \cite{Ishikawa:2011zz} is a magnetized tracker (drift chambers) that can reconstruct the trajectories of charged particles emitted from a \bb\ source foil. The momentum and kinetic energy are derived from the track curvature in the magnetic field. A prototype, DCBA-T2, has shown energy resolution of about 150 keV (FWHM) at 1 MeV, and the main source of background (\BI) has been identified. A new apparatus, DCBA-T3, with a more intense magnetic field is now under construction at KEK.
\subsubsection*{LUCIFER} The idea of LUCIFER \cite{Giuliani:2010zz, Ferroni:2011zz} is to join the bolometric technique proposed for the CUORE experiment with the bolometric light detection technique used in cryogenic dark matter experiments. Preliminary tests on several \bbonu\ detectors have clearly demonstrated the background rejection capabilities that arise from the simultaneous, independent, double readout (heat and scintillation). LUCIFER will consist of an array of ZnSe crystals operated at 20 mK. The proof of principle with about 10 kg of enriched Se is foreseen for 2014.
\subsubsection*{MOON} The MOON detector \cite{Ejiri:2010zz} is a stack of multi-layer modules, each one consisting of a scintillator plate for measuring energy and time, two thin detector layers for position and particle identification, and a thin \bb\ source film interleaved between them. At present, NaI(Tl) scintillators are considered as the candidates for the scintillator plates. Energy resolution around 3\% FWHM at 3 MeV has been achieved during the R\&D phase. For position-sensitive detectors, possible candidates are multi-wire proportional chambers (MWPCs) and Si-strip detectors. 
\subsubsection*{XMASS} XMASS \cite{Sekiya:2010bf, Takeda:2011zz} is a multi-purpose liquid xenon scintillator. Although optimized for dark matter searches, it will also investigate neutrinoless double beta decay and solar neutrinos. The detector, with about 800 kg of xenon, was installed in the Kamioka mine (Japan) in the fall of 2010. The excellent self-shielding capabilities of the liquid xenon will be used to define a virtually background-free inner volume. 

%% file: src/sensi.tex
In this section we try to assess the physics case of the new-generation double beta experiments described above\footnote{For the sensitivity computation, we restrict ourselves to experiments that involve at least a few kg of \bb\ emitter mass, that are approved, and that have been granted a significant financial support.}. We quote the experimental sensitivities to \mbb, assuming the standard light Majorana neutrino exchange as the dominant \bbonu\ mechanism. To perform this risky exercise we make use of the physics-motivated ranges for the NME values described in sect.~\ref{subsec:nme_pmr}, and of the set of experimental parameters summarized in table~\ref{tab:parameters}. A discussion motivating our choice of parameters in table~\ref{tab:parameters} is given in sect.~\ref{subsec:sensi_assumptions}.

\clearpage
\begin{sidewaystable}[H]
\begin{center}
\caption{Basic parameters for the different double beta experiments: \bb\ emitter mass \Mbb, \bbonu\ efficiency $\varepsilon$, FWHM energy resolution $\Delta E$, and background rate $c$ per unit energy, \bb\ isotope mass and time. The last column indicates the number of background events within the ROI, and is the product of the \Mbb, $\Delta E$ and $c$ columns. Comparison of different approaches is very difficult directly from the numbers in the table, but this information is fundamental to compute their sensitivity.} \label{tab:parameters}
\begin{tabular}{lcccccc}
\hline
Experiment  & \Mbb    & $\varepsilon$ & $\Delta E$ & $c$                  & Bgr/ROI   \\
            & (\kgbb) &               & (keV)      &  ($10^{-3}$ \ckkbby)  & (cts/yr)   \\ \hline
EXO-200     & 141     & 0.34 	      & 100        & 0.78--5              & 11--71  \\
GERDA-1     & 15.2    & 0.95          & 4.2        & 12--70               & 0.77--4.5  \\
GERDA-2     & 30.4    & 0.84          & 2          & 1.2--7               & 0.07--0.43 \\ 
CUORE-0     & 10.9    & 0.83          & 5          & 180--390             & 9.8--21.3  \\
CUORE 	    & 206     & 0.83          & 5    	   & 36--130              & 37.1--134  \\
KamLAND-Zen & 357     & 0.61          & 250 	   & 0.22--1.8            & 19.6--161  \\
MAJORANA Demonstrator   & 17.2    & 0.85          & 2          & 1.2--12              & 0.04--0.41  \\  
SNO+        & 44      & 0.50          & 220        & 9--70                & 87--680 \\
NEXT 	    & 89.2    & 0.33 	      & 18         & 0.2--1               & 0.32--1.6 \\
SuperNEMO Demonstrator  & 7       & 0.28          & 130        & 0.6--6               & 0.55--5.5 \\
 \hline
\end{tabular}
\end{center}
\end{sidewaystable}

Although different experimental aspects are relevant from the point of view of the feasibility of an experiment, the sensitivity can be computed using only a few parameters ---namely, effective mass of the isotope and background rate in the energy Region Of Interest (ROI)--- that can be extracted from the fundamental numbers in the design of the experiments. 

The sensitivity is calculated based on the Feldman-Cousins method \cite{Feldman:1997qc} for constructing confidence intervals, following the prescription given in \cite{GomezCadenas:2010gs}. For each experiment, we define a ROI centered at the \bb\ decay $Q$-value and extending for one FWHM of energy resolution, and we compute the experimental sensitivity at 90\% confidence level. We take into account the effect of the FWHM selection (corresponding to 76\% efficiency assuming gaussian resolution) as a multiplicative factor to the experimental efficiency reported in table~\ref{tab:parameters}.

 Our prescription assumes a counting experiment in the ROI with a perfectly known background rate. In other words, in our sensitivity computation, we neglect systematic uncertainties\footnote{With one exception for the SNO+ and KamLAND-Zen experiments, see sect.~\ref{subsec:sensi_assumptions}, where an attempt has been made to account for systematic effects affecting energy reconstruction.} and any energy shape information that may be present in the energy distribution of events. Systematic uncertainties may possibly affect the parameters listed above, especially the knowledge of the backgrounds, and deteriorate the sensitivity. On the other hand, use of additional information beyond the overall count rate of \bbonu\ candidates within the ROI may yield some sensitivity improvement. While important, both effects would be extremely difficult to incorporate in such a sensitivity comparison, given that most new-generation experiments discussed here have not even started their commissioning phase, yet.

\begin{figure}[ht]
\includegraphics[width=\textwidth]{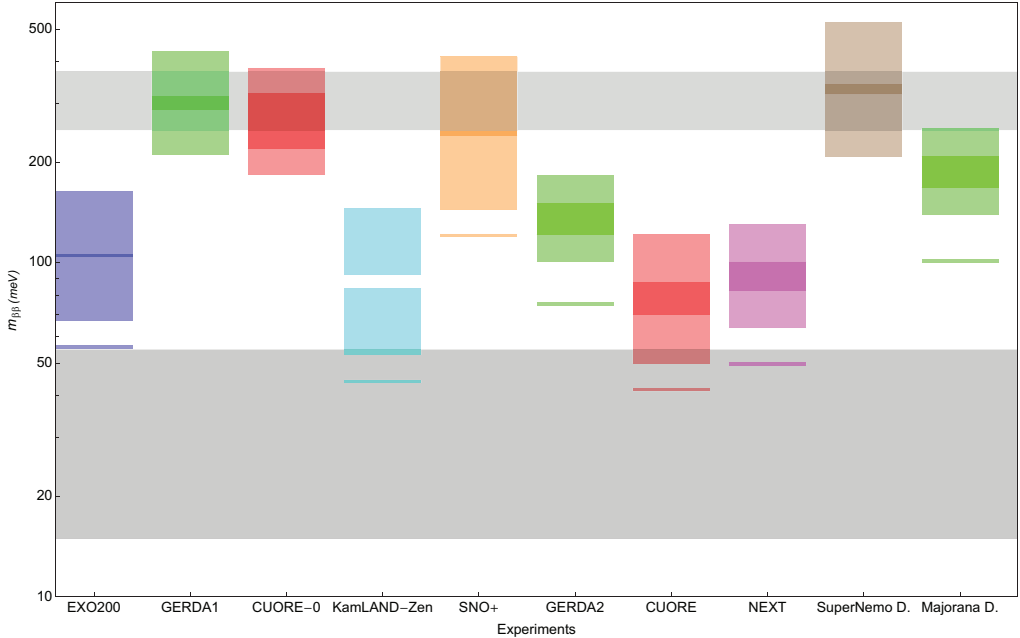}
\caption{Sensitivity of the different experiments to the neutrino effective mass \mbb\ computed assuming a 5 years exposure, the PMR intervals for the nuclear matrix elements (see sect.~\ref{subsec:nme_pmr}), and for both the optimistic and pessimistic experimental parameters of table~\ref{tab:parameters}. A statistical 90\% CL is computed according to the Feldman-Cousins method \cite{Feldman:1997qc}, assuming a signal region of one FWHM and the corresponding efficiency. For each experiment, the sensitivities for the two experimental parameter sets are drawn as overlapping rectangles. A sensitivity line corresponding to a 10 years exposure, and to the most optimistic NME and experimental parameter set, is also shown. The upper grey region represents the KK claim \cite{KlapdorKleingrothaus:2006ff} while the lower grey region represents the inverted hierarchy region (see fig.~\ref{fig:mbetabetavsmlight}).}
\label{fig:sens-pmr}
\end{figure}

In fig.~\ref{fig:sens-pmr} we show the \mbb\ sensitivities at 90\% CL for the new-generation proposals discussed above, assuming a 5 years exposure for all of them. The colored rectangles reflect the uncertainty coming from the PMR nuclear matrix elements, see fig.~\ref{fig:nme}. For each experiment, two rectangles are shown, corresponding to the most optimistic and most pessimistic background rate expectations given in tab.~\ref{tab:parameters}, respectively. For most experiments, the two rectangles overlap. For each experiment, the thin solid line at the lowest \mbb\ values is meant to give an idea of what can be gained by increased exposures. The line represents the \mbb\ sensitivity for the most optimistic NME values (within the PMR range of fig.~\ref{fig:nme}) and for the most optimistic background rate expectations (within the ranges of tab.~\ref{tab:parameters}) for a 10 years exposure. For reference, we also show the mass range of the standing KK claim of evidence for \bbonu\ in \GE\ \cite{KlapdorKleingrothaus:2006ff} (with a central value of 300 meV) and the mass region as predicted under the inverted hierarchy hypothesis (between 17 and 52 meV, see fig.~\ref{fig:mbetabetavsmlight}). The sensitivity of the various proposals after a 5 years exposure, both in terms of \Tonu\ and \mbb, is also shown in tab.~\ref{tab:sensitivity}, where the central value of the PMR interval for the nuclear matrix elements and the optimistic background rate expectations in tab.~\ref{tab:parameters} have been assumed.

\begin{table}[t!b!]
\caption{Sensitivity of the experiments at 90\% CL after a 5 years exposure, both in terms of half-life \Tonu\ and in terms of neutrino effective mass \mbb . These values are obtained from the optimistic background rate assumptions in tab.~\ref{tab:parameters}. The conversion from \Tonu\ to \mbb\ assumes the central value of the PMR interval for the nuclear matrix elements.}\label{tab:sensitivity}
\begin{center}
\begin{narrowtabular}{3cm}{lcr}
\hline
Experiment & \Tonu\ (years) & \mbb\ (meV) \\ \hline
CUORE-0 	& $8.67\times 10^{24}$ & 203 \\
CUORE 		& $8.86\times 10^{25}$ & 63\\
GERDA-1 	& $4.49\times 10^{25}$ & 252\\
GERDA-2 	& $1.37\times 10^{26}$ & 121\\
EXO200 		& $8.20\times 10^{25}$ & 82\\
NEXT 		& $9.13\times 10^{25}$ & 78\\
KamLAND-Zen 	& $1.32\times 10^{26}$ & 65\\
SNO+ 		& $5.38\times 10^{24}$ & 182\\
SuperNEMO Demonstrator 	& $9.15\times 10^{25}$ & 258\\
MAJORANA Demonstrator	& $7.19\times 10^{25}$ & 258\\
 \hline
\end{narrowtabular}
\end{center}
\end{table}

The first thing to remark from fig.~\ref{fig:sens-pmr} is that the KK claim should be unambiguously solved by several new-generation proposals using different isotopes. If our assumptions are correct, this will certainly be the case for \GE\ (GERDA, MAJORANA), \TE\ (CUORE) and \XE\ (EXO-200, KamLAND-Zen, NEXT), and possibly also for \SE\ (SuperNEMO) and \ND\ (SNO+). Multi-isotope determination of \bbonu\ may therefore become a real possibility within this decade. In this respect, it is important to note that GERDA and MAJORANA are the only experiments among those in fig.~\ref{fig:sens-pmr} using the same isotope of the HM experiment\footnote{Not only: we have seen that GERDA in its first phase re-uses the same detectors as in the HM experiment.}, and should therefore be able to provide the only truly model-independent confirmations of this claim.

On the other hand, several experiments appear to have a very good chance to reach a sensitivity of 100 meV or better, in particular CUORE, KamLAND-Zen, NEXT and EXO-200. In our most optimistic scenario concerning NME values and experimental performances, this target may also be reached by GERDA during its second phase. Given our uncertainties, we cannot predict which, among these 4--5 different experimental proposals, will provide the best \mbb\ sensitivity after a 5 years exposure. To this end, a better knowledge of the actual (as opposed to expected) values for the background rates, of the systematic uncertainties affecting the measurement, and of the NME values would be necessary for all experiments.

From fig.~\ref{fig:sens-pmr}, our expectation is that it will be almost impossible for the new-generation experiments discussed here to discover \bbonu\ after 5 years of exposure if the neutrino mass spectrum is hierarchical ($m_{\rm light}\simeq 0$) rather than degenerate, since essentially no overlap exists between the experimental sensitivities and the 17--52 meV inverted hierarchy region. Not only larger exposures, but also new (better) experimental proposals, would be needed to fully probe this mass region. As mentioned above, we expect experiments using \XE\ (EXO-200, KamLAND-Zen, NEXT) to provide for the first time during this decade a comparable or better sensitivity than \GE\ (GERDA, MAJORANA) and \TE\ (CUORE) experiments, which dominated the field over the past two decades. In perspective, if the low-background expectations of new-generation \XE\ proposals will be confirmed during this decade, \XE\ would be a particularly favorable isotope to use also in the longer term. This is because scalability to isotope masses in the 1-10 ton range are in this case more realistic than with any other isotope. 

\begin{figure}[ht]
\includegraphics[width=\textwidth]{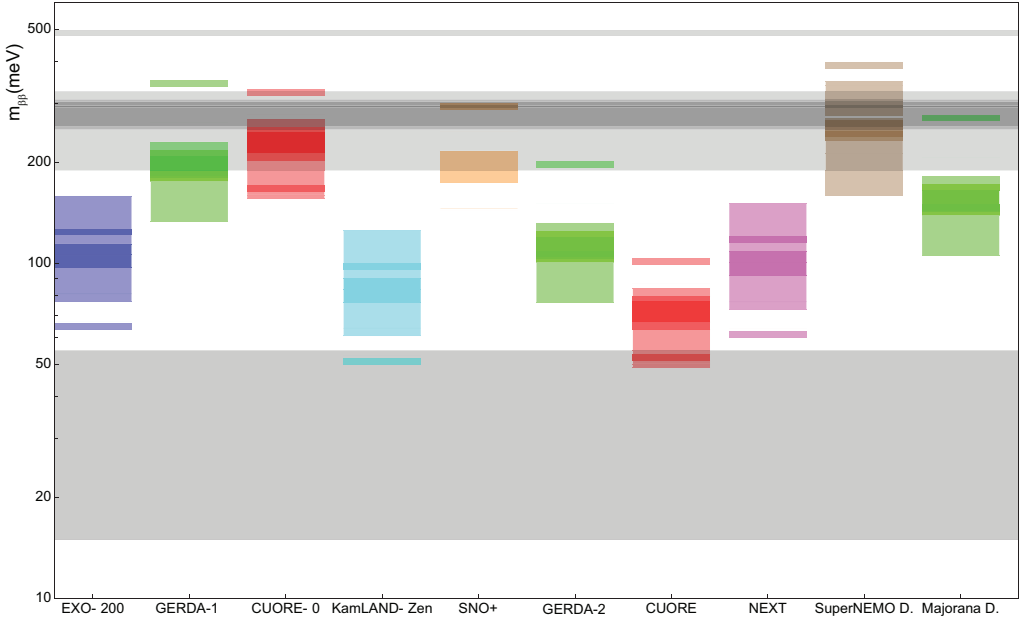}
\caption{Sensitivity of the different experiments to the neutrino effective mass \mbb\ computed assuming the optimistic experimental parameters of table~\ref{tab:parameters}. A statistical 90\% CL is computed according to the Feldman-Cousins method \cite{Feldman:1997qc}, assuming a signal region of one FWHM and the corresponding efficiency. Five different frameworks for NME calculations are considered, following reference \cite{Dueck:2011hu}, and are drawn as overlapping rectangles. The upper grey region represents the KK claim \cite{KlapdorKleingrothaus:2006ff} while the lower grey region represents the inverted hierarchy region (see fig.~\ref{fig:mbetabetavsmlight}).}
\label{fig:sens-models}
\end{figure}

Figure~\ref{fig:sens-pmr} represents our main result for the physics case comparison of new-generation \bbonu\ experiments. In this figure, a single, physics-motivated, NME uncertainty band is used for each isotope, following our discussion in sect.~\ref{subsec:nme_pmr}. For completeness, we have also repeated the same exercise for several other NME values or ranges, one for each theoretical framework considered in \cite{Dueck:2011hu}. The result is shown in fig.~\ref{fig:sens-models}. As mentioned above, the spread of the corresponding \mbb\ predictions most likely overestimates the theoretical uncertainty in the \Tonu\ $\to$ \mbb\ sensitivity conversion. For figure clarity, only nuclear theory assumptions are varied in fig.~\ref{fig:sens-models}, while the detector performance parameters are fixed to the most optimistic values of tab.~\ref{tab:parameters}. As can be seen in fig.~\ref{fig:sens-pmr}, our assumed uncertainties in the detector performance parameters would yield \mbb\ sensitivity changes of about the same size as the nuclear theory variations shown in fig.~\ref{fig:sens-models}.

%% file: src/sensi_assumptions.tex
The reliability of the sensitivity estimates given above critically depend on how realistic our choice of detector performance indicators is. In this section, we discuss how we have chosen the parameters reported in tab.~\ref{tab:parameters}. This discussion is mostly intended for the expert reader wishing to independently assess our choices, and to use his/her own judgment to modify them accordingly. Given that the largest uncertainty affecting the ultimate \mbb\ sensitivity of a proposal is almost always related the achievable background rates, we decide to quote a background rate range. For all other indicators, a single number rather than a range is used, see tab.~\ref{tab:parameters} 

The EXO-200 TPC is filled with about 175 kg of liquid xenon enriched to 80.6\% in the isotope \XE\ \cite{Ackerman:2011gz}, corresponding to a \bb\ mass of about 141 \kgbb. For the \bbonu\ efficiency around \Qbb, we take the efficiency assumed by the EXO Collaboration for their \bbtnu\ analysis \cite{Ackerman:2011gz}, corresponding to $\varepsilon = 0.34$ above the 720 keV analysis threshold. The inefficiencies are dominated by the fiducial volume cut, keeping 63 out of 175 kg of liquid xenon \cite{Ackerman:2011gz} (0.36 efficiency). A 6.3\% inefficiency introduced by vetoing $\beta$-$\alpha$ coincidences \cite{Ackerman:2011gz} has also been considered in our efficiency assumption. In \cite{Ackerman:2011gz}, the collaboration measured an energy resolution of $\sigma_E/E=4.5\%$ at 2615 keV for the EXO-200 detector. This value was obtained using a 376 V/cm drift field and ionization signals only. An improvement of up to a factor of 2.5 could be achieved with higher (1--4 kV/cm) drift fields and combining ionization with scintillation information, see \cite{Conti:2003av}. As a result, we assume a nearly-nominal, 4.2\% FWHM energy resolution at \Qbb, corresponding to about 100 keV. For our background rate lower limit, we consider the collaboration's goal of 20 radioactive background events per year in a $\pm 2\sigma$ interval around the \Qbb\ endpoint, for a $\sigma/E=1.6\%$ energy resolution at 2.5 MeV, a detector mass of 200 kg and a 80\% enrichment in \XE\ \cite{Hall:2010zzg}. From these numbers, we obtain a background rate of $c=0.78\times 10^{-3}$ \ckkbby. This nominal background rate prediction still remains to be updated based on real EXO-200 data. As a worst-case background rate scenario, we take the rate that has already been achieved: $4\times 10^{-3}$ \ckky, see \cite{Barbeau:taup2011}. This background level was obtained without full lead shielding, radon exclusion tent, radon trap or full 3-dim reconstruction, and might therefore be improved in the future \cite{Barbeau:taup2011}. This number corresponds to $c=5\times 10^{-3}$ \ckkbby\ per unit \bb\ mass.

GERDA-1 will use eight refurbished \GE\ diodes from the Heidelberg-Moscow and IGEX experiments, for a total active mass of 17.66 kg and 0.86 isotopic enrichment in \GE\ \cite{Knopfle:panic2011}, corresponding to a \bb\ mass of 15.2 \kgbb. As for the \bbonu\ efficiency, the actual value will ultimately depend on analysis details that are unknown at the moment, for example whether the collaboration will rely on pulse shape discrimination already in phase I to further reduce multi-site energy deposition events. In the absence of an updated number, we assume $\varepsilon =0.95$ as originally quoted by the collaboration in \cite{gerda_proposal}. The FWHM energy resolution for GERDA-1 diodes was measured to be between 3.6 and 6.0 keV at the 2615 keV gamma ray line from \TL\ \cite{Cattadori:taup2011}. Taking the central value of this interval (4.8 keV) and extrapolating to the \GE\ \Qbb\ value (2.039 MeV), we estimate $\Delta E= 4.2\ \text{keV}$. The optimistic background rate scenario is assumed to be the collaboration's goal of 0.01 \ckky\ \cite{gerda_proposal}, corresponding to 0.012 \ckkbby\ per unit \bb\ mass. GERDA started commissioning in mid-2010, and data obtained since then can be used to estimate a worst-case scenario for the achievable background rates. The best background rate measured with a string of 3 natural germanium detectors, refurbished from the Genius-TF experiment, is $0.06\pm 0.02$ \ckky\ \cite{Cattadori:taup2011}. Since mid-2011, the first enriched germanium detectors have been deployed on a second string arm, using the best detector configuration found so far. Preliminary data from the enriched germanium detectors indicate a background rate that is compatible with the one found with the natural germanium diodes \cite{Cattadori:taup2011}. As a consequence, we assume 0.06 \ckky\ as upper limit for the GERDA-1 background rate, translating into $c=0.07$ \ckkbby\ per unit \bb\ mass.

For the phase II of the experiment, the GERDA Collaboration purchased 37.5 kg of germanium with an isotopic abundance in \GE\ of 0.86. The material has already been purified into 35.4 kg of 6N germanium \cite{Junker:npa5}, corresponding to a \bb\ mass of 30.4 \kgbb. In order to reduce backgrounds, both sophisticated pulse-shape discrimination (PSD) techniques and additional instrumentation for the LAr veto are likely to be used by GERDA-2. We assume an overall \bbonu\ efficiency of $\varepsilon =0.84$, given by the product of the 0.86 efficiency for a PSD cut reported in \cite{Agostini:2010rd}, times the 0.973 efficiency for a LAr veto cut reported in \cite{Heisel_PhDthesis}. Compared to phase I detectors, a significantly improved FWHM energy resolution of 2 keV at \Qbb\ has been measured for the BEGe detectors to be used by GERDA-2, see \cite{Agostini:2010ke}. The lower limit on the background rate is taken to be the collaboration's goal of 0.001 \ckky\ \cite{gerda_proposal}, corresponding to $c=0.0012$ \ckkbby. Again, we use preliminary results from the GERDA commissioning runs to estimate an upper limit on the background rate. Commissioning data indicate that $\beta$ decays of $^{42}$K, that is in turn produced positively charged by the $^{42}$Ar decay within the LAr veto, can contribute very significantly to the \Qbb\ background rate. This $^{42}$K background can be most easily quantified by measuring the 1525 keV gamma ray line. On the one hand, the collaboration estimated via simulations that a background rate at \Qbb\ of up to $1.7\times 10^{-3}$ \ckky\ can be obtained by a homogeneous distribution of $^{42}$K around the detectors, for a 43.9 $\mu\text{Bq/kg}$ contamination in $^{42}$K \cite{Lehnert_PhDthesis}. On the other hand, a 1525 keV line about 20 times more intense than expected was observed during the first commissioning run. A significant fraction of this enhancement has been understood as due to a inhomogeneous $^{42}$K distribution caused by the field lines drifting the positively-charged $^{42}$K ions to the detector surface. To prevent the $^{42}$K ions to reach the detector surfaces, the collaboration deployed a copper shield called the \emph{``mini-shroud''}. The additional shroud reduced the counts at the 1525 keV line and at \Qbb\ by a factor of 4-5, and a preliminary measurement of the $^{42}$Ar specific activity of about 160 $\mu\text{Bq/kg}$ was obtained for almost field-free runs \cite{Knopfle:panic2011}. This measured value, combined with the simulation result of $1.7\times 10^{-3}$ \ckky\ for a 43.9 $\mu\text{Bq/kg}$ contamination in $^{42}$K, motivates our upper limit background rate assumption of about 0.006 \ckky, or $c=0.007$ \ckkbby\ per unit \bb\ mass.

CUORE-0 will make use of a single CUORE-like tower with 39 kg of natural TeO$_2$ crystals and with an isotopic abundance of 0.34167 of \TE\ \cite{Alessandria:2011rc}, corresponding to a \bb\ mass of 10.9 \kgbb. The overall signal efficiency has been estimated to be about $0.83$, as obtained for the big crystals of the CUORICINO experiment \cite{Andreotti:2010vj}. The main inefficiency source is the ``physical'' inefficiency due to beta particles escaping the detector and radiative processes. The expected FWHM energy resolution of the CUORE detectors is $\Delta E\simeq 5\ \text{keV}$ at the \bbonu\ transition energy \cite{Alessandria:2011rc}. As lower limit on the background rate, we assume 0.05 \ckky\ from the 2.615 keV gamma ray multi-Compton events coming from the irreducible \THORIUM\ contamination of the CUORICINO cryostat, as done in \cite{Alessandria:2011rc}. This rate corresponds to $c=0.18$ \ckkbby\ per unit \bb\ mass. As upper limit on the background rate, again we follow \cite{Alessandria:2011rc} and assume 0.11 \ckky, translating into $c=0.39$ \ckkbby\ per unit \bb\ mass. This latter number includes an additional background contribution from scaling the CUORICINO background in the conservative case of a factor of 2 improvement in \URANIUM\ and \THORIUM\ contamination of the copper and crystal surfaces. Such surface contamination results in degraded alphas that may mimic the \bb\ signal. This factor of 2 improvement is largely motivated by the background rates measured in the Three Towers Test (TTT), allowing to estimate the surface contamination of the copper detector holders, responsible for a large ($50\pm 20\%$ \cite{Alessandria:2011rc}) fraction of the CUORICINO backgrounds \cite{Alessandria:2011vj}. For such tests, crystals were dismounted from the CUORICINO detector, and repolished on the surfaces. Three different types of copper cleaning were tested. The copper treatment procedure selected by the collaboration proved to be able to reduce the copper surface contamination by at least a factor of 2 compared to CUORICINO \cite{Alessandria:2011rc,Alessandria:2011vj}.  

The CUORE total active mass will be 741 kg for the 988 envisaged cubic detectors \cite{Alessandria:2011rc}. As for CUORE-0, the natural TeO$_2$ crystals will have an isotopic abundance of 0.34167 of \TE\ \cite{Alessandria:2011rc}, resulting in about 206 \kgbb\ of \TE. The overall signal efficiency and the FWHM energy resolution at \Qbb\ are assumed to be equal to the CUORE-0 figures given above, $\varepsilon =0.83$ and $\Delta E\simeq 5\ \text{keV}$, respectively. The main improvements in background rate reduction compared to CUORE-0 are assumed to come from a much cleaner cryostat and from the different detector geometry. Under these assumptions, the CUORE background will be dominated by copper and crystal surface contaminations. As lower limit on the background rate, the 0.01 \ckky\ goal of the collaboration is assumed \cite{Alessandria:2011rc,Arnaboldi:2003tu}, corresponding to $c=0.036$ \ckkbby. As upper limit, an extrapolation of the currently available measurements \cite{Alessandria:2011vj} on copper and crystal surface contaminations to the CUORE geometry results in a 0.035 \ckky\ upper limit \cite{Gorla:taup2011}, resulting into $c=0.13$ \ckkbby\ per unit \bb\ mass.  

During the phase I of the experiment, KamLAND-Zen will use 389 kg of xenon \cite{Kozlov:taup2011}, enriched to 0.917 \cite{Koga:ichep2010} in the isotope \XE, for a \XE\ total mass of 357 \kgbb. Concerning \bbonu\ efficiency, the collaboration will make use of a fiducial volume cut to suppress backgrounds that are external to the mini-balloon, for example due to \URANIUM/\THORIUM\ contaminations in the outer liquid scintillator or the environmental gamma ray background. Depending on the \URANIUM/\THORIUM\ contamination levels of the mini-balloon materials themselves (Nylon film, supporting ropes and pipes), the collaboration is also considering a tighter fiducial volume cut to suppress backgrounds originating from the mini-balloon. In the following, we assume that a fiducial volume definition placed 3~$\sigma_r$ inward with respect to the mini-balloon surface will be used, where $\sigma_r$ is the radial position reconstruction accuracy. Assuming $\sigma_r=\mathrm{12.5~cm}/\sqrt{E\mathrm{(MeV)}}$ \cite{Kozlov:taup2011}, we estimate a corresponding fiducial volume efficiency of 0.61 for the 158 cm radius mini-balloon deployed. No other sources of inefficiency are considered in our estimate. Concerning energy reconstruction, the same performance measured in the previous phase of experiment is expected for KamLAND-Zen, given that the xenon-loaded liquid scintillator within the balloon has the same optical properties (light yield, transparency) as the KamLAND scintillator outside the balloon. An energy resolution of $\sigma_E/E=6.8\%/\sqrt{\textrm{E(MeV)}}$ is assumed \cite{Kozlov:taup2011}, scaling to 250 keV FWHM energy resolution at \Qbb. For the background rate lower limit, we take the latest collaboration's expectations, obtained via simulations. In \cite{Kozlov:taup2011}, 19.5 background events per year are expected. The three largest background sources are expected to be \bbtnu\ events entering the ROI, \BI\ events from the mini-balloon materials, and (to a lesser extent) $^{10}\textrm{C}$ events produced through cosmic muon spallation. Compared to previous estimates \cite{Koga:ichep2010}, a shorter \bbtnu\ half-life as measured in \cite{Ackerman:2011gz} is assumed ($T_{1/2}\simeq 2\times 10^{21}$ years), together with a less radiopure mini-balloon (\URANIUM/\THORIUM\ concentrations of $3\times 10^{-12}$ g/g). This estimate results in a background rate of $c=0.22\times 10^{-3}$ \ckkbby. Factors that may potentially result in a higher-than-expected background rate are a non-perfect knowledge of the reconstructed energy spectrum of \bbtnu\ events spilling over the ROI, a higher background contribution from mini-balloon materials, and a worse tagging of \BI\ and $^{10}\textrm{C}$ backgrounds (estimated tagging efficiencies of 66\% and 90\%, respectively). It is, however, rather difficult to quantitatively estimate what a ``pessimistic'' background rate might be observed, at this stage. As in the SNO+ discussion below, we assume (to a large extent in a arbitrary fashion) a 8 times higher-than-expected background rate as upper limit. We note that more information to revisit the KamLAND-Zen background model should become available soon, given that KamLAND-Zen data-taking has started.

The MAJORANA demonstrator will contain 40 kg of germanium, of which at least 20 kg and up to 30 kg will be enriched to 86\% in \GE\ \cite{Schubert:2011nm}. We follow the collaboration's baseline and assume 20 kg of enriched germanium \cite{Wilkerson:taup2011}, corresponding to 17.2 \kgbb\ of \GE. A total \bbonu\ efficiency of 71\% is estimated in \cite{majorana_precdr}, accounting for detector granularity, pulse-shape analysis (PSA), single-site time correlation (SSTC) and energy cuts. A 5\% loss due to edge effects and lost bremsstrahlung is also included in this number. We divide by the quoted 84\% efficiency of the energy cut alone to estimate a 85\% efficiency in tab.~\ref{tab:parameters}. For energy reconstruction, the 2 keV FWHM resolution measured at \Qbb\ in BEGe detectors \cite{Agostini:2010ke} is assumed, as for GERDA-2. For the background rate, we assume as lower limit the number quoted by the collaboration: 0.004 counts per year and kg in a 4 keV-wide ROI \cite{Schubert:2011nm}, corresponding to $c=1.2\times 10^{-3}$ \ckkbby. Main backgrounds responsible for this rate are expected to be prompt cosmogenics, and \TL/\BI\ contaminants in the cryostat and in the copper/lead shield \cite{Wilkerson:snolab}. A non-negligible contribution from $^{68}$Ge contaminants in the enriched germanium crystals is also expected. The material purity specifications are extremely challenging, with contaminant goals at the $<0.1 (<0.3)\ \mu\mathrm{Bq/kg}$ level in \TL\ (\BI) for the electroformed copper to be used for detector mounts, cryostat and the inner copper shield, and about one order of magnitude worse for the commercial high-purity copper and lead to be used for the outer shield \cite{majorana_precdr,Wilkerson:snolab}. Purities within one order of magnitude of such goals have already been demonstrated, see \cite{majorana_precdr}. We therefore assume a factor of 10 worse background rate than nominal as upper limit: $c=12\times 10^{-3}$ \ckkbby.

The SNO+ experiment will make use of 780 tonnes of liquid scintillator \cite{O'Keeffe:2011xi}, with natural neodymium loading at the 0.1\% (w/w) \cite{Wright_PhDthesis}. Given the 5.6\% isotopic abundance of \ND, this concentration corresponds to 44 \kgbb\ of \bb\ emitter mass. The experiment is expected to have a fiducial volume cut rejecting about 50\% of the active volume \cite{Wright_PhDthesis}. No other sources of inefficiency are considered in our estimate. For energy reconstruction, we assume a light output of 400 NHit/MeV \cite{Wright_PhDthesis}, or 1347 NHit at \Qbb, where NHit is the number of PMT hits. This number is highly sensitive to the amount of neodymium concentration, and corresponds to 0.1\% (w/w) loading. Such a light output implies a FWHM resolution of about 220 keV at \Qbb. For the background rate, of the order of 100 background events per kton of liquid scintillator and per year are expected via simulations in a 200 keV energy window around \Qbb\ \cite{Wright_PhDthesis}. This background level translates into a rate of $c=0.009$ \ckkbby, which we take as lower limit. The three dominant backgrounds are expected to be $^8\textrm{B}$ solar neutrinos, \TL\ decays and \bbtnu\ events. The background level due to $^8\textrm{B}$ solar neutrinos is well-known. The \TL\ background assumes a level of \THORIUM\ impurities at the level measured by the BOREXINO experiment, $8.3\times 10^{-18}$ g/g. The collaboration expects energy reconstruction systematic effects, such as non-gaussian resolution tails, on the shape of the \bbtnu\ energy spectrum near the ROI to affect the sensitivity the most, see for example the study in \cite{Wright_PhDthesis}. Preliminary studies conducted by the collaboration indicate that a \mbb\ sensitivity within a factor of $5/3$ worse than the purely statistical sensitivity can be preserved including this systematic effect \cite{Chen:DBD07}. In the large background approximation, which is valid for the SNO+ experiment, this systematic effect would therefore be equivalent to a background rate increase of up to a factor of $(5/3)^4\simeq 8$, resulting in a background rate upper limit of $c=0.07$ \ckkbby.

The NEXT detector will contain 99.14 kg of pressurized xenon gas in the TPC fiducial volume region, enriched to 0.90 in the isotope \XE\ \cite{Gomez:2011my}, corresponding to 89.2 \kgbb\ in \bb\ mass. A \bbonu\ efficiency of 0.25 has been estimated via simulations in \cite{Gomez:2011my}, accounting for inefficiencies in reconstructing the \bb\ track, and in imposing energy and topology cuts to suppress backgrounds. Given that the inefficiency introduced by the energy within ROI requirement is separately accounted for in our analysis, we assume an efficiency of $\varepsilon =0.25/0.76=0.33$ in tab.~\ref{tab:parameters}. Preliminary energy reconstruction measurements obtained with a kg-scale prototype yield a 4.6\% FWHM resolution at the 59.4 keV full energy peak for a $^{241}$Am calibration source \cite{Gomez:2011my}. This energy resolution extrapolates to 0.72\% FWHM resolution at \Qbb, or about 18 keV. As lower limit on the background rate, we take the collaboration's estimate of $0.2\times 10^{-3}$ \ckkbby, dominated by the \URANIUM/\THORIUM\ contamination of the titanium pressure vessel, conservatively assumed to be at the level of $200\ \mu\text{Bq/kg}$ for each isotope, see \cite{Gomez:2011my}. Such radiopurity assumptions are based upon the measured upper limits for the clean titanium used in the LUX cryostat \cite{Hall:2010zzlux}. As upper limit for the background rate, we take an independent (and more pessimistic) measurement of titanium radiopurity, at the level of $1.2\pm 0.4$ ($0.6\pm 0.3$) mBq/kg for \URANIUM\ (\THORIUM) using the Gator low-background counting facility at LNGS \cite{Baudis:taup2011,Baudis:2011am}. Contaminants in these amounts would translate into a background rate of about $10^{-3}$ \ckkbby\ at \Qbb. Also, the collaboration has started its own radiopurity R\&D campaign at LSC, with the goal of refining these assumptions in the near future.

For SuperNEMO, we assume a 7 kg mass in the isotope \SE, to be installed in the demonstrator module \cite{Shitov:2010nt}. A \bbonu\ efficiency of 0.28 has been estimated for a SuperNEMO module in a detailed study \cite{Novella_PhDthesis}, accounting for acceptance, reconstruction efficiency and event selection efficiency. We assume this number in our estimates, which is in fact quite similar to the collaboration's goal of $\varepsilon =0.30$ \cite{Freshville:2011zz}. Calorimeter R\&D efforts achieved a $7.7\%/\sqrt{\textrm{E(MeV)}}$ FWHM energy resolution using a PVT scintillator directly coupled to a 8 inch, high QE, Hamamatsu PMT, see \cite{Freshville:2011zz}. This energy resolution measurement extrapolates to 4.4\% (or about 130 keV) FWHM at \Qbb. For the background rate, we take $c= 6\times 10^{-3}$ \ckkbby\ as worst-case scenario. This number comes from the preliminary result on the \bbonu\ search in \SE\ with the NEMO-3 detector \cite{Simard:taup2011}: 14 events were observed (in agreement with background expectations) in the [2.6--3.2] MeV energy ROI, after 4.5 years of data-taking and 0.93 \kgbb\ of source foil. The backgrounds are dominated by \BI/\TL\ contamination of the foils (measured to be $530\pm 180\ \mu\mathrm{Bq/kg}$ and $340\pm 50\ \mu\mathrm{Bq/kg}$ in \SE, respectively, see \cite{Argyriades:2009vq}) and radon concentration in the tracking volume ($6.46\pm 0.02\ \mathrm{mBq/m}^3$ for the phase-2 of the experiment, see \cite{Argyriades:2009vq}). As background rate lower limit, we assume a factor of 10 improvement in both \TL/\BI\ radiopurity of the foils and in radon concentration in the tracker: $c=0.6\times 10^{-3}$ \ckkbby.

%% file: src/conclusions.tex
Double beta decay is a rare nuclear transition in which a nucleus with $Z$ protons decays into a nucleus with $Z+2$ protons and the same mass number $A$. The standard double beta decay mode, the two-neutrino mode \bbtnu, is a second-order weak transition producing two electrons and two antineutrinos, and is allowed in the Standard Model. Despite being a very slow process, it has been observed in a variety of even-even nuclei. The neutrinoless double beta decay mode \bbonu\ is instead a hypothetical process producing two electrons and no neutrinos. While several other processes have been investigated, \bbonu\ is the most promising probe we have in hand to test lepton number violation. Also, a positive result in the search for \bbonu\ would unavoidably indicate that neutrinos are Majorana particles, that is truly neutral particles.  

\begin{figure}[t!b!]
\begin{center}
\includegraphics[width=0.70\textwidth]{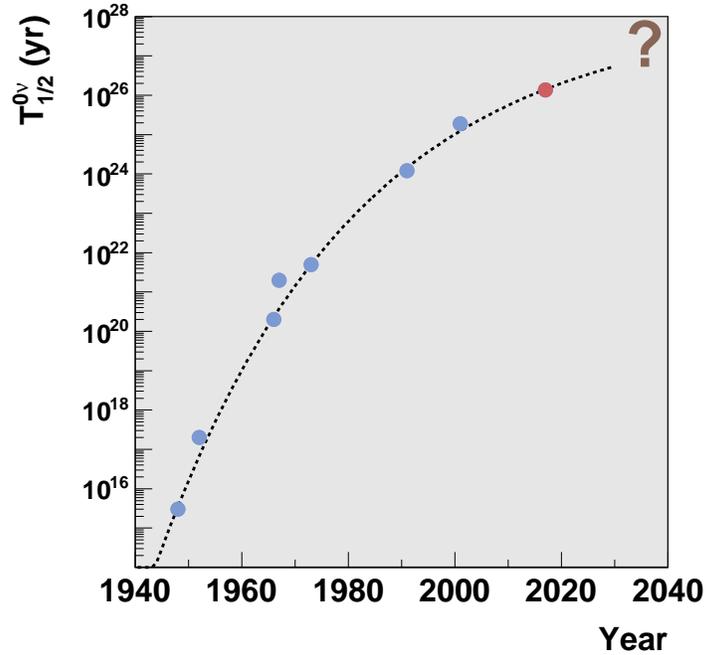}
\end{center}
\caption{Seventy years of direct \bbonu\ searches in perspective. Existing limits (shown in blue) are taken from \cite{Barabash:2011mf}. The sensitivity of new-generation proposals (shown in red) is based upon this review, see sect.~\ref{sec:experiments}.} \label{fig:bb0nusearches}
\end{figure}

The experimental exploration of \bbonu\ has a long history, see fig.~\ref{fig:bb0nusearches}. The first search was performed in 1948, where the half-life for \bbonu\ in $^{124}\text{Sn}$ was constrained to be longer than $3\times 10^{15}$ years \cite{Barabash:2011mf}. Half a century later, in 2001, the Heidelberg-Moscow Collaboration reported a half-life limit about ten orders of magnitude more stringent, using \GE\ as \bb\ emitter: $T_{1/2}^{0\nu}>1.9\times 10^{25}$ years \cite{KlapdorKleingrothaus:2000sn}. It is, also, a history plagued by frequent claimed discoveries (see, for example, \cite{discoveryclaims}) that have been later disproved by subsequent experiments. This observation alone reflects how difficult it is to search for \bbonu.

There is at present a diverse and healthy competition among a variety of experimental techniques to establish themselves as the best approach for \bbonu\ searches. The \bbonu\ field is now witnessing a \emph{golden age} in terms of experimental efforts. Why is that? Some reasons have been present all along during the era of \bbonu\ exploration:
\begin{itemize}
\item We have a fairly good idea of what to look for. While several mechanisms have been proposed to drive \bbonu, in most of them the two decay electrons are the only light particles emitted, therefore carrying most of the available energy. This can be contrasted with proton decay searches, where it is less clear which decay mode should be the focus of experimental investigation.
\item It is common belief that there is still ample room for improvement with respect to the most sensitive \bbonu\ searches performed to date, as can be guessed by the trend in fig.~\ref{fig:bb0nusearches}.
\end{itemize}
There are, however, additional reasons that are applicable to the present era: 
\begin{itemize}
\item Probably the most important reason has to do with the discovery of neutrino oscillations over the past two decades, implying that neutrinos are massive particles. If one assumes, as it is customarily done, that light Majorana neutrino exchange is the dominant contribution to \bbonu, there is a direct link between a measurable \bbonu\ rate on the one hand, and the absolute scale of neutrino masses scale and neutrino oscillations phenomenology on the other. In this context, one can also study what is the actual value of neutrino masses, and whether the neutrino mass spectrum exhibits some particular features (such as a hierarchical or a quasi-degenerate structure), via \bbonu. 
\item Searching for \bbonu\ is well motivated on theoretical grounds. On the one hand, there is no fundamental reason why total lepton number should be conserved. On the other hand, Majorana neutrinos provide natural explanations for both the smallness of neutrino masses and the baryon asymmetry of the Universe. As a consequence, theoretical prejudice in favor of Majorana neutrinos has gained widespread consensus. 
\item As it is well known, there is a $6\ \sigma$ evidence for \bbonu\ in \GE\ reported by part of the Heidelberg-Moscow Collaboration, $T_{1/2}^{0\nu}=(2.23^{+0.44}_{-0.31})\times 10^{25}$ years \cite{KlapdorKleingrothaus:2006ff}. It is also well known that this claim is highly controversial \cite{Aalseth:2002dt}. Consensus exists that the issue can only be definitely settled by new, and more sensitive, experiments.
\end{itemize}

The mapping of observed \bbonu\ rates into neutrino mass constraints not only requires assuming the standard \bbonu\ interpretation in terms of light Majorana neutrino exchange. It also requires precise nuclear physics knowledge, which can be factorized into the so-called nuclear matrix elements (NMEs). These NMEs cannot be measured, and need to be separately calculated for each \bb\ emitting isotope under consideration. Several calculations exist. While they share common ingredients, calculations differ in their treatment of nuclear structure. We argue that about a 20-30\% NME uncertainty exists for converting rates into neutrino masses. 

In this review, the different experimental aspects affecting \bbonu\ searches were extensively discussed. The requirements are often conflicting, and no new-generation experimental proposal is capable of optimizing all of them. Should we concentrate on approaches offering huge event rates, such as KamLAND-Zen? Are superior energy resolution techniques, such as GERDA or CUORE, the best way to go? Should background suppression focus more on radiopurity control, as in the EXO or KamLAND-Zen cases, or on powerful signal-background discrimination techniques, as in the SuperNEMO or NEXT approaches? We made an attempt at a quantitative comparison of the physics case of selected new-generation experimental approaches, which is summarized in fig.~\ref{fig:sens-pmr}. 

What about the longer-term future? How far can we go in \bbonu\ exploration? Nobody knows for sure. What we do know is that such future \bbonu\ searches will unavoidably need to involve experiments at the ton or multi-ton scale in \bb\ isotope mass. The diversity of experimental approaches we are currently witnessing will not be viable at that scale, and 2 or 3 approaches (most likely based on different isotopes) are going to be retained at most. However, an extrapolation of the trends from the past and the present may offer some qualitative clues. Figure~\ref{fig:bb0nusearches} seems to point to ``asymptotic'' limits of \bbonu\ half-life explorations in the $10^{28}$ years range. If the light Majorana neutrino exchange mechanism is realized in Nature, this would correspond to effective Majorana masses at the few meV scale. As can be appreciated in fig.~\ref{fig:mbetabetavsmlight}, such ultimate \bbonu\ sensitivities would give us good chances to detect \bbonu\ regardless of the value of the neutrino mass and mixing parameters. 

An unambiguous detection, either in the current-generation efforts starting now or in the longer-term future, would open up an even more exciting era for \bbonu\ searches, with the objective to actually understand what is the physics mechanism that is responsible for this elusive process.